# Introducing a model of pairing based on Base pair specific interactions between identical DNA sequences

Dominic J. (O') Lee[1,a.)]

[1]Department of Chemistry, Imperial College London, SW7 2AZ, London, UK

**Abstract**

At present, there have been suggested two types of physical mechanism that may facilitate preferential pairing between DNA molecules, with identical or similar base pair texts, without separation of base pairs. One solely relies on base pair specific patterns of helix distortion being the same on the two molecules, discussed extensively in the past. The other mechanism proposes that there are preferential interactions between base pairs of the same composition. We introduce a model, built on this second mechanism, where both thermal stretching and twisting fluctuations are included, as well as the base pair specific helix distortions. Firstly, we consider an approximation for weak pairing interactions, or short molecules. This yields a dependence of the energy on the square root of the molecular length, which could explain recent experimental data. However, analysis suggests that this approximation is no-longer valid at large DNA lengths.  In a second approximation, for long molecules, we define two adaptation lengths for twisting and stretching, over which the pairing interaction can limit the accumulation of helix disorder. When the pairing interaction is sufficiently strong, both adaptation lengths are finite; however, as we reduce pairing strength, the stretching adaptation length remains finite but the torsional one becomes infinite. This second state persists to arbitrarily weak values of the pairing strength; suggesting that, if the molecules are long enough, the pairing energy scales as length. To probe differences between the two pairing mechanisms, we also construct a model of similar form. However, now, pairing between identical sequences solely relies on the intrinsic helix distortion patterns. Between the two models, we see interesting qualitative differences. We discuss our findings, and suggest new work to distinguish between the two mechanisms.

## 1. Introduction

There is growing experimental evidence that there may be recognition forces between identical, and perhaps homologous, DNA molecules [1,2,3,4,5,6].  In particular, phase segregation was observed between DNA molecules of two different sequence types [2], as well as preferential pairing between identical sequences being observed in single molecule pulling experiments [6]. Such recognition forces might have an important role to play in initiating certain biological processes like homologous recombination [7,8] and the silencing of multiple copies of a gene [9,10], by allowing for homologous DNA molecules to associate.

The idea of such recognition forces is not a new one. It dates back to stem loop kissing model [11,12,13,14], where double helix is unravelled in loops on two molecules,  which recognize each other through the complementarity of base pairs at the loop ends. Though, now, it seems that such

---

[a.)] Electronic Mail: domolee@hotmail.com

a mechanism is rather unfeasible due to the required energetics for such a process to occur. Later, a mean field electrostatic model of DNA-DNA interactions was proposed [15,16], which led to the proposal of an alternate mechanism for DNA-DNA recognition [17]. This idea was based around the observation that DNA does not form an ideal helix, rather the helix is distorted due to imperfect stacking of the base pairs. This pattern of helix distortion depends on sequence [18,19,20]. Two identical sequences have the same pattern of distortions, while for two un-alike sequences the pattern of distortions is different. In the electrostatic model, when two rigid charge helices are commensurate with each other, so that phosphate charges on one molecule match the minor or major grooves of other molecule– where positively charged counterions also may bind or condense– the electrostatic energy is lowered. On the other hand, when the patterns of distortions are different for each helix, this commensurability is no longer preserved; and thus the electrostatic energy is higher. When DNA elasticity is taken into account [21], as well as thermal fluctuations [22,23,24,25], there can still be a sizable difference in electrostatic energy for non-alike base-pair texts compared with identical ones. The upshot is that similarities in the patterns of distortions may result in a pairing mechanism between identical base pair texts. This pairing mechanism does not exclusively rely on mean-field electrostatics; any interaction model that depends structurally on the helical shape of the molecule, but otherwise insensitive base pair sequence, will also result in the same mechanism.

However, recently, models have been proposed where recognition forces depend directly on base pair content through microscopic interactions [6,26,27,28]. In the work of [26], it has been speculated that identical base pairs can interact with each other through the formation of secondary hydrogen bonding between two DNA molecules. Another possible mechanism may be differences in van der Waals, or dispersion, forces between base pairs, which could lead to preferential pairing [27]. Also, in Ref. [28], it was suggested, due to the presence of small quantities of spermine, that there could be significantly more attraction between AT sequences on two molecules, as opposed to un-methylated GC ones, due to differences in where the spermine molecules position themselves. Lastly, in Ref. [6], an interaction mechanism was proposed that relies on the formation of ionic cages between molecules. This mechanism relies on the fact that monovalent counterions localize preferentially in the minor groove of AT sequences and in the major groove of GC sequences [29], and the possibility that such cages may form [30]. The idea is that there could be preferential pairing at the two minor grooves of AT sequences, on both molecules, facilitated by a counterion preferentially localized between them- a cage- and/or preferential pairing between the major grooves of GC sequences.

In this study, we consider such sequence dependent pairing mechanisms, and formulate an alternate large scale model to capture the difference in physics from the global mechanism of pairing. The key element in this formulation is phenomenological interaction potentials between base pairs which are base pair dependent. In the simplest case, one can consider relative interaction potentials between base pairs that vary only significantly from non-zero when molecules of the identical base pair text positioned in perfect alignment with each other. Already, a model that considers base pair specific pairing sites, spaced 3.4nm apart, along a thin rods has already been introduced [31]. This model also takes account of accidental base pair matches, which occur randomly in non-identical base pair texts. In Ref. [31] the kinetics of this pairing model was studied; where, amongst other attractive features, the model had the ability to sharply distinguish kinetically between the accidental base pair matches and homologous sequences. We follow on from this work by including the helical geometry of the molecule and pairing sites per base pair. Here, however, we will be solely interested in the equilibrium statistical mechanics. Indeed, the model we present could be extended to a dynamical one to study kinetics. Here, we neglect statistical matches between like base-pairs as a second order correction. As in previous studies [17,21,22,32], we build in both thermal and intrinsic base dependent distortions of the DNA helices. Also, for comparison purposes, we formulate a model based on the same interaction potentials, but now each base pair on one molecule is allowed to pair with any base pair on the other molecule, not just the same base pair. In this case, preferential pairing between identical base pair sequences relies on both molecules having the same pattern of helix distortion. The goal here is to find qualitative features that distinguish between the two mechanisms, which could form the basis for experiments.

In the next section, we'll introduce basic features of the model for local base specific pairing. In the first part, we'll start by considering rigid helices. Here, we'll write down the basic model assumptions and two phenomenological potentials that we use; we have chosen two different ones to show that qualitative features of the model are not sensitive to a particular choice of potential. For rigid molecules, we neglect thermal twisting and stretching fluctuations, but we do include intrinsic distortions due to structural misalignment of base-pairs. Next, in the second part, we consider thermal fluctuations for relatively short sequences, where the pairing interaction is considered not sufficiently strong enough to supress thermal fluctuations. What is interesting, here, is a dependence of the pairing free energy between identical sequences on the square root of the length of the molecule; such dependence, if it could persist over large enough lengths, could fit the experimental data of Ref. [6]. In the third part, we consider the case where the interactions are can be considered sufficiently strong to supress thermal fluctuation over large length scales. These length scales are set by two adaptation lengths, one for twisting fluctuations and the other for stretching. We find that, for long molecules, there are two states. For weak interactions, we always find that we always have a finite adaptation length for stretching fluctuations. Above this length scale, the pairing free energy scales again with the length of the molecule. However, there is no adaptation for twisting fluctuations. When interactions become sufficiently strong we enter a state where there is, now, a finite adaption length for the twisting fluctuations. Finally, in the last part of the model section, we build a local pairing model with the same phenomenological model to probe qualitative differences between the two mechanisms.

In the discussion and outlook section, we discuss our findings, especially in the context of Ref. [6], and suggest experiments that could distinguish between the two competing mechanisms. We also discuss what further theoretical developments are needed.

## 2. The Model

### 2.1 Introducing the model for Rigid Helices

We concern ourselves with a pairing model between DNA helices where each interaction between base pairs depends specifically on the base pair type. Let us consider two DNA molecules that share the same base pair text. The two DNA molecules may be modelled as helices (labelled 1 and 2), each made up of interaction sites that lie on each base pair. A site on one helix interacts with another site on the other helix through an interaction potential $V_{\text{int}}$, which depends on the distance between the two sites. We suppose that there is only one interaction site per base pair, though the model could be easily extended to consider multiple sites. The two helices are chosen to lie parallel to each other and in perfect parallel juxtaposition, so that both base pair sequences run the same direction and lie commensurate with each other. The centres of the two helices are then separated a position vector, that can be chosen to lie of the x-axis such that $\mathbf{R} = R\hat{\mathbf{i}}$. Then, we may write for such a model of DNA pairing

$$V_{local}(\mathbf{R}) = \sum_{j=-N}^{N} V_{int}\left(\mathbf{r}_{1,j} - \mathbf{r}_{2,j} - \mathbf{R}\right), \quad (2.1)$$

where $V_{int}\left(\mathbf{r}_{1,j} - \mathbf{r}_{2,j} - \mathbf{R}\right)$ is the interaction potential between two interaction sites (labelled $j$) lying at position vectors $\mathbf{r}_{1,j}$ and $\mathbf{r}_{2,j} + \mathbf{R}$. Here, $\mathbf{r}_{1,j}$ and $\mathbf{r}_{2,j}$ are the position vectors of the interaction sites measured from the centre of each helix (see Fig. 1). We suppose that there are $2N+1$ base pairs making up each helix. In writing Eq. (2.1), we have assumed that each base pair can only interact with only the base pair at the same position within the same sequence. It could be conceivable to have favourable interactions due to accidental matches of base pairs [31], but we neglect this as a small contribution to Eq. (2.1). The model(s) described by Eq. (2.1) we call locally base pair specific, where the interaction sites interact in such away, as opposed to a global helix distortion model where all each interaction sites on one helix are allowed to interact with all of the sites on the other helix. We should point out that, in addition to this pairing interaction, one should consider the electrostatic repulsion between the DNA molecules, where the molecules may be modelled as uniformly charged cylinders.

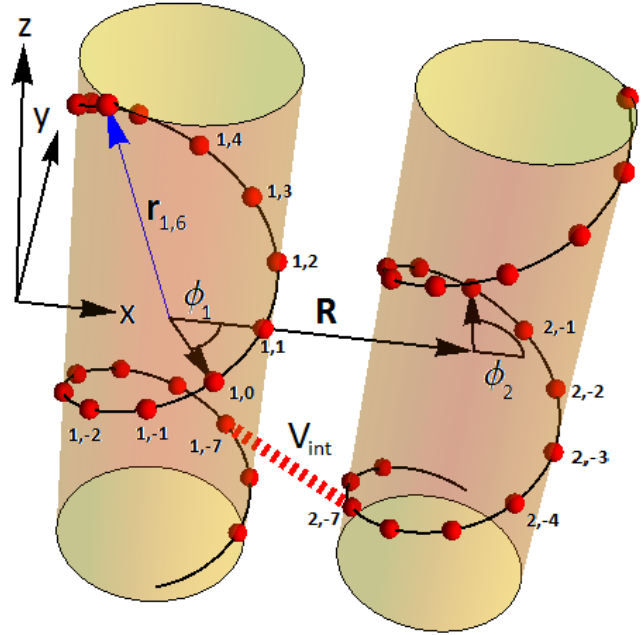

Fig. 1 Schematic picture of the model of local base pair specific pairing between DNA helices. We suppose that there is one interaction site per base pair on each helix these are shown by the small red spheres. Each interaction site is labelled with the index $\{\mu, j\}$, where $\mu$ is the index that refers to the helix the site sits on, and $j$ labels the position of each site on the helix. Only sites with the same value of $j$ are allowed to pair. The position of each site a helix is defined relative its centre through the position vector $\mathbf{r}_{\mu,j}$. As a particular example, the position vector $\mathbf{r}_{1,6}$ to site $\{1,6\}$ is shown. We choose the position vector connecting the two helix centres $\mathbf{R}$ to lie on the x-axis and the principle axes of the two cylinders lie parallel to the z-axis. The azimuthal orientations of the two helices are defined through the angles $\phi_1$ and $\phi_2$ these are measured in the x-y plane from the x-axis to lines connecting the interaction sites $\{1,0\}$ and $\{2,0\}$, respectively. We suppose that one interaction site lies at the midpoint of the principle axis of each helix.

We have considered two possible phenomenological forms for the interaction potential; these are two typical choices of potential. The first we consider is a Debye-Huckel form, for which we write

$$V_{int}(\mathbf{r}_1 - \mathbf{r}_2) = -\frac{\gamma_{eff}}{4\pi|\mathbf{r}_1 - \mathbf{r}_2|}\exp\left(-\frac{|\mathbf{r}_1 - \mathbf{r}_2|}{\lambda_{eff}}\right). \tag{2.2}$$

between two sites at $\mathbf{r}_1$ and $\mathbf{r}_2$. In Eq. (2.2), there are two parameters: $\lambda_{eff}$ is the effective range of the interaction, and $\gamma_{eff}$ is the effective interaction strength. The second type that we consider is a potential of Morse form

$$V_{int}(\mathbf{r}_1 - \mathbf{r}_2) = \frac{\tilde{\gamma}_{eff}}{8\pi}\left(\exp\left(-\frac{2|\mathbf{r}_1 - \mathbf{r}_2|}{\lambda_{eff}}\right) - 2\exp\left(-\frac{|\mathbf{r}_1 - \mathbf{r}_2|}{\lambda_{eff}}\right)\right), \tag{2.3}$$

where, again, we have an effective range $\lambda_{eff}$ and an effective interaction strength $\tilde{\gamma}_{eff}$.

We will start by considering the helices as rigid, before later considering thermal fluctuations. In this case, the position vectors for the interaction sites lying on the two helices, $\mathbf{r}_{1,j}$ and $\mathbf{r}_{2,j}$ may be described by the expression ($\mu = 1, 2$)

$$\mathbf{r}_{\mu,j} = a\cos(ghj + \phi_\mu + \delta\phi_j)\hat{\mathbf{i}} + a\sin(ghj + \phi_\mu + \delta\phi_j)\hat{\mathbf{j}} + (hj + \delta s_j)\hat{\mathbf{k}}. \tag{2.4}$$

Eq. (2.4) describes distorted helices of fixed radius $a$ and mean pitch $H = 2\pi/g$, where $h$ is the average spacing between base pairs. The angles $\phi_\mu$ are the angles that the two helices make at $s = 0$ (the centre of their principle axes) with the line connecting their two centres (see Fig. 1). We can define the azimuthal orientation of the two helices through the difference $\Delta\phi = \phi_1 - \phi_2$. The distortion allows for the fact that the angles and distances between adjacent base pairs are not constant in DNA, but deviate away from average values [18,19,20]. In Eq. (2.4), this distortion is characterized by $\delta\phi_j$ and $\delta s_j$, accounting for deviations in the angles and vertical distances between base pairs, respectively. Both $\delta\phi_j$ and $\delta s_j$ can be written as

$$\delta\phi_j = \sum_{j'=\text{sgn}(j)}^{j}\delta\Omega_{j'}, \qquad \delta s_j = \sum_{j'=\text{sgn}(j)}^{j}\delta h_{j'}, \tag{2.5}$$

where we are free to choose $\delta\phi_0 = \delta s_0 = 0$. Here, $\delta\Omega_j$ is the deviation in the twist angle between two adjacent base pairs $j$ and $j-1$ away from its average value of $gh$, and $\delta h_j$ is deviation in vertical distance between the same two base pairs from the average value $h$. For perfect helices, we would have that $\delta\Omega_j = \delta h_j = 0$, for all values of $j$. For DNA, the particular values that $\delta\Omega_j$ and $\delta h_j$ take depend on the base pair sequence [18,19,20]. Here, we will deal with these base pair dependant variations by considering an ensemble average of the interaction energy between the two helices over all base pair realizations, namely $\langle V_{helix}(\mathbf{R}) \rangle_\Omega$. We assume that $\delta\Omega_j$ and $\delta h_j$ are Gaussian distributed and uncorrelated so that (for a justification of this see Ref. [20])

$$\langle \delta\Omega_j \delta\Omega_k \rangle_\Omega = \frac{h}{\lambda_{tw}^{(0)}} \delta_{j,k} \quad \text{and} \quad \langle \delta h_j \delta h_k \rangle_\Omega = \frac{h}{g^2 \lambda_{st}^{(0)}} \delta_{j,k}, \tag{2.6}$$

where the subscript $\Omega$ on the averaging bracket denotes that this is the ensemble average, not a thermal one. Here, $\lambda_{tw}^{(0)}$ and $\lambda_{st}^{(0)}$ are measures of helix non-ideality; they are, essentially, the distances over which azimuthal and axial orientations of the base pairs deviate significantly from their positions in forming an ideal helix, respectively, when $\delta\phi_j$ and $\delta s_j$ are allowed to accumulate freely.

To analytically perform the ensemble average over base pairs and obtain a useful form for the interaction energy, it is first useful to perform some mathematical manipulations. Using $\tilde{V}_{int}(\mathbf{k})$, the Fourier transform of $V_{int}(\mathbf{r}_1 - \mathbf{r}_2)$, we can write Eq. (2.1) so that

$$\langle V_{local}(\mathbf{R}) \rangle_\Omega = \frac{1}{(2\pi)^3} \sum_{j=-N}^{N} \int d^3k \left\langle \exp\left(i\mathbf{k}\left(\mathbf{r}_{1,j} - \mathbf{r}_{2,j} - \mathbf{R}\right)\right) \right\rangle_\Omega \tilde{V}_{int}(\mathbf{k}). \tag{2.7}$$

The Fourier transforms of both the Debye Huckle and Morse potentials are given by

$$\tilde{V}_{int}^{DH}(\mathbf{k}) = -\frac{\gamma_{eff}}{\mathbf{k}^2 + \kappa_{eff}^2} \tag{2.8}$$

and

$$\tilde{V}_{int}^{M}(\mathbf{k}) = -\tilde{\gamma}_{eff} \frac{\partial}{\partial \kappa_{eff}} \left[ \frac{1}{4} \frac{1}{\mathbf{k}^2 + 4\kappa_{eff}^2} - \frac{1}{\mathbf{k}^2 + \kappa_{eff}^2} \right], \tag{2.9}$$

where $\kappa_{eff} = 1/\lambda_{eff}$.

In this initial study, we will make (as was initially done in the helix distortion theory of Ref. [15,16]) make a continuum approximation. In such an approximation, we replace the sum over base pairs with an integral over length $s$, and so write

$$\langle V_{local}(\mathbf{R}) \rangle_\Omega \approx \frac{1}{(2\pi)^3 h} \int_{-L/2}^{L/2} ds \int d^3k \left\langle \exp\left(i\mathbf{k}\left(\mathbf{r}_1(s) - \mathbf{r}_2(s) - \mathbf{R}\right)\right) \right\rangle_\Omega \tilde{V}_{int}(\mathbf{k}), \tag{2.10}$$

with the length of the molecules being $L$. The position vectors of the helices are now given through

$$\mathbf{r}_\mu(s) = a\cos\left(gs + \phi_\mu + \delta\phi(s)\right)\hat{\mathbf{i}} + a\sin\left(gs + \phi_\mu + \delta\phi(s)\right)\hat{\mathbf{j}} + (s + \delta s(s))\hat{\mathbf{k}}, \qquad (2.11)$$

with

$$\delta\phi(s) = \frac{1}{h}\int_0^s ds'\,\delta\Omega(s'), \qquad \delta s(s) = \frac{1}{h}\int_0^s ds'\,\delta h(s'). \qquad (2.12)$$

Here, $\delta\Omega(s)$ and $\delta h(s)$ are continuum representations of $\delta\Omega_j$ and $\delta h_j$, respectively. The continuum approximation neglects Fourier modes in k-space due to the discreteness of base pairs. For flexible helices (considered later), sufficiently large thermal fluctuations, as well as including the base pair dependent helix distortions– these matter significantly for modes due to the discreteness of base pairs– the contribution from discrete modes is expected to be negligible over a large range of interaction strengths. Also, considering the interaction sites as discrete rather complicates the analysis, for a first time study. However, if the interaction potentials were sufficiently strong enough these modes may become more significant. Therefore, in later work, we should investigate above what interaction strength discrete modes might have to be included, and what effect they might have on the results presented here.

Following a series of mathematical steps (see Appendix A of supplemental material) we can recast Eq.(2.10) into the form

$$\langle V_{local}(R)\rangle_\Omega = -\Lambda \int_{-L/2}^{L/2} ds \sum_{n=-\infty}^{\infty} \sum_{n'=-\infty}^{\infty} \exp\left(i(n-n')gs\right)\exp\left(i(\phi_1 n - \phi_2 n')\right)\left\langle \exp\left(i\delta\phi(s)(n-n')\right)\right\rangle_\Omega$$
$$\int_{-\infty}^{\infty} dk_z (-1)^{n'} G_{n,n'}(R/a, a\kappa_{eff}, k_z). \qquad (2.13)$$

Here, $\Lambda$ is the effective strength of the interactions between the two helices that is either $\Lambda = \gamma_{eff}/ha(2\pi)^2$ or $\Lambda = \tilde{\gamma}_{eff}/ha(2\pi)^2$, for Debye-Huckle and Morse forms respectively. The specific form of $G_{n,n'}(R/a, a\kappa_{eff})$ depends on the form of the interaction potential used. For the Debye Huckel form, it is given by

$$G^{DH}_{n,n'}(R/a, a\kappa_{eff}, k_z) = I_n\left(\sqrt{k_z^2 + \kappa_{eff}^2 a^2}\right) I_{n'}\left(\sqrt{k_z^2 + \kappa_{eff}^2 a^2}\right) K_{n-n'}\left(\frac{R}{a}\sqrt{k_z^2 + \kappa_{eff}^2 a^2}\right). \qquad (2.14)$$

For the More potential, the form of $G_{n,n'}$ is more complicated and is given by

$$G^M_{n,n'}(R/a, a\kappa_{eff}, k_z) = a^2 \kappa_{eff}\left(H_{n,n'}(R/a, a\kappa_{eff}, k_z) - H_{n,n'}(R/a, 2a\kappa_{eff}, k_z)\right), \qquad (2.15)$$

where

$$H_{n,n'}(R/a, a\kappa_{eff}, k_z) = -\frac{1}{\sqrt{k_z^2 + \kappa_{eff}^2 a^2}} \Bigg[ \Big[ I_n'\left(\sqrt{k_z^2 + \kappa_{eff}^2 a^2}\right) I_{n'}\left(\sqrt{k_z^2 + \kappa_{eff}^2 a^2}\right)$$

$$+ I_n\left(\sqrt{k_z^2 + \kappa_{eff}^2 a^2}\right) I_{n'}'\left(\sqrt{k_z^2 + \kappa_{eff}^2 a^2}\right) \Big] K_{n-n'}\left(\frac{R}{a}\sqrt{k_z^2 + \kappa_{eff}^2 a^2}\right) \quad (2.16)$$

$$+ \frac{R}{a} I_n\left(\sqrt{k_z^2 + \kappa_{eff}^2 a^2}\right) I_{n'}\left(\sqrt{k_z^2 + \kappa_{eff}^2 a^2}\right) K_{n-n'}'\left(\frac{R}{a}\sqrt{k_z^2 + \kappa_{eff}^2 a^2}\right) \Bigg].$$

Note here that we have rescaled $k_z$ by $a$, for it to be dimensionless. It is also worth pointing out that the form of Eq. (2.13) can be utilized for any interaction potential that can be expressed in terms of a Laplace transform, i.e. as a sum of decaying exponentials (for details and a generalized expression for $G_{n,n'}$ see Appendix A of the supplemental material). Using the continuum version of Eq. (2.6), we can evaluate the average in Eq. (2.13) and perform the integration over $s$. We find that, with $\lambda_{tw}^{(0)} g \gg 1$ and $Lg \gg 1$ (see Appendix A of Supplemental Material), Eq. (2.13) is completely dominated by diagonal $n=n'$ modes that don't depend on $\delta\phi(s)$, so that $\langle V_{local}(R) \rangle_\Omega \simeq V_{local}(R)$. This result is not that surprising: the base pairs on two helices with a shared pattern of distortion should remain commentate with each other in both vertical and azimuthal directions, so the strength of the interaction should not be much affected much by helix distortions. Note that if we would go on to consider arrays of homology and non-homology, as was considered in Ref. [10], this expectation that $\langle V_{local}(R) \rangle_\Omega \simeq V_{local}(R)$ may not, in fact, be satisfied.

Considering only the dominant $n=n'$ modes, we can write now write down the interaction in its final form for two identical rigid helices

$$V_{local}(R) \simeq -\Lambda L \sum_{n=-\infty}^{\infty} \exp(in(\phi_1 - \phi_2)) \int_{-\infty}^{\infty} dk_z (-1)^n G_{n,n}(R/a, a\kappa_{eff}, k_z). \quad (2.17)$$

In Eq. (2.17), it is a very good approximation to truncate all sums to $|n| \leq 4$.

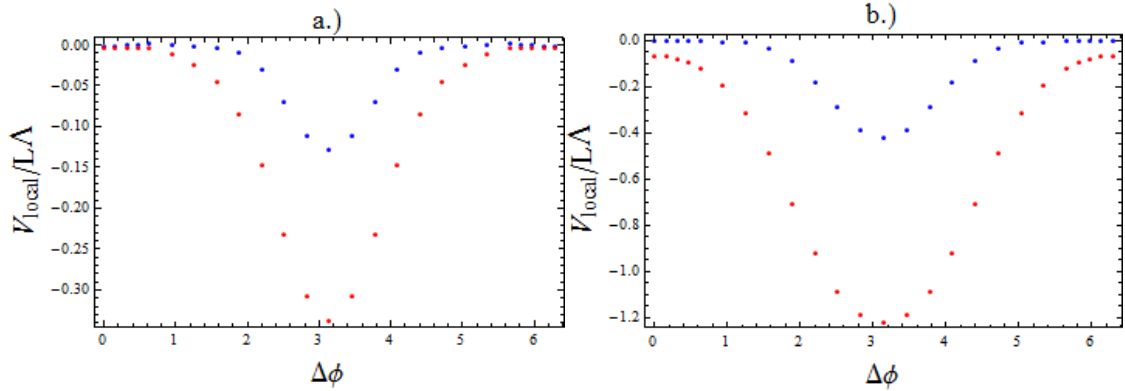

Fig. 2. The azimuthal dependence of $V_{local}(R)/L\Lambda$. We fix $R = 25\text{Å}$ and investigate $V_{local}(R)/L\Lambda$ given by Eq. (2.17) as a function of $\Delta\phi = \phi_1 - \phi_2$, where we have truncated the sum to $|n| \leq 4$, which is shown to be a good approximation for the cases considered. In panel a.) we show $V_{local}(R)/L\Lambda$ calculated using the Debye-Huckel form for $G_{n,n}$ given by Eq. (2.14), whereas in b.) we use the form for the Morse potential given by Eq. (2.15). For the red points we use an effective decay length of $\lambda_{eff} = 4\text{Å}$ and for the blue points we use a decay length of $\lambda_{eff} = 2\text{Å}$.

In Fig.2 we investigate the azimuthal dependence of the interaction described by Eq. (2.17). We see that the interaction is most attractive, when $\Delta\phi = \phi_1 - \phi_2 = \pi$. Again, this result is not that surprising, as this orientation allows interaction sites, at each helical pitch, to face each other. The decay range in the interactions very much influences for what values of $\Delta\phi$ the interaction is effectively negligible and the width of the well about $\Delta\phi = \pi$. As we reduce the decay range, within a narrower region around $\Delta\phi = \pi$ is the pairing interaction significant. The choice of potential also influences the width of this well of attraction. Generally, using the Morse potential gives broader potential wells about $\Delta\phi = \pi$ than using the Debye-Huckel form. This can be explained by the fact that the Debye-Huckel potential effectively has a shorter range due to the $1/|\mathbf{r}_1 - \mathbf{r}_2|$ factor that stands in Eq. (2.2).

## 2.2 Thermal fluctuations for short identical base pair tracts in perfect juxtaposition

We now consider the helices to be flexible. This now allows for thermal fluctuations in both twisting and stretching. First of all, to describe a thermally fluctuating helix, we modify Eq. (2.11) to

$$\mathbf{r}_\mu(s) = a\cos\left(gs + \phi_1 + \delta\phi(s) + \delta\phi_{\mu,T}(s)\right)\hat{\mathbf{i}} + a\sin\left(gs + \phi_1 + \delta\phi(s) + \delta\phi_{\mu,T}(s)\right)\hat{\mathbf{j}} + (s + \delta s(s) + \delta s_{\mu,T}(s))\hat{\mathbf{k}},$$
(2.18)

where $\delta\phi_{\mu,T}(s)$ and $\delta s_{\mu,T}(s)$ are thermal twisting and stretching distortions for each of the two helices. Here, in the main text, we again consider only the dominant $n = n'$ modes (see Appendix B of supplemental material for justification of this). Eq. (2.18) allows us now to write a functional for $V_{local}$

$$V_{local}[\delta\phi_{1,T}(s), \delta\phi_{2,T}(s), \delta s_{1,T}(s), \delta s_{2,T}(s)] \approx -\Lambda \int_{-L/2}^{L/2} ds \sum_{n=-\infty}^{\infty} \exp(i(n-n')gs)\exp(in(\phi_1 - \phi_2))$$
$$\exp\left(in\left(\delta\phi_{1,T}(s) - \delta\phi_{2,T}(s)\right)\right) \int_{-\infty}^{\infty} dk_z (-1)^n G_{n,n}(R/a, a\kappa_{eff}, k_z) \exp\left(ik_z\left(\delta s_{1,T}(s) - \delta s_{2,T}(s)\right)\right).$$
(2.19)

Allowing for the helices to be flexible, we must now include elastic energy terms for both stretching and twisting. In the elastic rod model, these energy functional contributions are given by

$$E_{tw}[\delta\phi_{1,T}(s), \delta\phi_{2,T}(s)] = \frac{k_B T l_{tw}}{2} \int_{-L/2}^{L/2} ds \left[\left(\frac{d\delta\phi_{1,T}(s)}{ds}\right)^2 + \left(\frac{d\delta\phi_{2,T}(s)}{ds}\right)^2\right],$$
(2.20)

and

$$E_{st}[\delta s_{1,T}(s), \delta s_{2,T}(s)] = \frac{k_B T l_{st}}{2} \int_{-L/2}^{L/2} ds \left[ \left( \frac{d\delta s_{1,T}(s)}{ds} \right)^2 + \left( \frac{d\delta s_{2,T}(s)}{ds} \right)^2 \right], \tag{2.21}$$

where $l_{tw}$ and $l_{st}$ are the thermal twisting and stretching persistence lengths, that measure the DNA stiffness in regards to twisting and stretching, respectively. We take the values $l_{tw} = 1000 \text{Å}$ and $l_{st} = 700 \text{Å}$ taken from Refs. [33] and [34].

Thus, we can write the partition function of the two interacting helices as

$$Z = \int D\delta\phi_{1,T}(s) \int D\delta\phi_{2,T}(s) \int D\delta s_{1,T}(s) \int D\delta s_{2,T}(s) \exp\left( -\frac{E_{el}[\delta\phi_{1,T}(s), \delta\phi_{2,T}(s), \delta s_{1,T}(s), \delta s_{2,T}(s)]}{k_B T} \right)$$
$$\exp\left( -\frac{V_{local}[\delta\phi_{1,T}(s), \delta\phi_{2,T}(s), \delta s_{1,T}(s), \delta s_{2,T}(s)]}{k_B T} \right), \tag{2.22}$$

where the total elastic energy functional $E_{el}$ is the sum of elastic stretching and twisting energies, i.e. $E_{el} = E_{tw} + E_{st}$, where both $E_{tw}$ and $E_{st}$ are given by Eqs.(2.20) and (2.21). In Eq. (2.22), the functional integrals (denoted by measure $Df(s)$ for functions $f(s)$) sum over all possible realizations of the functions $\delta\phi_{\mu,T}(s)$ and $\delta s_{\mu,T}(s)$, required in writing the partition function. For the unacquainted, functional integration is simply the continuum limit of multiple integrations over the distortions at each site $j$, $\delta\phi_{\mu,j}$ and $\delta h_{\mu,j}$.

Firstly, we will suppose the molecules sufficiently short, or the interactions sufficiently weak, for the influence of $V_{local}$ on the partition function to be weak. Thus, in this case, we make the expansion

$$Z = \sum_{n=0}^{\infty} Z^{(n)}, \tag{2.23}$$

where

$$Z^{(n)} = \frac{(-1)^n}{n!} \int D\delta\phi_{1,T}(s) \int D\delta\phi_{2,T}(s) \int D\delta s_{1,T}(s) \int D\delta s_{2,T}(s)$$
$$\left( \frac{V_{local}[\delta\phi_{1,T}(s), \delta\phi_{2,T}(s), \delta s_{1,T}(s), \delta s_{2,T}(s)]}{k_B T} \right)^n \exp\left( -\frac{E_{el}[\delta\phi_{1,T}(s), \delta\phi_{2,T}(s), \delta s_{1,T}(s), \delta s_{2,T}(s)]}{k_B T} \right).$$
$$\tag{2.24}$$

Here, in the main text, we will simply examine the series (Eq. (2.23)) up to the next to leading order term in the expansion, namely $Z^{(1)}$. In Appendix B of the supplemental material, we will consider what effect including the correction $Z^{(2)}$ has on the free energy. Retaining only $Z^{(0)}$ and $Z^{(1)}$, we may write for the free energy

$$F \approx -k_B T \ln\left(Z^{(0)} + Z^{(1)}\right) \approx -k_B T \ln Z^{(0)} + \left\langle V_{local}[\delta\phi_{1,T}(s), \delta\phi_{2,T}(s), \delta s_{1,T}(s), \delta s_{2,T}(s)] \right\rangle_0, \quad (2.25)$$

where the thermal average of the interaction potential is given by

$$\left\langle V_{local}[\delta\phi_{1,T}(s), \delta\phi_{2,T}(s), \delta s_{1,T}(s), \delta s_{2,T}(s)] \right\rangle_0 = -k_B T Z^{(1)} / Z^{(0)}. \quad (2.26)$$

The first term in Eq. (2.25) is a rather unimportant constant, and can be discarded by subtracting off the free energy at $R = \infty$, $F_0 = -k_B T \ln Z^{(0)}$. Evaluation of the second term in Eq.(2.25) is straightforward (see Appendix B of supplemental material for details) and yields the expression

$$F - F_0 \approx \left\langle V_{local}[\delta\phi_{1,T}(s), \delta\phi_{2,T}(s), \delta s_{1,T}(s), \delta s_{2,T}(s)] \right\rangle_0$$

$$\approx -2\Lambda l_{tw} \sum_{n=-\infty}^{\infty} (-1)^n \exp\left(in(\phi_1 - \phi_2)\right) \int_{-\infty}^{\infty} dk_z \frac{G_{n,n}(R/a, a\kappa_{eff}, k_z)}{n^2 + \frac{k_z^2}{a^2 g^2}\frac{l_{tw}}{l_{st}}} \left(1 - \exp\left(-\frac{L}{2l_{tw}}\left(n^2 + \frac{k_z^2}{a^2 g^2}\frac{l_{tw}}{l_{st}}\right)\right)\right).$$

(2.27)

It is instructive to look at limiting cases of Eq. (2.27). In the case where $L$ is small we simply get back the result Eq. (2.17); thermal fluctuations in the twisting and stretching have not yet accumulated enough to effect $V_{local}$. This limiting case is a valid approximation when $L \ll l_{tw}, l_{st}$. The large $L$ limit ($L \gg l_{tw}, l_{st}$) is more interesting. In this limit, it is useful to separate the $n = 0$ mode, and so start by writing for large $L$

$$\left\langle V_{local}[\delta\phi_{1,T}(s), \delta\phi_{2,T}(s), \delta s_{1,T}(s), \delta s_{2,T}(s)] \right\rangle_0 \approx E_0 - 4\Lambda l_{tw} \sum_{n=1}^{\infty} (-1)^n \cos\left(n(\phi_1 - \phi_2)\right)$$

$$\int_{-\infty}^{\infty} dk_z \frac{1}{n^2 + \frac{k_z^2}{a^2 g^2}\frac{l_{tw}}{l_{st}}} G_{n,n}(R/a, a\kappa_{eff}, k_z), \quad (2.28)$$

where $E_0$ is the contribution from $n = 0$ mode. In general, $E_0$ is given by

$$E_0 \approx -2\Lambda l_{st} a^2 g^2 \int_{-\infty}^{\infty} dk_z \frac{1}{k_z^2}\left(1 - \exp\left(-\frac{L}{2l_{st}}\frac{k_z^2}{a^2 g^2}\right)\right) G_{0,0}(R/a, a\kappa_{eff}, k_z). \quad (2.29)$$

Special care must be taken with Eq. (2.29) due to the $1/k_z^2$ pole; in the integrand, the limit $L \to \infty$ cannot simply be taken, as this results in $k_z$ integral becoming singular. The large $L$ behaviour can be obtained by expanding $G_{0,0}(R/a, a\kappa_{eff}, k_z)$ about $k_z = 0$, and then one can perform the resulting $k_z$ integration analytically. This yields, for large $L$, the limiting behaviour

$$E_0 \approx -2\Lambda a g \sqrt{2\pi l_{st} L} G_{0,0}(R/a, a\kappa_{eff}, 0). \quad (2.30)$$

Eq. (2.30) is a rather surprising and non-trivial result, the accumulation of thermal stretching fluctuations leads to a $\sqrt{L}$ dependence in the energy. At large lengths, terms that depend on $\Delta\phi$ in Eq. (2.28) effectively get washed out over large length scales, as they tend to relatively small constants with respect to $L$; but the $n=0$ mode still remains significant. The scaling with respect to length of the $n=0$ mode is altered, due to the random walk nature of the stretching fluctuations. The fact that the effective strength of the interaction term is increasing with $L$ as $\Lambda\sqrt{l_{st}L}$ (c.f. Eq.(2.30) ) suggests that, for very long ideal helices in perfect juxtaposition, we must deal with $n=0$ mode in a different way than with short helices. This inference is further supported by the analysis of Appendix B of the supplemental material, where the relative size of the next to leading order term in the expansion, Eq. (2.23), scales as $L$ for large lengths. Thus, at large $L$ it seems more appropriate to deal with the $n=0$ mode in a variational approximation, which is discussed in the next subsection, regardless of the size of $\Lambda$.

In Fig. 3, we plot $\langle V_{local}\rangle_0 / \Lambda L$ as a function of $\Delta\phi$, given by Eq. (2.27), using the $G_{n,n}$ for Debye-Huckel form of the interaction potential (Eq. (2.14)). The results using the Morse potential are qualitatively similar and are displayed in Appendix B of the supplemental material. We see in Fig.3 that, as the length is increased, the sensitivity of the interaction to $\Delta\phi$ diminishes. The features of this is seen by a reduction in the well at $\Delta\phi=\pi$ as well as an increase in the strength of the interaction near $\Delta\phi=0$. If we keep increasing the length, $\langle V_{local}\rangle_0 / \Lambda L$ will eventually become roughly constant with respect to $\Delta\phi$, at the value $E_0/\Lambda L$. The cause of this effect is the following: though, at $s=0$, interaction sites (at the centre of each molecule) are in optimal azimuthal alignment when $\Delta\phi=\pi$ (facing each other), for other alignments ($\Delta\phi\neq\pi$) thermal fluctuations in stretching and twisting allow for the opportunity for interaction sites to face each other in optimal orientations further along the molecule. If the molecules are sufficiently long it does not matter what orientation the molecules are at $s=0$, as thermal fluctuations accumulate enough for information about the starting ($s=0$) location to be completely lost.

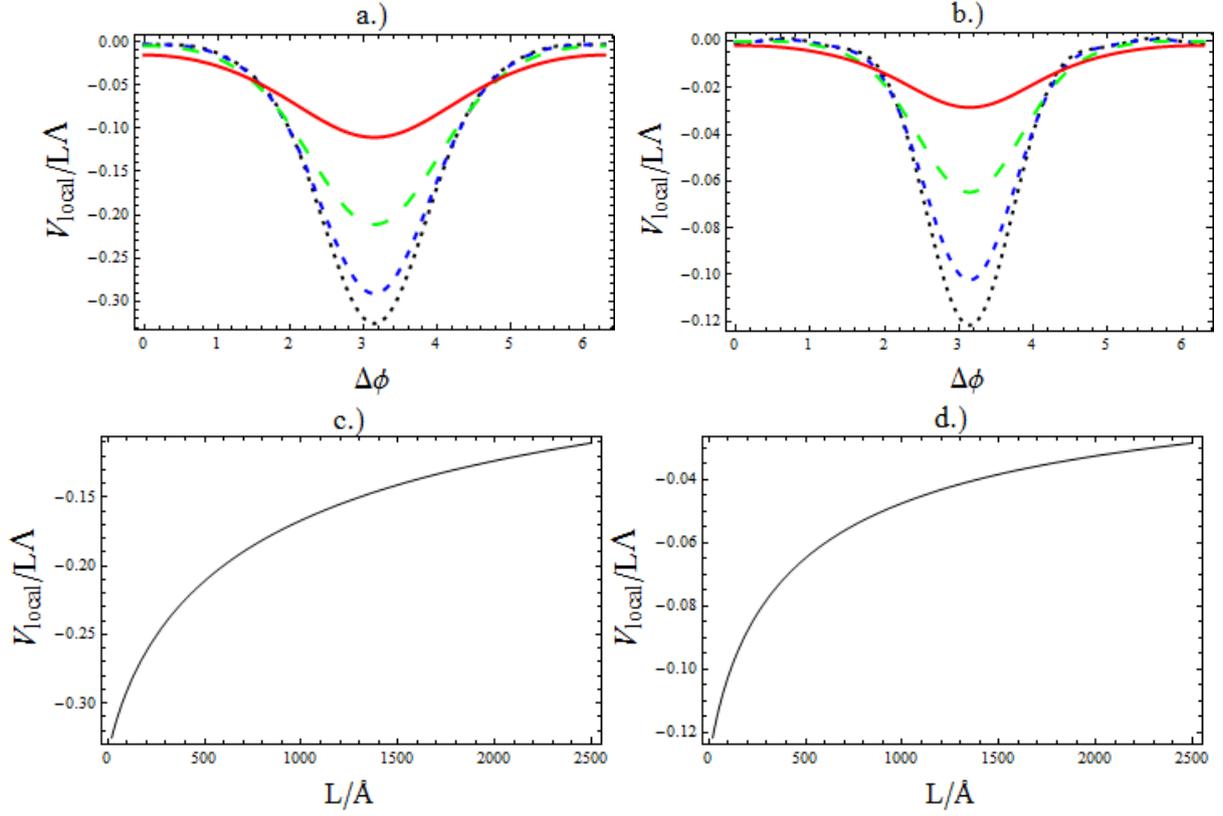

Fig.3. The azimuthal dependence of the thermal average of $V_{local}$ and the 'recognition energy' per unit length, where we have supposed thermal fluctuations are unconstrained by the interaction energy. In generating the plots presented in this figure, both Eqs. (2.14) and (2.27) are employed, supposing a Debye Huckel like potential between interaction sites. In top panels, we show the azimuthal dependence of $V_{local}/\Lambda L$ on $\Delta \phi = \phi_1 - \phi_2$ for a.) $\lambda_{eff} = 4\text{Å}$ and b.) $\lambda_{eff} = 2\text{Å}$, for various values of helix length $L$. In both plots, the dotted black, small dashed blue, long dashed green and solid red curves correspond to the values of $L = 20\text{Å}, 100\text{Å}, 500\text{Å}, 2500\text{Å}$, respectively. We also plot $V_{local}/\Lambda L$ at $\Delta \phi = \pi$, which we term the recognition energy (see main text below) as a function of $L$ for c.) $\lambda_{eff} = 4\text{Å}$ and d.) $\lambda_{eff} = 2\text{Å}$. In performing the calculations the values $R = 25\text{Å}$, $a = 11.2\text{Å}$, $l_{tw} = 1000\text{Å}$ and $l_{st} = 700\text{Å}$ have been used.

Also, in Fig. 3, we plot what we term the recognition energy, the difference in energy between two alike DNA sequences in perfect alignment ($\Delta \phi = \pi$) from two non-alike sequences. For the latter case, we assume that $V_{local} = 0$; although, as stated, there could be a small contribution from accidental base pair matches, but this would be further diminished by helix distortions. We see that as the length of the molecules increases, the recognition energy per unit length decreases due to the $\sqrt{L}$ behaviour. However, when the length becomes sufficiently large we may pass into another regime discussed below.

## 2.3 Thermal fluctuations for long identical base pair tracts in juxtaposition

We now consider the case where the helices are sufficiently long, or the interaction strength is sufficiently strong, for the interaction energy to limit the accumulation of thermal fluctuations along the helices. In this case, we consider a variational approximation where we utilize a trial functional to describe how the fluctuations are limited. In writing such a trial functional, we suppose that the fluctuations can be described in a harmonic way centred around some mean field state, which reduces to the harmonic approximation of $V_{local}$ for very large $\Lambda$. For both the $\delta\phi_{1,T}(s) - \delta\phi_{2,T}(s)$ and $\delta h_{1,T}(s) - \delta h_{2,T}(s)$ fluctuations, we write the trial functional of the form

$$E_{tr}[\delta\phi_{1,T}(s), \delta\phi_{2,T}(s), \delta s_{1,T}(s), \delta s_{2,T}(s)] = E_{tr,\phi}[\delta\phi_{1,T}(s), \delta\phi_{2,T}(s)] + E_{tr,s}[\delta s_{1,T}(s), \delta s_{2,T}(s)], \quad (2.31)$$

where

$$\frac{E_{tr,\phi}[\delta\phi_{1,T}(s), \delta\phi_{2,T}(s)]}{k_B T} = \int_{-L/2}^{L/2} ds \left\{ \frac{l_{tw}}{2} \left[ \left( \frac{d\delta\phi_{1,T}(s)}{ds} \right)^2 + \left( \frac{d\delta\phi_{2,T}(s)}{ds} \right)^2 \right] + \frac{\alpha}{2} \left( \delta\phi_{1,T}(s) - \delta\phi_{2,T}(s) \right)^2 \right\},$$

(2.32)

and

$$\frac{E_{tr,s}[\delta s_{1,T}(s), \delta s_{2,T}(s)]}{k_B T} = \int_{-L/2}^{L/2} ds \left\{ \frac{l_{st} g^2}{2} \left[ \left( \frac{d\delta s_{1,T}(s)}{ds} \right)^2 + \left( \frac{d\delta s_{2,T}(s)}{ds} \right)^2 \right] + \frac{\beta}{2} \left( \delta s_{1,T}(s) - \delta s_{2,T}(s) \right)^2 \right\}.$$

(2.33)

We then write down an approximate variational free energy of the form

$$F_T = -k_B T \ln Z_{tr} + \left\langle E_{total}[\delta\phi_{1,T}(s), \delta\phi_{2,T}(s), \delta s_{1,T}(s), \delta s_{2,T}(s)] - E_{tr}[\delta\phi_{1,T}(s), \delta\phi_{2,T}(s), \delta s_{1,T}(s), \delta s_{2,T}(s)] \right\rangle_{tr},$$

(2.34)

where $E_{total} = E_{el} + V_{local}$, and we have for the thermal average

$$\left\langle E_{total}[\delta\phi_{1,T}(s), \delta\phi_{2,T}(s), \delta s_{1,T}(s), \delta s_{2,T}(s)] - E_{tr}[\delta\phi_{1,T}(s), \delta\phi_{2,T}(s), \delta s_{1,T}(s), \delta s_{2,T}(s)] \right\rangle_{tr}$$

$$= \frac{1}{Z_{tr}} \int D\delta\phi_{1,T}(s) \int D\delta\phi_{2,T}(s) \int D\delta s_{1,T}(s) \int D\delta s_{2,T}(s) \exp\left( -\frac{E_{tr}[\delta\phi_{1,T}(s), \delta\phi_{2,T}(s), \delta s_{1,T}(s), \delta s_{2,T}(s)]}{k_B T} \right)$$

$$(E_{total}[\delta\phi_{1,T}(s), \delta\phi_{2,T}(s), \delta s_{1,T}(s), \delta s_{2,T}(s)] - E_{tr}[\delta\phi_{1,T}(s), \delta\phi_{2,T}(s), \delta s_{1,T}(s), \delta s_{2,T}(s)]),$$

(2.35)

as well as

$$Z_{tr} = \int D\delta\phi_{1,T}(s) \int D\delta\phi_{2,T}(s) \int D\delta s_{1,T}(s) \int D\delta s_{2,T}(s) \exp\left( -\frac{E_{tr}[\delta\phi_{1,T}(s), \delta\phi_{2,T}(s), \delta s_{1,T}(s), \delta s_{2,T}(s)]}{k_B T} \right).$$

(2.36)

It is important to realize that here, when writing Eq. (2.19) for $V_{local}$ within Eq. (2.34), that now both $\phi_1$ and $\phi_2$ correspond to the thermally averaged values of $\phi_1 + \delta\phi_{1,T}(s)$ and $\phi_2 + \delta\phi_{2,T}(s)$, not necessarily the angles that the helices make at $s=0$ to the line connecting them (when the helices are very long). Note that the variational free energy, given by Eqs. (2.34)-(2.36), is always more positive than the exact free energy. Therefore, to get the best approximation to the exact free energy, we choose the values of the variational parameters $\alpha$ and $\beta$ that minimize the free energy, Eq. (2.34), as well as minimizing with respect to $\phi_1 - \phi_2$. In Appendix C of the supplemental material we show that, using the trial functional provided by Eqs. (2.31)-(2.33), the variational free energy defined by Eq. (2.34) evaluates to

$$\frac{F_T}{k_B T} = L\left[\frac{1}{2\lambda_{tw}} + \frac{1}{2\lambda_{st}}\right] - \Lambda' L \sum_{n=-\infty}^{\infty} \exp\left(-\frac{n^2 \lambda_{tw}}{4l_{tw}}\right) \int_{-\infty}^{\infty} dk_z G_{n,n}(R/a, a\kappa_{eff}, k_z) \exp\left(-\frac{k_z^2 \lambda_{st}}{4a^2 g^2 l_{st}}\right),$$

(2.37)

where the adaptation lengths $\lambda_{tw}$ and $\lambda_{st}$ are defined by $\lambda_{tw} = (2l_{tw}/\alpha)^{1/2}$ and $\lambda_{st} = (2g^2 l_{st}/\beta)^{1/2}$; as we have that $\Lambda' = \Lambda/k_B T$. We will discuss why $\lambda_{tw}$ and $\lambda_{st}$ are termed adaptation lengths below. In writing Eq. (2.37), we have already used the fact that for when $\lambda_{tw}$ is finite the mean orientation of the helices is $\langle \Delta\phi \rangle = \pi$. One should be aware that minimization of Eq. (2.37) with respect to $\lambda_{tw}$ and $\lambda_{st}$ is equivalent to minimization with respect to $\alpha$ and $\beta$. Minimization with respect to $\lambda_{tw}$ and $\lambda_{st}$ yields the coupled equations

$$\frac{1}{\lambda_{tw}^2} = \frac{\Lambda'}{l_{tw}} \sum_{n=1}^{\infty} n^2 \exp\left(-\frac{n^2 \lambda_{tw}}{4l_{tw}}\right) \int_{-\infty}^{\infty} dk_z G_{n,n}(R/a, a\kappa_{eff}, k_z) \exp\left(-\frac{k_z^2 \lambda_{st}}{4a^2 g^2 l_{st}}\right),$$ (2.38)

$$\frac{1}{\lambda_{st}^2} = \frac{\Lambda'}{a^2 g^2 l_{st}} \sum_{n=1}^{\infty} \exp\left(-\frac{n^2 \lambda_{tw}}{4l_{tw}}\right) \int_{-\infty}^{\infty} dk_z k_z^2 G_{n,n}(R/a, a\kappa_{eff}, k_z) \exp\left(-\frac{k_z^2 \lambda_{st}}{4a^2 g^2 l_{st}}\right)$$

$$+ \frac{\Lambda'}{2a^2 g^2 l_{st}} \int_{-\infty}^{\infty} dk_z k_z^2 G_{0,0}(R/a, a\kappa_{eff}, k_z) \exp\left(-\frac{k_z^2 \lambda_{st}}{4a^2 g^2 l_{st}}\right).$$

(2.39)

There are, in fact, two states that Eqs. (2.37)-(2.39) may describe. When the local pairing interaction is sufficiently strong ($\Lambda$ sufficiently large), the state that minimizes the free energy corresponds to both $\lambda_{tw}$ and $\lambda_{st}$ being finite, with no trivial roots to both Eqs. (2.38) and (2.39). In such a state, the preferred average azimuthal orientation is indeed $\langle \Delta\phi \rangle = \pi$. However, for small $\Lambda$, it is possible to have another state that minimizes the free energy where $\lambda_{tw} = \infty$, which is indeed a trivial solution of Eq. (2.38). From Eq. (2.39), when $\lambda_{tw} = \infty$, $\lambda_{st}$ is determined through

$$\frac{1}{\lambda_{st}^2} = \frac{\Lambda'}{2a^2 g^2 l_{st}} \int_{-\infty}^{\infty} dk_z k_z^2 G_{0,0}(R/a, a\kappa_{eff}, k_z) \exp\left(-\frac{k_z^2 \lambda_{st}}{4a^2 g^2 l_{st}}\right),$$ (2.40)

In this state, if fact, there is no preferred value of $\langle \Delta\phi \rangle$; the helices can freely rotate relative to each other.

Let us further analyse the $\lambda_{tw} = \infty$ state. It must be the case, physically, that in the limit $\Lambda \to 0$ we must have that $\lambda_{st} \to \infty$. This is because when $\Lambda \to 0$, there is no interaction to constrain the stretching fluctuations, so that we have $\beta = 0$, and simply random walk behaviour. This suggests that, for small $\Lambda$, we should be able obtain an analytical expression for $\lambda_{st}$ from Eq. (2.40) which diverges in the limit $\Lambda \to 0$, if a state described by a finite value of $\lambda_{st}$ persists for small $\Lambda$. To obtain such an approximate solution, we again expand $G_{0,0}(R/a, a\kappa_{eff}, k_z)$ about $k_z = 0$. This yields the expression for $\lambda_{st}$ at small $\Lambda$

$$\lambda_{st} \approx \frac{1}{4\pi\Lambda'^2 a^2 g^2 l_{st} G_{0,0}(R/a, a\kappa_{eff}, 0)^2} - \frac{12a^2 g^2 l_{st}}{G_{0,0}(R/a, a\kappa_{eff}, 0)} G'_{0,0}(R/a, a\kappa_{eff}, 0), \qquad (2.41)$$

where

$$G'_{0,0}(R/a, a\kappa_{eff}, k_z) = \frac{d}{dk_z^2} G_{0,0}(R/a, a\kappa_{eff}, k_z). \qquad (2.42)$$

Note that $G_{0,0}(R/a, a\kappa_{eff}, k_z)$ is a pure function of $k_z^2$, thus we need only take the derivative with respect to $k_z^2$ to generate a pure power series in $k_z^2$.

In the state where $\lambda_{tw} = \infty$, we have for the free energy

$$\frac{F_T}{k_B T L} = \frac{1}{2\lambda_{st}} - \Lambda' \int_{-\infty}^{\infty} dk_z G_{0,0}(R/a, a\kappa_{eff}, k_z) \exp\left(-\frac{k_z^2 \lambda_{st}}{4a^2 g^2 l_{st}}\right). \qquad (2.43)$$

For small $\Lambda$, we may again expand out $G_{0,0}(R/a, a\kappa_{eff}, k_z)$ about $k_z = 0$, yielding the approximate expression

$$\frac{F_T}{k_B T L} \approx -2\pi a^2 g^2 l_{st} \Lambda'^2 G_{0,0}(R/a, a\kappa_{eff}, 0)^2 - 32\pi^2 a^6 g^6 l_{st}^3 \Lambda'^4 G_{0,0}(R/a, a\kappa_{eff}, 0)^3 G'_{0,0}(R/a, a\kappa_{eff}, 0).$$

(2.44)

The forms of Eqs. (2.41) and (2.44) indeed suggest that there is a smooth continuous increase to $\lambda_{st} = \infty$ as $\Lambda \to 0$ and no sudden jump to $\lambda_{st} = \infty$ at some finite value of $\Lambda$, which is indeed evidenced in a numerical calculation using Eqs. (2.43) and (2.40).

To understand the physical meaning of the lengths $\lambda_{tw}$ and $\lambda_{st}$, as adaptation lengths, it is useful to consider the correlation functions that describe the accumulation in both twisting and stretching distortions due to thermal fluctuations. Namely, these are $\langle (\Delta h_T(s) - \Delta h_T(s'))^2 \rangle_{tr}$ and $\langle (\Delta \phi_T(s) - \Delta \phi_T(s'))^2 \rangle_{tr}$, where $\Delta h_T(s) = \delta h_{1,T}(s) - \delta h_{2,T}(s)$ and $\Delta \phi_T(s) = \delta \phi_{1,T}(s) - \delta \phi_{2,T}(s)$. These are found to evaluate to (using Eqs. (2.31)-(2.33), also see Appendix C of the supplemental material for details)

$$\left\langle (\Delta h_T(s) - \Delta h_T(s'))^2 \right\rangle_{tr} = \frac{\lambda_{st}}{l_{st}g^2}\left(1 - \exp\left(-\frac{2|s-s'|}{\lambda_{st}}\right)\right), \qquad (2.45)$$

$$\left\langle (\Delta \phi_T(s) - \Delta \phi_T(s'))^2 \right\rangle_{tr} = \frac{\lambda_{tw}}{l_{tw}}\left(1 - \exp\left(-\frac{2|s-s'|}{\lambda_{tw}}\right)\right). \qquad (2.46)$$

If we examine Eq. (2.45), we find that when $|s-s'| \ll \lambda_{st}$ we have a random walk behaviour of the form $\left\langle (\Delta h_T(s) - \Delta h_T(s'))^2 \right\rangle_{tr} \approx 2|s-s'|/l_{st}g^2$; on the other hand, when $|s-s'| \gg \lambda_{st}$, we have that $\left\langle (\Delta h_T(s) - \Delta h_T(s'))^2 \right\rangle_{tr} \approx \lambda_{tw}/l_{st}g^2$. The upshot is that, at sufficiently long length scales larger than $\lambda_{st}$, the accumulation of the fluctuations in $\Delta h_T(s)$ saturates to a constant value. We see that $\lambda_{st}$ is a characteristic length over which the pairing interactions are able to limit accumulation of fluctuations in $\Delta h_T(s)$ and adjust the helices; this is what we mean by an adaptation length. When we consider Eq. (2.46), again for length scales smaller than $\lambda_{tw}$, we again have random walk behaviour $\left\langle (\Delta \phi_T(s) - \Delta \phi_T(s'))^2 \right\rangle_{tr} \approx 2|s-s'|/l_{tw}$. Here, when we sit at length scales larger than $\lambda_{tw}$ the accumulation saturates at the value $\left\langle (\Delta \phi_T(s) - \Delta \phi_T(s'))^2 \right\rangle_{tr} \approx \lambda_{tw}/l_{tw}$. Thus, we see that $\lambda_{tw}$ is a characteristic length over which the local pairing interactions limit the accumulation in fluctuations $\Delta \phi_T(s)$.

Finally, in this subsection, we'll consider one final piece of analysis. Let us consider small values of $\Lambda$. We may ask ourselves what is the criteria for either of the two regimes for weak $\Lambda$ to apply; the long helix length one discussed here, and the short helix length one discussed in the previous subsection. We first note that for small values of $\Lambda$ we can write Eq. (2.44) as $F_T \approx -L/2\lambda_{st}$, and also we can write Eq. (2.30) as $E_0 = \sqrt{2L/\lambda_{st}}$. Thus, we can estimate the crossover length between the small $L$ regime and the large $L$ regime (discussed in this subsection), by equating $F_T$ with $E_0$. This yields $L \sim \lambda_{st}$, and so suggests that $\lambda_{st}$ is the characteristic length at which the crossover between the two regimes occurs. The consequence is that for $L \gg \lambda_{st}$ the regime described by Eqs. (2.40), (2.41), (2.43) and (2.44) applies, and for $L \ll \lambda_{st}, \lambda_{tw}$ the regime described by the expansion for weak $V_{local}$ applies, i.e. Eq. (2.27), (2.28) and (2.29). This suggests that, at large $L$, the free energy always scales as $L$. Supposing the free energy of two unalike helices to be zero, this yields a recognition energy per unit length which is constant with respect to $L$, when the molecules are in perfect juxtaposition.

In Fig. 4 we show plots for both $\lambda_{st}$ and $\lambda_{tw}$ as a function of the interaction strength $\Lambda$, using the Debye-Huckel form (Eq. (2.14)). The critical value $\Lambda$, at which there is a transition between the $\lambda_{tw} = \infty$ state and finite $\lambda_{tw}$ state, is determined when the free energies described by Eqs. (2.37) and (2.43) are equal; or alternatively, using the small $\Lambda$ approximation, when Eqs. (2.37) and (2.44) are equal. At the point of transition there is a discontinuous jump in $\lambda_{st}$; with increasing $\Lambda$, $\lambda_{st}$ suddenly drops to much smaller values when we move across the transition. Looking at the graphs for $\lambda_{st}$, we see that the approximate solution, Eq. (2.41) works quite well in describing $\lambda_{st}$. Though, for a value of $\lambda_{eff} = 4\text{Å}$, it doesn't capture the jump very well; but, for $\lambda_{eff} = 2\text{Å}$, Eq. (2.41) is an excellent approximation right up to the transition. In Fig. 4, $\lambda_{tw}$ becomes finite at the point of transition and then monotonically decreases with increasing $\Lambda$. This makes sense, as stronger interactions should limit thermal twisting fluctuations over shorter lengths. We see that the value of $\lambda_{tw}$, at the transition point for $\lambda_{eff} = 2\text{Å}$, is smaller than that for $\lambda_{eff} = 4\text{Å}$.

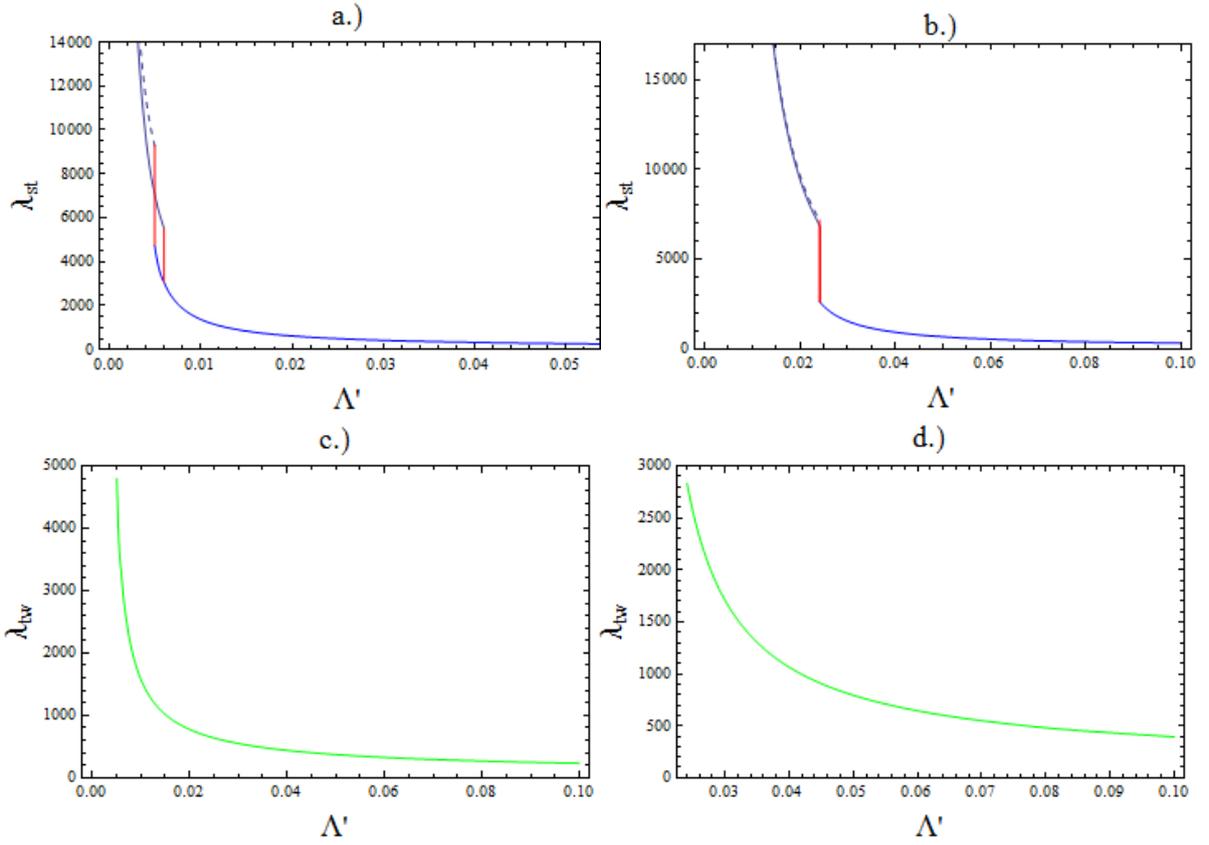

Fig. 4. Plots showing the adaptation lengths $\lambda_{st}$ and $\lambda_{tw}$ calculated using the Debye Huckel Form (Eq. (2.14)). In panels a.) and b.) we show $\lambda_{st}$, as a function of the interaction strength $\Lambda' = \Lambda/k_B T$, for $\lambda_{eff} = 4\text{Å}$ and $\lambda_{eff} = 2\text{Å}$, respectively. The red lines represent a transition from the state where there is no adaptation in twisting fluctuations (effectively $\lambda_{tw} = \infty$) to a state where $\lambda_{tw}$ is finite, which marks the discontinuity in $\lambda_{st}$. To the left of the transition (smaller $\Lambda'$), the solid line is the solution to the full equation (Eq. (2.40)), while the dashed line is the small $\Lambda$ solution (Eq. (2.41)). To the right of the transition (larger $\Lambda'$ values), we have the value of $\lambda_{st}$ that solves both Eqs. (2.38) and (2.39). In panels c.) and d.), we show $\lambda_{tw}$, as function of the interaction strength $\Lambda'$, for $\lambda_{eff} = 4\text{Å}$ and $\lambda_{eff} = 2\text{Å}$ respectively. These values solve both Eqs. (2.38) and (2.39), and $\lambda_{tw}$ is only finite above the transition. In the calculations the values $R = 25\text{Å}$, $a = 11.2\text{Å}$, $l_{tw} = 1000\text{Å}$ and $l_{st} = 700\text{Å}$ are used.

In Appendix D of the supplemental material, we have worked out the leading order corrections to both Eqs. (2.43) and (2.44), expressions for the free energy, from residual correlations in the relative azimuthal angle between the two helices, i.e. $\Delta\phi(s) = \delta\phi_{1,T}(s) - \delta\phi_{2,T}(s)$ (this is a modified form of the expansion given by Eq. (2.23) in concert with the variational approximation utilised here for $\lambda_{st}$) and examine how this affects the transition between the finite $\lambda_{tw}$ state and $\lambda_{tw} = \infty$ state. This analysis suggests that higher order corrections should become small at the point of transition, but more terms in the expansion are needed to get a reliable correction to the crossing point in free energy between the two states. In addition, the analysis suggests that we expect that qualitative behaviour given in Fig. 4 is not much affected, and the transition point in $\Lambda$ will be not much affected when higher order corrections are taken into account.

## 2.4 Comparison with analogous models with non-base pair specific pairing interactions

To put the previously discussed pairing model in context and to identify generic differences between this and a pairing mechanism based purely on the patterns of helix distortions, it will be useful to construct a locally non-base pair specific analogue to the model described by Eq. (2.1). This will be especially important for subsequent work. The interaction energy for such an analogue is simply is given by

$$V_{global}(\mathbf{R}) = \sum_{j'=-N}^{N} \sum_{j=-N}^{N} V_{int}\left(\mathbf{r}_{1,j} - \mathbf{r}_{2,j'} - \mathbf{R}\right). \qquad (2.47)$$

The important difference between this and the local model described by Eq. (2.1) is the double sum. Here, the important difference is that we now allow for interaction site on one helix to interact with all the sites on other helix. In this case, the interaction mechanism does not discriminate between base pairs, and so helices with non-alike sequences must interact through the same potentials between base pairs. Indeed, here, the crucial difference that distinguishes two non-alike molecules from two like ones is the pattern of helix distortions. For non-alike helices it is harder to maintain commensurability between the interaction sites; therefore we expect a less negative interaction free energy than identical helices. It is this difference that results in a recognition energy (an energy difference that make the interaction between identical helices more favourable), and not the interaction discriminating between base pairs, as in the previous model.

For rigid helices, again making the continuum approximation and going through similar steps, we obtain for Eq. (2.47), the following expression for identical sequences (see Appendix E of Supplemental Material)

$$V_{global}(R) \simeq -\Lambda_g L \sum_{n=-\infty}^{\infty} \exp(in(\phi_1 - \phi_2))(-1)^n G_{n,n}(R/a, a\kappa_{eff}, -ang). \qquad (2.48)$$

where $\Lambda_g = \gamma_{eff} / h^2(2\pi)$. Here, again, $G_{n,n'}$ can be chosen to be either of the forms given by Eqs. (2.14) and (2.15), or something quite different [35]. Again, Eq. (2.48) can also be generalized to any interaction between sites that can be expressed in terms of a Laplace transformation.

We are interested in the ways the two types of pairing mechanism give rise to different results. Thus, for comparison purposes, it is useful to define a new, normalized interaction strength $\Lambda_c = \Lambda_g V_{global}(R;\Delta\phi=\pi)/V_{local}(R;\Delta\phi=\pi)$. This insures that the same values of $\Lambda$ and $\Lambda_c$, in the absence of thermal fluctuations, both models give the same value of interaction energy at the optimal orientation of $\Delta\phi=\pi$.

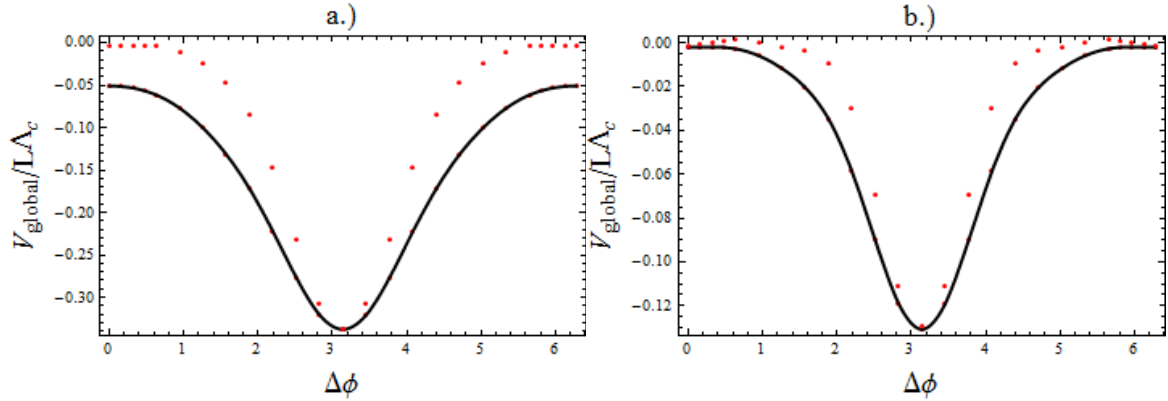

Fig. 5. How the azimuthal dependence of $V_{global}$ compares with that of $V_{local}$ for rigid helices. Here we use Eq. (2.14) for $G_{n,n}(R/a,a\kappa_{eff},-ang)$. We plot $V_{global}/L\Lambda_c$ as a function of $\Delta\phi=\phi_1-\phi_2$ and compare it against $V_{local}/L\Lambda$. The red points are values calculated for $V_{local}/L\Lambda_c$, while the solid black curves are for $V_{global}/L\Lambda$. The values of $\lambda_{eff}$ used are in a.) $\lambda_{eff}=4\text{Å}$ and in b.) $\lambda_{eff}=4\text{Å}$.

In Fig. 5 we compare the azimuthal dependence of $V_{global}/L\Lambda_c$ with $V_{local}/L\Lambda$ for rigid helices using the Debye-Huckel form for $V_{int}$ (Eq. (2.2)). Two qualitative differences emerge. The first is that the potential well centred around $\Delta\phi=\pi$ is much broader for the global model. The second is that if the decay length is sufficiently long, as in the case of $\lambda_{eff}=4\text{Å}$, the interaction is no longer negligible at $\Delta\phi=0$. This is because all interactions sites can interact with each other, so when the helices are in the $\Delta\phi=0$ orientation the interaction sites can still find directly facing interaction sites to interact with on the other molecules separated by distance $H/2$.

One important difference between the non-locally base pair specific helix distortion model and the local base pair specific model is the selection rule $k_z=-ang$. This arises from a corkscrew symmetry: the interaction energy remains invariant under the transformations $\phi_1-\phi_2\to\phi_1-\phi_2-g\Delta s$ and $s\to s+\Delta s$, i.e. corkscrew motion (strictly speaking, this is in the case when $L\to\infty$; though, for helices for which $L/H\gg 1$, the symmetry is approximate– see Appendix E of supplemental material). The local base pair specific pairing model lacks this symmetry, as corkscrew motion causes the favourable interactions to diminish. When we consider thermal fluctuations, provided that $gl_{tw},gl_{st}\gg 1$ (indeed the case for the values considered), this symmetry is approximately maintained (see Appendix F of supplemental material). This means that for weak helix dependent forces, for the model described by Eq. (2.47), the free energy described by Eq. (2.27) is now replaced by for identical sequences

$$F_g - F_0 \approx \langle V_{global}[\delta\phi_{1,T}(s), \delta\phi_{2,T}(s), \delta s_{1,T}(s), \delta s_{2,T}(s)] \rangle_0$$
$$\approx -2\Lambda_g l_h \sum_{n=-\infty}^{\infty} (-1)^n \exp(in(\phi_1 - \phi_2)) \frac{1}{n^2}\left(1 - \exp\left(-\frac{Ln^2}{2l_h}\right)\right) G_{n,n}(R/a, a\kappa_{eff}, -ang), \quad (2.49)$$

where, now, we have a combined helical persistence length

$$l_h = \frac{l_{tw} l_{st}}{l_{tw} + l_{st}}. \quad (2.50)$$

Using the values of $l_{tw} = 1000\text{Å}$ and $l_{st} = 700\text{Å}$, we estimate that $l_h \approx 410\text{Å}$. We notice that the limiting behaviour of Eq. (2.49) is now quite different for large $L$. We have that

$$F_g - F_0 \approx -\Lambda_g L G_{0,0}(R/a, a\kappa_{eff}, 0) - 4\Lambda_g l_h \sum_{n=1}^{\infty} \frac{(-1)^n}{n^2} \cos(n(\phi_1 - \phi_2)) G_{n,n}(R/a, a\kappa_{eff}, -ang).$$

(2.51)

We see that the in the limit of large $L$, the dominant contribution to the energy scales as $L$, as opposed to $\sqrt{L}$. As we are interested in the recognition energy, it is also worth asking what the free energy is, here, for non-alike helices. We find that (Appendix F of supplemental material) this free energy is given by

$$F_g^{NL} - F_0 \approx \langle V_{global}^{NL}[\delta\phi_{1,T}(s), \delta\phi_{2,T}(s), \delta s_{1,T}(s), \delta s_{2,T}(s)] \rangle_0$$
$$\approx -2\Lambda_g \lambda_c \sum_{n=-\infty}^{\infty} \frac{(-1)^n}{n^2} \exp(in(\phi_1 - \phi_2))\left(1 - \exp\left(-\frac{Ln^2}{2\lambda_c}\right)\right) G_{n,n}(R/a, a\kappa_{eff}, -ang). \quad (2.52)$$

where

$$\lambda_c = \frac{l_h \lambda_c^{(0)}}{l_h + \lambda_c^{(0)}}. \quad (2.53)$$

The length $\lambda_c^{(0)}$ is the contribution to $\lambda_c$ from helix distortions that depend on base pair sequence, rather than to do with thermal fluctuations (taken into account through $l_h$); it is estimated to be $\lambda_c^{(0)} \approx 150\text{Å}$ (see Ref. [20]). The length $\lambda_c^{(0)}$ depends on both $\lambda_{tw}^{(0)}$ and $\lambda_{st}^{(0)}$, but also correlations between the two types of distortion [20]. From Eq. (2.53), this yields a value of $\lambda_c \approx 110\text{Å}$ this value falls close to the value used to fit X-ray diffraction data in Ref. [36]. These expressions (Eqs. (2.51) and (2.52)) would suggest that the recognition energy per unit length, $(F - F_{NA})/L$, would go as $1/L$, as the length of the molecules is increased much above $l_h$.

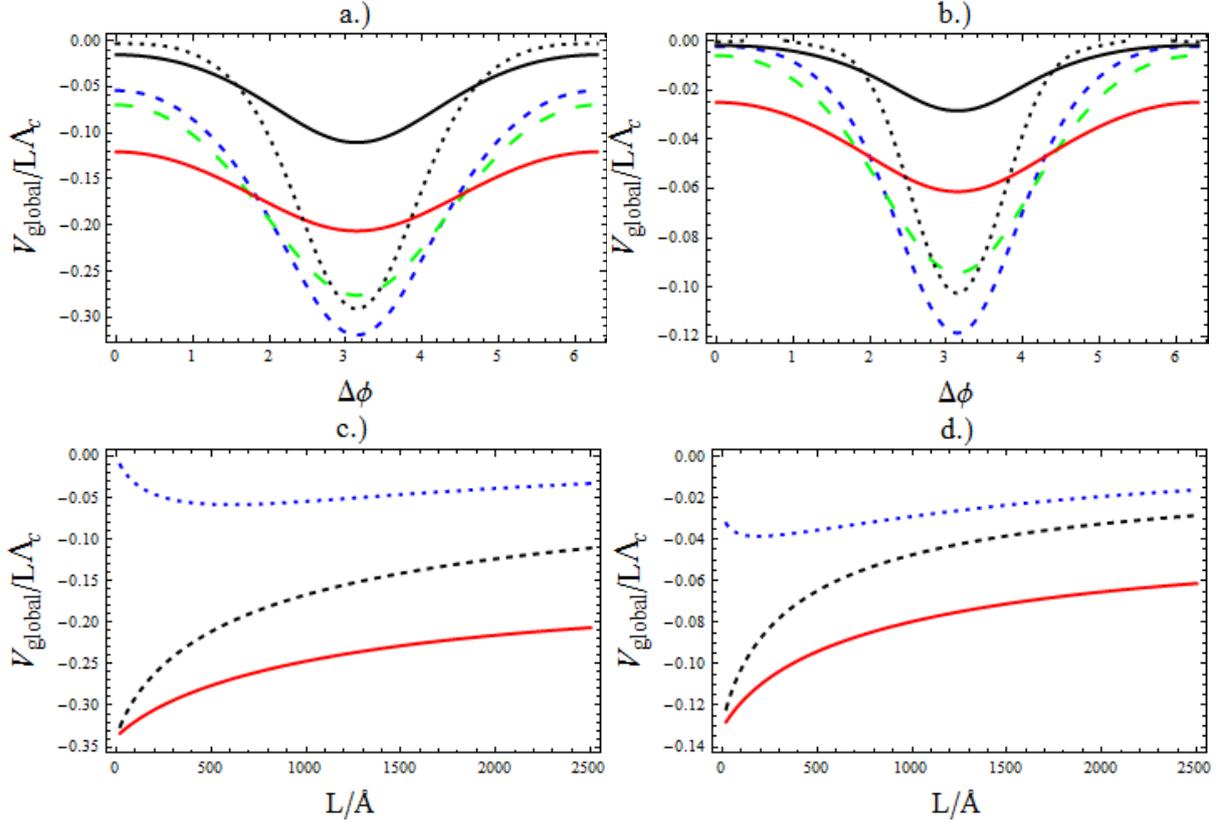

Fig. 6. The azimuthal dependence of the thermal average of $V_{global}$ and recognition energy per unit length, where thermal fluctuations are unconstrained by the interaction energy. In calculating the plots for $V_{global}$ both Eqs. (2.14) and (2.49) are employed, supposing a Debye Huckel like potential. In top panels, we show the azimuthal dependence of $V_{global}/\Lambda_c L$, calculated for identical molecules, for a.) $\lambda_{eff} = 4\text{Å}$ and b.) $\lambda_{eff} = 2\text{Å}$ for various values of length $L$ and compare them with values of $V_{local}/\Lambda L$. In both plots, small dashed blue, long dashed green and solid red curves correspond to $V_{global}/\Lambda_c L$, at the values of $L = 100\text{Å}, 500\text{Å}, 2500\text{Å}$, respectively. In addition dotted and solid black lines correspond to $V_{local}/\Lambda L$, at the values $L = 100\text{Å}$ and $2500\text{Å}$, respectively. We also plot the average value of $V_{global}/\Lambda_c L$ (where we set $\Delta\phi = \phi_1 - \phi_2 = \pi$) for identical sequences as a function of $L$ and compare with the average value $V_{local}/\Lambda L$ for c.) $\lambda_{eff} = 4\text{Å}$ and d.) $\lambda_{eff} = 2\text{Å}$. The solid red curves correspond to $V_{global}/\Lambda_c L$ and the dashed black ones to $V_{local}/\Lambda L$. In panels c.) and d.) the recognition energy for global pairing is also plotted, which unlike the local pairing is quite different from $V_{global}/\Lambda_c L$, as one has to subtract off the average energy for non-alike helices. This is shown by a dotted blue curve. In performing the calculations the values $R = 25\text{Å}$, $a = 11.2\text{Å}$, $l_c = 410\text{Å}$ and $\lambda_c \approx 110\text{Å}$ are all used.

Let us now plot Eq. (2.49) and compare it with plots of Eq. (2.27), using Eq. (2.14) for $G_{n,n}(R/a, a\kappa_{eff}, -ang)$, as is shown in Fig. 6. When we compare the dependence on $\Delta\phi = \phi_1 - \phi_2$ between the two, allowing all the base pairs to pair with each other leads to a broader well about $\Delta\phi = \pi$. We also see a much larger contribution from the $n = 0$ mode (which is independent of $\Delta\phi$) at large $L$, due to scaling as $L$, and not $\sqrt{L}$. What are more interesting are the differences in the recognition energy per unit length, the difference in free energy of a pair of identical helices and that of two non-identical ones. We see that the trend with increasing length is quite different from pairing between identical base pairs. Initially, here, the recognition energy per unit length increases up from zero, this is because of the random walk accumulation of helix distortions on both identical and non-alike helices. When the length of the helices is such that $L \ll \lambda_c$, the recognition energy per unit length is very small. However, when $L$ becomes comparable to $\lambda_c$, the accumulation of helical distortions is much larger for non-alike helices than alike ones, resulting in a less negative free energy for the former. Thus, at this point the magnitude of the recognition energy goes up. However, when $L$ becomes larger, to become comparable to $l_h$, the free energy per unit length for identical helices also starts to diminish, and the recognition energy goes down.

Though, this is not the whole story; the next to leading order corrections to Eqs. (2.49) and (2.52), for the free energy have been computed in Appendix F of the supplemental material . This correction is found to scale as $L$, for large values of $L$, and to be proportional to $\Lambda_c^2$, and different for identical and non-alike sequences. This leads similar scaling of the recognition energy, at large $L$, as the local base pair specific pairing model(s) (Eq.(2.44) ). For long helices and sufficiently strong $\Lambda_c$, also we need to calculate the recognition energy a different way, when the corrections to Eqs. (2.49) and (2.51) become too large, discussed below. Nevertheless, the robust features of this analysis are that for local pairing the recognition energy per unit length diminishes with increasing $L$ and that for global pairing it always initially increases.

Let us now examine the case where the molecules are long and the interactions are sufficiently strong. Due to the approximate symmetry the fluctuations $\Delta\phi_T(s) = \delta\phi_{1,T}(s) - \delta\phi_{2,T}(s)$ and $\Delta h_T(s) = \delta h_{1,T}(s) - \delta h_{2,T}(s)$ can be combined into fluctuations of what we term the helical phase difference, namely $\Delta\Phi_T(s) = \Delta\phi_T(s) - g\Delta h_T(s)$ (see Appendix F of supplemental material). These are the only fluctuation modes that matter for non-specific base pair pairing. In this case, we can repeat the variational approximation, but now the choice of trial function for the model described by Eq. (2.47) is

$$\frac{E_{tr}[\Delta\Phi_T(s)]}{k_B T} = \int_{-L/2}^{L/2} ds \left\{ \frac{l_{tw}}{4}\left(\frac{d\Delta\Phi_T(s)}{ds}\right)^2 + \frac{\alpha_g}{2}\left(\Delta\Phi_T(s)\right)^2 \right\}. \tag{2.54}$$

Utilizing the variational approximation outlined previously, we arrive at free energy (for non-identical helices)

$$\frac{\langle F_{g,T}^{NL}\rangle_\Omega}{k_B TL} = \frac{(l_c + \lambda_c)^2}{16\lambda_h^* \lambda_c l_h} - \Lambda_g' \sum_{n=-\infty}^{\infty} \exp(in(\phi_1 - \phi_2))\exp\left(-\frac{n^2 \lambda_h^*}{2\lambda_c}\right)(-1)^n \bar{G}_{n,n}(R/a, -ng), \tag{2.55}$$

for details see Appendix G. Here, $\Lambda_g'$ is $\Lambda_g / k_B T$. The free energy for identical helices is simply got by setting $\lambda_c = l_h$. The situation, here, is quite different from the local theory. For large $\Lambda_g$ we are in a state where $\lambda_h^*$ is finite, and for $G_{n,n}$ described by either Eqs. (2.14) or (2.15); the optimal value of $\Delta\phi$ is still $\pi$. Eq. (2.55) is minimized by

$$\frac{(l_c + \lambda_c)^2}{8(\lambda_h^*)^2 l_h} = \tilde{\Lambda}_g \sum_{n=-\infty}^{\infty} n^2 \exp\left(-\frac{n^2 \lambda_h^*}{2\lambda_c}\right) \bar{G}_{n,n}(R/a, -ng), \qquad (2.56)$$

with again setting $\lambda_c = l_h$ for helices with identical base pair texts. However, as we reduce $\Lambda$, we eventually jump to a state where $\lambda_h^* = \infty$ where

$$\frac{F_{T,global}}{k_B T} = -\Lambda_g L G_{0,0}(R/a, a\kappa_{eff}, 0). \qquad (2.57)$$

This is simply the dominant ($n = 0$) term we had previously in large $L$ limit of Eq. (2.49), i.e. Eq. (2.51). There is no adaptation for this state, i.e. limitation of accumulation of helix distortion due to the pairing interaction. Thus, there is no crossover between two regimes for arbitrarily small $\Lambda_g$, as seen with local pairing, as one increases $L$. However in the $\lambda_h^* = \infty$ state, as discussed previously, there are corrections to Eq. (2.57) from residual correlations in $\Delta\Phi_T(s)$, leading to a $\Lambda_g^2$ correction to Eq. (2.57) (for a discussion of such terms see Appendices F and G of supplemental material).

In Fig. 7. we present plots of the adaptation length $\lambda_h^*$ calculated for the model described by Eq. (2.47), utilizing Eq. (2.14) for a Debye-Huckel potential between interaction sites. Presented in Fig. 7 are the adaptation lengths for alike and non-alike helices. As for local base pair specific pairing, the adaptation length governs the regime one sits in; for $L \ll \lambda_h^*$ the free energy is described either by Eq. (2.49) or Eq. (2.52) (plus corrections, the leading order of which are considered in Appendix F); and for $L \gg \lambda_h^*$ the free energy is described by Eq. (2.55), with either $\lambda_c$ described by Eq. (2.53) or setting $\lambda_c = l_h$. There are distinct features that are worth commenting on. The first, is unlike specific base pair pairing, where there is always adaptation in the stretching fluctuations for small $\Lambda_g$, we indeed see, in the numerical calculations, for the model described by Eq. (2.47) there exists the state where there is no adaptation at all. However, for helix distortion pattern pairing model (Eq. (2.47)), the $n = 0$ mode is unaffected by both stretching and twisting fluctuations, while for local base pair specific pairing it clearly indeed is. A second feature is that a pairing energy that scales as $L$ (for $L \gg \lambda_h^*$) also exists for unalike molecules, with a finite value of $\lambda_h^*$. For unalike molecules, the transition between the state for which $\lambda_h^* = \infty$ to that where $\lambda_h^*$ is finite happens at a larger value of $\Lambda_c$ than for identical helices. This is to be expected, as the strength of the pairing will be weakened more by increased helical distortions, in the case of non-alike molecules. We note that we have a smaller value of $\lambda_h^*$ for non-alike helices; however the free energy still remains higher than that for identical helices.

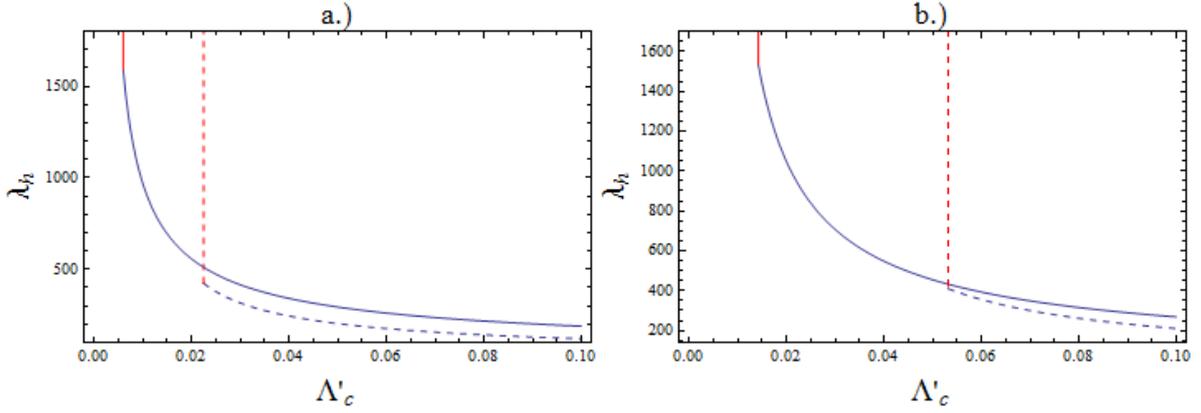

Fig. 7. Plots showing the adaptation length $\lambda_h^*$ calculated using the Debye Huckel Form for the interaction between sites (Eq. (2.14)), for the locally non-base specific pairing described by Eq. (2.47). We calculate it for the interaction between alike and non-alike helices. In panels a.) and b.) we show $\lambda_h^*$ (calculated using Eq. (2.56)), as a function of the interaction strength $\Lambda'_c = \Lambda_c / k_B T$, for $\lambda_{eff} = 4\text{Å}$ and $\lambda_{eff} = 2\text{Å}$, respectively. The red lines represent a transition from the state where there is no helical adaptation (effectively $\lambda_h^* = \infty$) to a state where $\lambda_h^*$ is finite. The solid lines are plots of $\lambda_h^*$ for identical helices and the dashed lines are for helices that are not alike. In the calculations, the values $R = 25\text{Å}$, $a = 11.2\text{Å}$, $l_c = 410\text{Å}$ and $\lambda_c = 109\text{Å}$ are used.

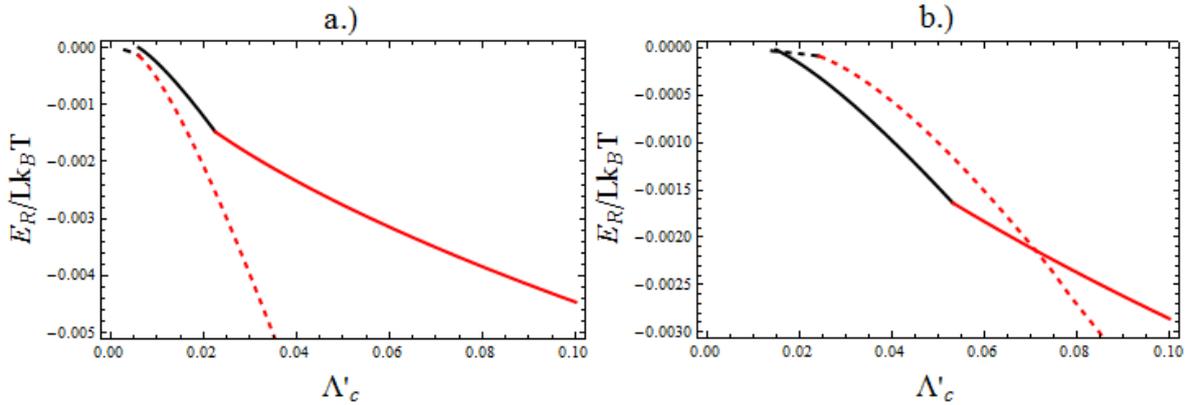

Fig. 8. Plots of the recognition energy per unit length $E_R / L$ for very long helices as a function of the interaction strength $\Lambda'_c = \Lambda_c / k_B T$ ($\Lambda$ for local base pair specific pairing). The recognition energy is calculated using the Debye-Huckel form for the interaction between sites (utilizing Eq.(2.14)). The solid lines refer to local non-base pair specific pairing of homology recognition described by Eq. (2.47), where the dashed lines correspond to local base pair specific pairing (Eq. (2.1)). The solid red curves are for the recognition energy for local non base pair specific pairing, where both identical helices and un-alike helices are in the state for which $\lambda_h^*$ is finite. The solid black lines are for the local non base pair specific pairing, as well, but in this case, the pairing interactions for non-alike helices are too weak for adaptation. The red dashed lines refer to the state where $\lambda_{tw}$ is finite, and the dashed black lines the state where $\lambda_{tw}$ is infinite. For the local base pair specific pairing there is always adaptation in the stretching fluctuations and the pairing free energy for non-alike helices is effectively zero. In the calculations, the values $R = 25\text{Å}$, $a = 11.2\text{Å}$, $l_{tw} = 1000\text{Å}$ and $l_{st} = 700\text{Å}$, $\lambda_c = 109\text{Å}$ are used.

In Fig. 8, we plot the recognition energy per unit length for very long helices. Here, it is defined for long molecules as $E_R/L = (F - F^{NL})/L$, where $F$ is the free energy for helices with two identical base pair sequences and $F^{NL}$ is the free energy for two helices with different sequences. In the local base pair specific pairing mechanism, we have that $F^{NL} = 0$ and $F$ is determined through Eq. (2.37) (or Eq. (2.43)). In Fig. 8, we compare $E_R/L$ using the two mechanisms of pairing as a function of $\Lambda_c$ ($\Lambda$ for the local base pair specific pairing model). In both calculations we use Eq. (2.14), presuming Debye-Huckel interactions between sites. We see notable qualitative differences. The most significant is difference is in the gradient $dE_R/d\Lambda$. For local base pair specific pairing, as $\Lambda$ is increased, $dE_R/d\Lambda$ always increases, as the free energy $F_{local}$ falls to more negative values in a non-linear fashion in respect to $\Lambda$ due to both $\lambda_{st}$ and $\lambda_{tw}$. A significant increase in $dE_R/d\Lambda$ is seen when we move into the state for which $\lambda_{tw}$ is finite, as we increase $\Lambda$. For the other model, we see something quite different. When there is no adaptation for non-alike helices, we see an increase in $dE_R/d\Lambda_c$, like the local pairing. However, when the interaction becomes strong enough, adaptation starts occur between non-alike helices; at this point $dE_R/d\Lambda_c$ suddenly diminishes and gets smaller as $\Lambda_c$ increases. This is because, for global pairing, $F_g^{NL}$ starts to compete with $F_g$ in magnitude, for $\Lambda_c$, above this point.

A minor difference between the two pairing mechanisms is that, in the case where $\Lambda_c$ is too small for there to be any adaptation at all, we have $E_R = 0$ for local non base pair specific pairing. For the alternate model $E_R$ is never zero, except at the value $\Lambda = 0$. However, this difference should be an artefact of not including residual correlations in the azimuthal orientations of the helices to Eq. (2.51) (as well as Eq. (2.52)). In Appendix F and G we have considered the leading order corrections; however we found to obtain a transition to the finite $\lambda_h^*$ state, higher order corrections than just the next to leading order one (that we focus on in Appendices F and G) need to be considered. Nevertheless, it is not expected that accounting for such correlations will affect most significant differences between the two mechanisms of pairing, discussed above. One last feature worth pointing out is that the recognition energy of the local base pair specific pairing seems to be far more sensitive to the value of $\lambda_{eff}$, the decay length, than for that of the other mechanism.

## 3. Discussion and Outlook

In this work, we have modelled an alternate mechanism for the possible pairing of DNA molecules with identical base pair texts. This mechanism has supposed that actual the pairing of alike sequences, evidenced in experiments [6], comes from microscopic interactions between base pairs sensitive to base pair type. Two possible candidates for such interactions might be the secondary hydrogen bonding, as studied in Ref. [26], or ion mediated interactions that are discussed in Ref. [6]. This mechanism differs considerably from the pairing mechanism originally proposed in Refs. [15,16,17]. In this previous pairing description, it is the base pair dependent distortions of the DNA helices that take central role in distinguishing between like and non-alike sequences. It is similarities in the helical structure of two DNA molecules, with the same base text, that leads to preferential pairing between them. For identical molecules, the commensurateness between interaction sites is better maintained here than for two DNA molecules with different texts, whose base pair dependent patterns of distortions are different.

With generic models describing the two types of pairing mechanism, local specific and non-specific base pairing, our objective is to compare and distinguish differences in the effective interaction between helices, when the statistical mechanics of elastic rod fluctuations is considered. The goal of investigations is to look for disparities between two mechanisms that could be detected in specially designed experiments. To this end, as well the local base pair specific pairing model, we have presented an analogous pairing model for which the microscopic interactions are not base pair specific. In both models, we have used the same form of the interaction potential between sites, instead of a more complicated form of interaction like the KL theory [15,16] in the second model.

Already in this initial study, we see notable differences between the two types of pairing. For the local base pair specific pairing, what is particularly interesting is when the interaction strength is insufficient to limit thermal fluctuations, we see that the pairing free energy between two parallel DNA helices scales as $\sqrt{L}$, for long helices, where $L$ is their length. No such dependence is seen when the microscopic pairing interaction is base pair specific. In light of the experiments of Ref. [6], this result is rather intriguing, as a free energy between molecules of the form $F = \varepsilon L - \alpha\sqrt{L}$ can indeed explain the smooth continuous transformation between two sequences forming a paired loop and a fully extended molecule seen in those experiments. This behaviour can also fit the experimental data of Ref. [6].

This, however, is not the whole picture. Further analysis suggests that, if the helices are made long enough, the pairing interactions may accumulate enough strength to limit stretching fluctuation no matter what the strength of the interaction between pairing sites may be. There is a length scale $\lambda_{st}$, an adaptation length for stretching fluctuations, at which this starts to occur. For helices much longer than this length, the free energy should behave as $F \approx \varepsilon' L$. But, if the pairing interactions are sufficiently weak, then this length scale can be of the order of thousands of angstroms and this might indeed account for the observed behaviour in Ref. [6]. However, there are two distinct problems here.

One problem is that $\Lambda$ needs to be large to overcome large electrostatic repulsion, if there are no small amounts of spermine facilitating an overall average attraction per base pair [28]. Modelling the electrostatic repulsion as that between uniform charged cylinders and taking account of image charge repulsion (see Ref. [15])– if we suppose that the DNA surface charge is 70% neutralized by condensed ions, and a typical Debye screening length of $\lambda_D = 7\text{Å}$ – we obtain roughly a positive contribution to the free energy of $F_{elst} \approx 0.07 k_B T / \text{Å}$, at $25\text{Å}$ interaxial separation, or roughly $70 k_B T$ per Kuln length. For stable pairing or thermally accessible metastable pairing this suggests that, using the effective Debye interaction between sites, we would require a value $\Lambda \gtrsim 0.3 k_B T / \text{Å}$ for $\lambda_{eff} = 4\text{Å}$, and $\Lambda \gtrsim 0.8 k_B T / \text{Å}$ for $\lambda_{eff} = 2\text{Å}$. For these values of $\Lambda$, the value of $\lambda_{st}$ is of the order of tens of angstroms, not thousands. The other problem is that the results of Ref. [6] suggest that pairing persists over a large range of force values. This means that a large value of $\alpha$ (multiplying the $\sqrt{L}$ dependence) is required to fit the data, and this, in turn, should also correspond to a relatively large value of $\Lambda$.

Though we should point out that in this current study, we have considered only straight helices and not those which are bent due to bending fluctuations. It could well be that bending severely disrupts the ability of the local pairing interaction to limit the size of stretching fluctuations. If the free energy per unit length of pairing lies close to the value of electrostatic energy of repulsion, bending may also facilitate the formation of loop bubbles [6], where the paired segments come apart. This may allow for both stretching and bending fluctuations to accumulate in a random walk fashion over the entire loop, and so produce a $\sqrt{L}$ behaviour for local pairing interactions. The reason why such loop bubbles may be formed is that bending may cause the two helices to fall out of alignment within loops of the paired molecules, where one of them has a larger arc-length than the other; local pairing is likely to very sensitive to this misalignment. Thus, through such a mechanism with bending, the $\sqrt{L}$ behaviour might persist up to much larger values of $\Lambda$ and over longer lengths of DNA. All in all, the idea that local pairing interactions could, in fact, be the reason around the continuous transition observed in Ref. [6] is an appealing one, but needs further investigation.

Another important difference, highlighted in this study, is in how the recognition energy varies with respect to the interaction strength, between the two pairing mechanisms. This is inherently linked to the nature of the pairing. For local base pair specific pairing, the attractive component of the interaction between DNA helices is only (significantly) non-zero when the base pair texts are identical. When the interactions between base pairs do not depend on base pair type, non-alike helices may also have an attractive force component between them that depends on helix structure; but this is more disrupted by helix distortions than for like ones. In the latter case, to calculate the recognition energy, the pairing energy of non-alike helices needs to be subtracted away from that of identical ones. This leads to a qualitative difference in how both recognition energies change with respect to $\Lambda$. In the case of local base pair specific pairing, as $\Lambda$ is increased, the rate of increase in the recognition energy always increases. In the case of the other pairing mechanism, the behaviour is more complex. At relatively low values of $\Lambda$, the pairing interaction between non-alike helices is washed out by thermal fluctuations , and recognition energy increases in the same manner as local pairing with increasing $\Lambda$. However, when $\Lambda$ becomes sufficiently large, the pairing interaction between non-alike helices becomes strong enough to limit thermal fluctuations and quite a different behaviour is seen. Here, with further increasing $\Lambda$, the rate of change of recognition energy now diminishes. This is because we have to now subtract off the growing attractive pairing interaction between non-alike helices.

In principle, it might be possible to probe such differences experimentally, as the size of $\Lambda$ might be controlled through different ion species and concentrations, as well as temperature. However, somehow the relationship between $\Lambda$ and these factors needs to be established. One possible way there may be to establish it is through experiments that consider relatively short molecules where thermal elastic model fluctuations are not that important, where there is expected a linear relationship between the pairing energy and $\Lambda$ regardless of mechanism; if the molecules are able to pair. Still, one has to factor in bending fluctuations into the theoretical analysis and design the appropriate experiments.

What might potentially yield much clearer cut results is considering the interaction between two special periodic constructs. These two constructs, as opposed to simply just two identical sequences, would contain short tracts of a fixed number of base pairs that are identical for both constructs. Yet, these would now be separated by short tracts of a fixed number of base pairs, which are different between the two constructs. These sets of identical and non-alike sequences would then be repeated in periodic fashion. In such constructs there are two factors that can influence the pairing strength. One is the overall amount of homology, and the other is the period over which these arrays repeat between identical and non-identical sequences. DNA constructs of this form have already been considered in-vivo [10]. In the study of Ref. [10], it seemed that the periodicity was an important factor, but as the experiments were conducted in-vivo there could be many other factors involved. The sensitivity to the periodicity of the constructs, could be an indicator that the local base pair specific pairing mechanism is mainly responsible, as was inferred in Ref. [10]; but more careful investigation is needed, both experimental and theoretical. On the experimental side, one needs reconsider these constructs in specially controlled in-vitro experiments, to see if indeed the results of Ref. [10] hold there and to further quantify them. A detailed theoretical investigation, using this initial study as its starting point, is also needed to see exactly what qualitative differences the two pairing mechanisms produce for these arrays, when thermal fluctuations and helix distortions are both considered. This is to be the subject of future work.

Other aspects that deserve further investigation, for base pair specific local pairing, are the shape of the recognition well and the role of discreteness between pairing sites (relaxing the continuum approximation used to write Eq.(2.10)). The recognition well is the recognition energy as a function of $\Delta z$, the distance one helix is slid with respect to the other in the vertical parallel to their principal axes ($z$-axis). This dependence has been studied in Refs. [23,24,25], when only helix distortions matter in pairing. The shape of recognition well profile may have consequence in the how quickly two identical base pair sequences might be able recognise each other. Also, it could also give an indication how sensitive the pairing might be to the bending fluctuations, as the two sequence will fall out of alignment. In the case of rigid ideal helices, for local base pair specific pairing, the homology recognition well should be very narrow, with expected width $\lambda_{eff}$, the range of the base pair interactions. This is in marked contrast to pairing interactions allowed between all base pairs, where the width of the well would be infinite, in the case of ideal helices. However, for two long identical helices stretching fluctuations and elasticity may change this. With the former, identical bases can find each other eventually through a random walk, even though the centres of the two helices are shifted by $\Delta z$. Therefore, the recognition profile might, in fact, in the local base pair specific pairing model, be less trivial for such pairing of long molecules than naive expectations. Still, major qualitative differences may exist between the two mechanisms in the shape of the recognition well; thus, this is also worth investigating. If the pairing interaction is sufficiently weak, the effects of the discreteness between pairing sites does not matter, as the interaction modes, in $k$-space, associated with relaxing continuum approximation are washed out by thermal fluctuations [32]. However, when $\Lambda$ becomes sufficiently strong, we would expect that these modes will start to become more important; and in fully understanding the local pairing mechanism, the study of these modes may still be useful. Again, both of these investigations have been left for later work.

Another avenue of research would be to build on the kinetic studies of Ref. [31], by now including the helical structure of the molecule in simulations, along with the pattern of base pair intrinsic distortions, as given by Eqs. (2.1)-(2.5). Here, it would be interesting to what effects of helicity, hard core interactions between rods and intrinsic distortions, as well as helix flexibility, have on the kinetics of finding an exact match between identical or homologous sequences.

To round up: there is increasing evidence, in vitro, of pairing between identical DNA; and, now, some emphasis should lie on understanding its possible nature. Importantly, does the pairing exclusively depend on differences in structure that depend on base pair text, or is it actual interactions between base pairs dependent on local base pair text that matter? This paper is hopefully the first in a series of papers that compares theoretically the two modes of pairing, searching for differences that can be resolved experimentally.

## Acknowledgements

D.J. Lee would like to acknowledge stimulating discussions with A. A. Kornyshev of Imperial College; as well as those with M. Prentiss of Harvard University, who inspired this work and he is particularly grateful to. He is also thankful of the continuing support of the Department of Chemistry at Imperial College.

[35] We should point out that if we had chosen, instead of one interaction site per base pair, one for each phosphate and smeared positive charge in the DNA grooves, as well as Debye-Huckle interactions (with $\lambda_{eff}$ being the Debye screening length), but taking account of image charges on the DNA surfaces, we would recover the mean-field electrostatic model discussed in [A. Kornyshev, S. Leikin, *J. Chem. Phys*., **107**, 3656 (1997)]. The $n=0$ mode would then be the uniform cylinder electrostatic repulsion. However, as we are interested in studying generic differences between the two models, the simpler choice of either Eq. (2.14) or (2.15) is a more suitable one. This simple choice would correspond to the origin of such forces as not being mean field electrostatic in origin due to the overall attraction, for positive $\Lambda_g$. Ionic correlations [D. J. Lee, J. Phys. Cond. Matter. **23** 105102 (2011)] would correspond better to the choices of Eq. (2.14) or (2.15), unless we were to change the sign of $\Lambda_g$. But in fact, choosing negative $\Lambda_g$ shouldn't affect some of the more important qualitative features, discussed here, to do with $n \neq 0$ modes (the parts of the interaction that are helix structure dependent). On particular difference would be that the preferred orientation would be $\Delta\phi = 0$ instead of $\Delta\phi = \pi$. In this study, we stick with positive values of $\Lambda_g$. Though, at a later stage, a comparison with negative values of $\Lambda_g$ could also be considered.

# Supplemental Material

## Part 1 Local base pair specific models of homology pairing

### Appendix A Rigid helices

Our starting point, here, is Eq.(2.10) of the main text. This can be rewritten as

$$\langle V_{local}(R) \rangle_\Omega = \frac{1}{h} \int_{-L/2}^{L/2} ds \int \frac{d^3k}{(2\pi)^3} \tilde{V}_{int}(\mathbf{k}) \exp(-Rk_r \cos\phi_k)$$
$$\langle \exp(-iak_r \cos(gs + \phi_2 + \delta\phi(s) - \phi_k)) \exp(iak_r \cos(gs + \phi_1 + \delta\phi(s) - \phi_k)) \rangle_\Omega ,$$
(A.1)

where we use the cylindrical coordinates $k_r$, $\phi_k$ and $k_z$ in $k$-space and choose $\mathbf{R} = R\hat{\mathbf{i}}$. Here, we will also generalize the form for $\tilde{V}_{int}(\mathbf{k})$ by considering interaction potentials with the generalized forms

$$V_{int}^{(1)}(\mathbf{r}_1 - \mathbf{r}_2) = -\frac{\gamma_{eff}}{4\pi|\mathbf{r}_1 - \mathbf{r}_2|} \int_0^\infty d\kappa \Upsilon^{(1)}(\kappa) \exp(-\kappa|\mathbf{r}_1 - \mathbf{r}_2|),$$
(A.2)

$$V_{int}^{(2)}(\mathbf{r}_1 - \mathbf{r}_2) = -\frac{\gamma_{eff}}{4\pi} \int_0^\infty d\kappa \Upsilon^{(2)}(\kappa) \exp(-\kappa|\mathbf{r}_1 - \mathbf{r}_2|),$$
(A.3)

where the choices of $\Upsilon^{(1)}(\kappa)$ and $\Upsilon^{(2)}(\kappa)$ are arbitrary, and can be chosen to best represent the interactions between base pair sites. For the specific choices of the Debye-Huckel form and Morse potential, given by Eqs. (2.2) and (2.3) of the main text, we set the functions $\Upsilon^{(1)}(\kappa) = \delta(\kappa - 1/\lambda_{eff})$ and $\Upsilon^{(2)}(\kappa) = \delta(\kappa - 1/\lambda_{eff}) - 1/2\delta(\kappa - 2/\lambda_{eff})$ we recover those expressions. The Fourier transforms of Eq. (A.2) and (A.3) can be written as

$$\tilde{V}_{int}^{(1)}(\mathbf{k}) = -\gamma_{eff} \int_0^\infty d\kappa \frac{\Upsilon^{(1)}(\kappa)}{\mathbf{k}^2 + \kappa^2},$$
(A.4)

and

$$\tilde{V}_{int}^{(2)}(\mathbf{k}) = \gamma_{eff} \int_0^\infty d\kappa \Upsilon^{(2)}(\kappa) \frac{\partial}{\partial \kappa}\left[\frac{1}{\mathbf{k}^2 + \kappa^2}\right].$$
(A.5)

Now, to evaluate Eq. (A.1) further, it is possible to use the following mathematical identities

$$\exp(iak_r \cos(gs + \phi_1 + \delta\phi(s) - \phi_k)) \equiv \sum_{n=-\infty}^\infty i^n J_n(ak_r) \exp(in(gs + \phi_1 + \delta\phi(s) - \phi_k)),$$
(A.6)

$$\exp(-iRk_r \cos\phi_k) \equiv \sum_{m=-\infty}^\infty i^{-m} J_m(Rk_r) \exp(-im\phi_k),$$
(A.7)

$$\exp\left(-iak_r \cos\left(gs + \phi_2 + \delta\phi(s) - \phi_k\right)\right) \equiv \sum_{n'=-\infty}^{\infty} i^{-n'} J_{n'}(ak_r) \exp\left(-in\left(gs + \phi_2 + \delta\phi(s) - \phi_k\right)\right), \quad (A.8)$$

where $J_n(ak_r)$ is a Bessel function of the first kind of order $n$. This allows us to perform the $\phi_k$ integration and write either

$$\langle V_{local}(R)\rangle_\Omega = -\frac{\gamma_{eff}}{(2\pi)^2} \int_{-L/2}^{L/2} ds \int_{-\infty}^{\infty} dk_z \int_0^{\infty} k_r dk_r \sum_{n=-\infty}^{\infty} \sum_{n'=-\infty}^{\infty} J_{n-n'}(Rk_r) J_n(ak_r) J_{n'}(ak_r)$$

$$\exp(i(n-n')gs) \exp(i(\phi_1 n - \phi_2 n')) \exp(i\delta\phi(s)(n-n')) \int_0^{\infty} d\kappa \frac{\Upsilon^{(1)}(\kappa)}{k_z^2 + k_r^2 + \kappa^2}, \quad (A.9)$$

or

$$\langle V_{local}(R)\rangle_\Omega = -\frac{\gamma_{eff}}{(2\pi)^2} \int_{-L/2}^{L/2} ds \int_{-\infty}^{\infty} dk_z \int_0^{\infty} k_r dk_r \sum_{n=-\infty}^{\infty} \sum_{n'=-\infty}^{\infty} \exp(i(n-n')gs) \exp(i(\phi_1 n - \phi_2 n'))$$

$$\exp(i\delta\phi(s)(n-n')) \int_0^{\infty} d\kappa \, \Upsilon^{(2)}(\kappa) \frac{\partial}{\partial \kappa} \left[\frac{J_{n-n'}(Rk_r) J_n(ak_r) J_{n'}(ak_r)}{k_z^2 + k_r^2 + \kappa^2}\right], \quad (A.10)$$

depending on the choice of interaction potential, Eq. (A.4) and (A.5). The $k_r$ integrals can be evaluated using the formula

$$\int_0^{\infty} k_r dk_r \frac{J_{n-n'}(Rk_r) J_n(ak_r) J_{n'}(ak_r)}{k_z^2 + k_r^2 + \kappa^2} = (-1)^{n'} K_{n-n'}\left(R\sqrt{k_z^2 + \kappa^2}\right) I_n\left(a\sqrt{k_z^2 + \kappa^2}\right) I_{n'}\left(a\sqrt{k_z^2 + \kappa^2}\right),$$

(A.11)

where $I_n(x)$ and $K_n(x)$ are order $n$ modified Bessel functions of the first and second kinds, respectively. This yields the generic form for the ensemble averaged interaction energy

$$\langle V_{local}(R)\rangle_\Omega = -\Lambda \int_{-L/2}^{L/2} ds \sum_{n=-\infty}^{\infty} \sum_{n'=-\infty}^{\infty} \exp(i(n-n')gs) \exp(i(\phi_1 n - \phi_2 n')) \langle \exp(i\delta\phi(s)(n-n'))\rangle_\Omega$$

$$\int_{-\infty}^{\infty} dk_z (-1)^{n'} \bar{G}_{n,n'}(R/a, k_z), \quad (A.12)$$

with either the form

$$\bar{G}_{n,n'}(R/a, k_z) = \bar{G}^{(1)}_{n,n'}(R/a, k_z)$$

$$= \int_0^{\infty} d\kappa \, \Upsilon^{(1)}(\kappa) K_{n-n'}\left(\frac{R}{a}\sqrt{k_z^2 + a^2\kappa^2}\right) I_n\left(\sqrt{k_z^2 + a^2\kappa^2}\right) I_{n'}\left(\sqrt{k_z^2 + a^2\kappa^2}\right), \quad (A.13)$$

or

$$\bar{G}_{n,n'}(R/a, k_z) = \bar{G}_{n,n'}^{(2)}(R/a, k_z)$$

$$= -\int_0^\infty d\kappa \frac{a^2 \kappa \Upsilon^{(2)}(\kappa)}{\sqrt{k_z^2 + a^2\kappa^2}} \left[ \left[ I_n'\left(\sqrt{k_z^2 + \kappa^2 a^2}\right) I_{n'}\left(\sqrt{k_z^2 + \kappa^2 a^2}\right) \right. \right.$$

$$\left. + I_n\left(\sqrt{k_z^2 + \kappa^2 a^2}\right) I_{n'}'\left(\sqrt{k_z^2 + \kappa^2 a^2}\right) \right] K_{n-n'}\left(\frac{R}{a}\sqrt{k_z^2 + \kappa^2 a^2}\right) \quad (A.14)$$

$$\left. + \frac{R}{a} I_n\left(\sqrt{k_z^2 + \kappa^2 a^2}\right) I_{n'}\left(\sqrt{k_z^2 + \kappa^2 a^2}\right) K_{n-n'}'\left(\frac{R}{a}\sqrt{k_z^2 + \kappa^2 a^2}\right) \right],$$

where $\Lambda = \gamma_{eff} / ha(2\pi)^2$. Here, $I_n'(x)$ and $K_n'(x)$ are the derivatives with respect to argument of the modified Bessel functions. With the choices $\Upsilon^{(1)}(\kappa) = \delta(\kappa - 1/\lambda_{eff})$ and $\Upsilon^{(2)}(\kappa) = \delta(\kappa - 1/\lambda_{eff}) - 1/2\delta(\kappa - 2/\lambda_{eff})$, Eqs. (A.12)-(A.14) reduce to Eqs. (2.13)-(2.16) of the main text. Now we consider calculating the ensemble average in Eq. (A.12). This is defined as

$$\left\langle \exp(i\delta\phi(s)(n-n')) \right\rangle_\Omega = \frac{\int D\delta\Omega(s) P[\delta\Omega(s)] \exp\left( -\frac{(n-n')}{h} \int_0^s ds' \delta\Omega(s') \right)}{\int D\delta\Omega(s) P[\delta\Omega(s)]}. \quad (A.15)$$

We assume that patterns of twist distortions, $\delta\Omega(s)$ to be Gaussian distributed over all base pair realizations, with Eq. (2.6) holding. Thus, we may write

$$P[\delta\Omega(s)] = \exp\left( -\frac{\lambda_{tw}^{(0)}}{2h^2} \int_{-L/2}^{L/2} ds \delta\Omega(s)^2 \right). \quad (A.16)$$

The functional integrals over $\delta\Omega(s)$, in Eq. (A.15), sum over all possible realizations of base pair sequences. Evaluation of Eq. (A.15) yields

$$\left\langle \exp(i\delta\phi(s)(n-n')) \right\rangle_\Omega = \exp\left( -\frac{(n-n')^2 |s|}{2\lambda_{tw}^{(0)}} \right), \quad (A.17)$$

and, thus, we may write

$$\left\langle V_{local}(R) \right\rangle_\Omega = -\Lambda \int_{-L/2}^{L/2} ds \sum_{n=-\infty}^{\infty} \sum_{n'=-\infty}^{\infty} \exp(i(n-n')gs) \exp(i(\phi_1 n - \phi_2 n')) \exp\left( -\frac{(n-n')^2 |s|}{2\lambda_{tw}^{(0)}} \right)$$

$$\int_{-\infty}^{\infty} dk_z (-1)^{n'} \bar{G}_{n,n'}(R/a, k_z). \quad (A.18)$$

We can then perform the $s$ integration yielding

$$\left\langle V_{helix}(R) \right\rangle_\Omega = -\frac{\Lambda}{g} \sum_{n=-\infty}^{\infty} \sum_{n'=-\infty}^{\infty} \exp(i(\phi_1 n - \phi_2 n')) \int_{-\infty}^{\infty} dk_z (-1)^{n'} l_{n,n'}(\Gamma, Lg; k_z) \bar{G}_{n,n'}(R/a, k_z), \quad (A.19)$$

where

$$\Gamma = \frac{(n-n')^2}{\lambda_{tw}^{(0)} g}, \tag{A.20}$$

and

$$l_{n,n'}(\Gamma, Lg; k_z) = \frac{\Gamma}{\left(\frac{\Gamma}{2}\right)^2 + (n-n')^2} \left[1 - \exp\left(-\frac{Lg\Gamma}{4}\right)\cos\left(\frac{Lg}{2}(n-n')\right)\right]$$
$$+ \frac{2(n-n')}{\left(\frac{\Gamma}{2}\right)^2 + (n-n')^2} \exp\left(-\frac{Lg\Gamma}{4}\right)\sin\left(\frac{Lg}{2}(n-n')\right). \tag{A.21}$$

For $\lambda_{tw}^{(0)} g \gg 1$, we approximate Eq. (A.21) with

$$l_{n,n'}(\Gamma, Lg; k_z) \approx \frac{2}{(n-n')}\sin\left(\frac{Lg}{2}(n-n')\right). \tag{A.22}$$

So when $Lg \gg 1$, the $n = n'$ modes dominates and we have that $l_{n,n}(\Gamma, Lg; k_z) \approx Lg$. Thus, Eq.(A.19) simply reduces to Eq. (2.17) of the main text. In the main text, we make the choices $\bar{G}_{n,n'}(R/a, k_z) = G_{n,n'}^{DH}(R/a, a\kappa_{eff}, k_z)$ and $\bar{G}_{n,n'}(R/a, k_z) = G_{n,n'}^{M}(R/a, a\kappa_{eff}, k_z)$ (where $G_{n,n'}^{DH}(R/a, a\kappa_{eff}, k_z)$ and $G_{n,n'}^{M}(R/a, a\kappa_{eff}, k_z)$ are given by Eqs. (2.14)-(2.16) of the main text respectively), but we may also use the form of Eq. (2.17) with the general expressions given by Eqs. (A.13) and (A.14).

## Appendix B Thermal fluctuations for weak local pairing interaction of short helices

Here, we examine in detail the expansion given by Eqs. (2.23) and (2.24) of the main text. Namely,

$$Z = \sum_{n=0}^{\infty} Z^{(n)}, \tag{B.1}$$

with

$$Z^{(n)} = \frac{(-1)^n}{n!} \int D\delta\phi_{1,T}(s) \int D\delta\phi_{2,T}(s) \int D\delta s_{1,T}(s) \int D\delta s_{2,T}(s)$$
$$\left(\frac{V_{local}[\delta\phi_{1,T}(s), \delta\phi_{2,T}(s), \delta s_{1,T}(s), \delta s_{2,T}(s)]}{k_B T}\right)^n \exp\left(-\frac{E_{el}[\delta\phi_{1,T}(s), \delta\phi_{2,T}(s), \delta s_{1,T}(s), \delta s_{2,T}(s)]}{k_B T}\right).$$

(B.2)

Here, we retain terms in the expansion (Eq. (B.1)) up to $n = 2$. Thus, for the free energy, we obtain the expression

$$F_{local} \approx -k_B T \ln\left(Z^{(0)} + Z^{(1)} + Z^{(2)}\right) \approx -k_B T \ln Z^{(0)} + \left\langle V_{local}[\delta\phi_{1,T}(s), \delta\phi_{2,T}(s), \delta s_{1,T}(s), \delta s_{2,T}(s)]\right\rangle_0$$
$$-\frac{1}{2k_B T}\left[\left\langle V_{local}[\delta\phi_{1,T}(s), \delta\phi_{2,T}(s), \delta s_{1,T}(s), \delta s_{2,T}(s)]^2\right\rangle_0 - \left\langle V_{local}[\delta\phi_{1,T}(s), \delta\phi_{2,T}(s), \delta s_{1,T}(s), \delta s_{2,T}(s)]\right\rangle_0^2\right],$$
(B.3)

where the thermal averages are given by

$$\left\langle V_{local}[\delta\phi_{1,T}(s), \delta\phi_{2,T}(s), \delta s_{1,T}(s), \delta s_{2,T}(s)]\right\rangle_0 = -k_B T Z^{(1)}/Z^{(0)} \tag{B.4}$$

and

$$\left\langle V_{local}[\delta\phi_{1,T}(s), \delta\phi_{2,T}(s), \delta s_{1,T}(s), \delta s_{2,T}(s)]^2\right\rangle_0 = (k_B T)^2 Z^{(2)}/Z^{(0)} \tag{B.5}$$

First, let us consider the ensemble average of the first thermal average given by Eq. (B.4). Firstly, using Eqs. (2.10) and (2.18) of the main text, this may be written as

$$\left\langle\left\langle V_{local}[\delta\phi_{1,T}(s), \delta\phi_{2,T}(s), \delta s_{1,T}(s), \delta s_{2,T}(s)]\right\rangle_0\right\rangle_\Omega = \frac{1}{h}\int_{-L/2}^{L/2} ds \int \frac{d^3k}{(2\pi)^3}\tilde{V}_{int}(\mathbf{k})\exp(-iRk_r \cos\phi_k)$$
$$\left\langle\left\langle \exp\left(ik_z\left(\delta s_{1,T}(s) - \delta s_{2,T}(s)\right)\right)\exp\left(-iak_r \cos\left(gs + \phi_2 + \delta\phi(s) + \delta\phi_{2,T}(s) - \phi_k\right)\right)\right.\right.$$
$$\left.\left.\exp\left(iak_r \cos\left(gs + \phi_1 + \delta\phi(s) + \delta\phi_{1,T}(s) - \phi_k\right)\right)\right\rangle_0\right\rangle_\Omega.$$

(B.6)

Following similar steps, as in the Appendix A, one is able to re-express Eq.(B.6) as

$$\left\langle\left\langle V_{local}[\delta\phi_{1,T}(s), \delta\phi_{2,T}(s), \delta s_{1,T}(s), \delta s_{2,T}(s)]\right\rangle_0\right\rangle_\Omega = -\Lambda \int_{-L/2}^{L/2} ds \sum_{n=-\infty}^{\infty}\sum_{n'=-\infty}^{\infty}\exp(i(n-n')gs)\exp(i(\phi_1 n - \phi_2 n'))$$
$$\left\langle\exp(i\delta\phi(s)(n-n'))\right\rangle_\Omega \left\langle\exp\left(\frac{ik_z}{a}\left(\delta s_{1,T}(s) - \delta s_{2,T}(s)\right)\right)\exp(in\delta\phi_{1,T}(s) - in'\delta\phi_{2,T}(s))\right\rangle \int_0^\infty dk_z (-1)^{n'} \bar{G}_{n,n'}(R/a, k_z).$$
(B.7)

where $k_z$ has been rescaled by $a$ to be dimensionless. Note that the energy functional presented in the main text (Eq. (2.19)), corresponds only to contribution to the modes where $n = n'$, which are shown to be dominant in Eq. (B.7) in the analysis below. Such 'diagonal' modes are also expected to be dominant for each successive term in the expansion, Eq. (B.1). This means we will always make the approximation that $\left\langle V_{local}(R)\right\rangle_\Omega \approx V_{local}(R)$; i.e. the ensemble average over base realizations in Eq. (B.3) is assumed not to matter.

The ensemble average is again given by Eq. (A.17), whereas the thermal average may be expressed as

$$\left\langle \exp\left(\frac{ik_z}{a}\left(\delta s_{1,T}(s) - \delta s_{2,T}(s)\right)\right) \exp\left(in\delta\phi_{1,T}(s) - in'\delta\phi_{2,T}(s)\right) \right\rangle_0$$
$$= \left\langle \exp\left(\frac{ik_z}{a}\left(\delta s_{1,T}(s) - \delta s_{2,T}(s)\right)\right) \right\rangle_s \left\langle \exp\left(in\delta\phi_{1,T}(s) - in'\delta\phi_{2,T}(s)\right) \right\rangle_\phi,$$

(B.8)

where

$$\left\langle \exp\left(\frac{ik_z}{a}\left(\delta s_{1,T}(s) - \delta s_{2,T}(s)\right)\right) \right\rangle_s = \frac{1}{Z_s} \int D\delta s_{1,T}(s) \int D\delta s_{2,T}(s) \exp\left(\frac{ik_z}{a}\left(\delta s_{1,T}(s) - \delta s_{2,T}(s)\right)\right)$$
$$\exp\left(-\frac{E_{st}[\delta s_{1,T}(s), \delta s_{2,T}(s)]}{k_B T}\right),$$

(B.9)

$$\left\langle \exp\left(in\delta\phi_{1,T}(s) - in'\delta\phi_{2,T}(s)\right) \right\rangle_\phi = \frac{1}{Z_\phi} \int D\delta\phi_{1,T}(s) \int D\delta\phi_{2,T}(s) \exp\left(i\left(n\delta\phi_{1,T}(s) - n'\delta\phi_{2,T}(s)\right)\right)$$
$$\exp\left(-\frac{E_{tw}[\delta\phi_{1,T}(s), \delta\phi_{2,T}(s)]}{k_B T}\right),$$

(B.10)

and

$$Z_s = \int D\delta s_{1,T}(s) \int D\delta s_{2,T}(s) \exp\left(-\frac{E_{st}[\delta s_{1,T}(s), \delta s_{2,T}(s)]}{k_B T}\right), \quad (B.11)$$

$$Z_\phi = \int D\delta\phi_{1,T}(s) \int D\delta\phi_{2,T}(s) \exp\left(-\frac{E_{tw}[\delta\phi_{1,T}(s), \delta\phi_{2,T}(s)]}{k_B T}\right). \quad (B.12)$$

The elastic energy functionals $E_{st}$ and $E_{tw}$ are given by Eqs. (2.20) and (2.21) of the main text. One can make the functional change of variables from $\delta\phi_{\mu,T}(s)$ and $\delta s_{\mu,T}(s)$ to $\delta\dot\phi_{\mu,T}(s)$ and $\delta\dot s_{\mu,T}(s)$ (the derivatives of $\delta\phi_{\mu,T}(s)$ and $\delta s_{\mu,T}(s)$ with respect to their argument) so that we may write (an alternate method is through Fourier transforms of $\delta\phi_{\mu,T}(s)$ and $\delta s_{\mu,T}(s)$)

$$\left\langle \exp\left(in\delta\phi_{1,T}(s) - in'\delta\phi_{2,T}(s)\right) \right\rangle_\phi = \frac{1}{\tilde Z_\phi} \int D\delta\dot\phi_{1,T}(s) \int D\delta\dot\phi_{2,T}(s) \exp\left(i\left(\int_0^s ds'\left[n\delta\dot\phi_{1,T}(s') - n'\delta\dot\phi_{2,T}(s')\right]\right)\right)$$
$$\exp\left(-\frac{l_{tw}}{2} \int_{-L/2}^{L/2} ds'\left[\delta\dot\phi_{1,T}(s')^2 + \delta\dot\phi_{2,T}(s')^2\right]\right),$$

(B.13)

$$\left\langle \exp\left(\frac{ik_z}{a}\left(\delta s_{1,T}(s) - \delta s_{2,T}(s)\right)\right)\right\rangle_s = \frac{1}{\tilde{Z}_s}\int D\delta\dot{s}_{1,T}(s)\int D\delta\dot{s}_{2,T}(s)\exp\left(\frac{ik_z}{a}\int_0^s ds'\left(\delta\dot{s}_{1,T}(s') - \delta\dot{s}_{2,T}(s')\right)\right)$$

$$\exp\left(-\frac{l_{st}}{2}\int_{-L/2}^{L/2} ds'\left[\delta\dot{s}_{1,T}(s')^2 + \delta\dot{s}_{2,T}(s')^2\right]\right),$$

(B.14)

$$\tilde{Z}_\phi = \int D\delta\dot{\phi}_{1,T}(s)\int D\delta\dot{\phi}_{2,T}(s)\exp\left(-\frac{l_{tw}}{2}\int_{-L/2}^{L/2} ds'\left[\delta\dot{\phi}_{1,T}(s')^2 + \delta\dot{\phi}_{2,T}(s')^2\right]\right),$$

(B.15)

$$\tilde{Z}_s = \int D\delta\dot{s}_{1,T}(s)\int D\delta\dot{s}_{2,T}(s)\exp\left(-\frac{l_{st}}{2}\int_{-L/2}^{L/2} ds'\left[\delta\dot{s}_{1,T}(s')^2 + \delta\dot{s}_{2,T}(s')^2\right]\right).$$

(B.16)

where we make the choice to fix $\delta\phi_{1,T}(0) = \delta\phi_{2,T}(0) = \delta s_{1,T}(0) = \delta s_{2,T}(0)$ so that $\phi_1$ and $\phi_2$ are still chosen to be the azimuthal orientations of the helices at the centres. The functional integrations are then readily performed yielding

$$\left\langle \exp\left(in\delta\phi_{1,T}(s) - in'\delta\phi_{2,T}(s)\right)\right\rangle_\phi = \exp\left(-\frac{(n^2+n'^2)|s|}{2l_{tw}}\right),$$

(B.17)

and

$$\left\langle \exp\left(i\frac{\left(\delta s_{1,T}(s) - \delta s_{2,T}(s)\right)}{a}k_z\right)\right\rangle_s = \exp\left(-\frac{k_z^2}{a^2g^2}\frac{|s|}{l_{st}}\right).$$

(B.18)

Using Eqs. (A.17), (B.17) and (B.18) we may write Eq. (B.7) as

$$\left\langle V_{local}[\delta\phi_{1,T}(s), \delta\phi_{2,T}(s), \delta s_{1,T}(s), \delta s_{2,T}(s)]\right\rangle_0 = -\frac{\Lambda}{g}\int_{-Lg/2}^{Lg/2} dx \sum_{n=-\infty}^{\infty}\sum_{n'=-\infty}^{\infty} \exp(i(n-n')x)\exp(i(\phi_1 n - \phi_2 n'))\int_{-\infty}^{\infty} dk_z(-1)^{n'}$$

$$\bar{G}_{n,n'}(R/a, k_z)\exp\left(-\frac{\tilde{\Gamma}}{2}|x|\right),$$

(B.19)

where

$$\tilde{\Gamma} = \frac{n^2 + n'^2}{l_{tw}g} + \frac{(n-n')^2}{\lambda_{tw}^{(0)}g} + \frac{2k_z^2}{g^3 l_{st}a^2}.$$

(B.20)

On performing integration over $x$, Eq. (B.19) evaluates to

$$\langle V_{local}[\delta\phi_{1,T}(s),\delta\phi_{2,T}(s),\delta s_{1,T}(s),\delta s_{2,T}(s)]\rangle_0 =$$
$$-\frac{\Lambda}{g}\sum_{n=-\infty}^{\infty}\sum_{n'=-\infty}^{\infty}\exp\left(i(\phi_1 n-\phi_2 n')\right)\int_{-\infty}^{\infty}dk_z(-1)^{n'}l_{n,n'}(\tilde{\Gamma},Lg;k_z)\bar{G}_{n,n'}(R/a,k_z), \quad (B.21)$$

where the form of $l_{n,n'}(\tilde{\Gamma},Lg;k_z)$ is again given by Eq. (A.21), but now with $\tilde{\Gamma}$ replacing $\Gamma$. By inspecting Eq. (A.21), we see that, provided that $g^3 l_{st} a^2 \gg 1$ as well as $l_{tw} g \gg 1$, we may neglect again $n \neq n'$ modes. In this case $l_{n,n}(\tilde{\Gamma},Lg;k_z)$ reduces to

$$l_{n,n}(\tilde{\Gamma},Lg;k_z) \approx \frac{2l_{tw}g}{n^2+\frac{k_z^2 l_{tw}}{a^2 g^2 l_{st}}}\left[1-\exp\left(-\frac{L}{2l_{tw}}\left(n^2+\frac{l_{tw}k_z^2}{g^2 l_{st}a^2}\right)\right)\right]. \quad (B.22)$$

Thus, substituting Eq. (B.22) into Eq. (B.19), we obtain Eq. (2.27) of the main text.

Next, we consider the next to leading order contribution contained in Eq. (B.3), where now we focus on the dominant $n=n'$ modes, so that $\langle V_{helix}(R)\rangle_\Omega \approx V_{helix}(R)$. Thus, we can write

$$\langle V_{local}[\delta\phi_{1,T}(s),\delta\phi_{2,T}(s),\delta h_{1,T}(s),\delta h_{2,T}(s)]^2\rangle_0 = \Lambda^2 \int_{-L/2}^{L/2}ds\int_{-L/2}^{L/2}ds'\sum_{n=-\infty}^{\infty}\sum_{m=-\infty}^{\infty}\exp(im(\phi_1-\phi_2))\exp(in(\phi_1-\phi_2))$$
$$\langle\exp(in(\delta\phi_{1,T}(s)-\delta\phi_{2,T}(s)))\exp(im(\delta\phi_{1,T}(s')-\delta\phi_{2,T}(s')))\rangle_\phi$$
$$\int_{-\infty}^{\infty}dk_z\int_{-\infty}^{\infty}dk_z'(-1)^{n+m}\bar{G}_{n,n}(R/a,k_z)\bar{G}_{m,m}(R/a,k_z')\left\langle\exp\left(i(\delta s_{1,T}(s)-\delta s_{2,T}(s))\frac{k_z}{a}\right)\exp\left(i(\delta s_{1,T}(s')-\delta s_{2,T}(s'))\frac{k_z'}{a}\right)\right\rangle_s.$$
$$(B.23)$$

Thus, we need to evaluate the averages

$$\langle\exp(in(\delta\phi_{1,T}(s)-\delta\phi_{2,T}(s)))\exp(im(\delta\phi_{1,T}(s')-\delta\phi_{2,T}(s')))\rangle_\phi$$
$$=\frac{1}{Z_\phi}\int D\delta\phi_{1,T}(s)\int D\delta\phi_{2,T}(s)\exp(in(\delta\phi_{1,T}(s)-\delta\phi_{2,T}(s)))\exp(im(\delta\phi_{1,T}(s')-\delta\phi_{2,T}(s'))) \quad (B.24)$$
$$\exp\left(-\frac{E_{tw}[\delta\phi_{1,T}(s),\delta\phi_{2,T}(s)]}{k_B T}\right),$$

$$\left\langle\exp\left(i(\delta s_{1,T}(s)-\delta s_{2,T}(s))\frac{k_z}{a}\right)\exp\left(i(\delta s_{1,T}(s')-\delta s_{2,T}(s'))\frac{k_z'}{a}\right)\right\rangle_s$$
$$=\frac{1}{Z_s}\int D\delta s_{1,T}(s)\int D\delta s_{2,T}(s)\exp\left(i\frac{k_z}{a}(\delta s_{1,T}(s)-\delta s_{2,T}(s))\right)\exp\left(i\frac{k_z'}{a}(\delta s_{1,T}(s')-\delta s_{2,T}(s'))\right)$$
$$\exp\left(-\frac{E_{st}[\delta s_{1,T}(s),\delta s_{2,T}(s)]}{k_B T}\right),$$

$$(B.25)$$

We can express the averages in terms of the functions $\Delta\dot\phi_T(s) = \dfrac{d\delta\phi_{1,T}(s)}{ds} - \dfrac{d\delta\phi_{2,T}(s)}{ds}$ and $\Delta\dot s_T(s) = \dfrac{d\delta s_{1,T}(s)}{ds} - \dfrac{d\delta s_{2,T}(s)}{ds}$, thus rewriting Eqs. (B.24) and (B.25) as

$$\left\langle \exp\left(in\left(\delta\phi_{1,T}(s) - \delta\phi_{2,T}(s)\right)\right)\exp\left(im\left(\delta\phi_{1,T}(s') - \delta\phi_{2,T}(s')\right)\right)\right\rangle_\phi$$

$$= \frac{1}{\bar Z_\phi}\int D\Delta\dot\phi_T(s)\left[\exp\left(in\int_0^s \Delta\dot\phi_T(s'')ds''\right)\exp\left(im\int_0^{s'}\Delta\dot\phi_T(s'')ds''\right)\theta(s)\theta(s')\right.$$

$$+\exp\left(-in\int_s^0\Delta\dot\phi_T(s'')ds''\right)\exp\left(im\int_0^{s'}\Delta\dot\phi_T(s'')ds''\right)\theta(-s)\theta(s')$$

$$+\exp\left(in\int_0^s\Delta\dot\phi_T(s'')ds''\right)\exp\left(-im\int_{s'}^0\Delta\dot\phi_T(s'')ds''\right)\theta(s)\theta(-s')$$

$$\left.+\exp\left(-in\int_s^0\Delta\dot\phi_T(s'')ds''\right)\exp\left(-im\int_{s'}^0\Delta\dot\phi_T(s'')ds''\right)\theta(-s)\theta(-s')\right]\exp\left(-\frac{l_{tw}}{4}\int_{-L/2}^{L/2}\Delta\dot\phi_T(s'')^2 ds''\right),$$

(B.26)

and

$$\left\langle \exp\left(\frac{ik_z}{a}\left(\delta s_{1,T}(s) - \delta s_{2,T}(s)\right)\right)\exp\left(\frac{ik'_z}{a}\left(\delta s_{1,T}(s') - \delta s_{2,T}(s')\right)\right)\right\rangle_s$$

$$= \frac{1}{\bar Z_s}\int D\Delta\dot s_T(s)\left[\exp\left(\frac{ik_z}{a}\int_0^s\Delta\dot s_T(s'')ds''\right)\exp\left(\frac{ik'_z}{a}\int_0^{s'}\Delta\dot s_T(s'')ds''\right)\theta(s)\theta(s')\right.$$

$$+\exp\left(-\frac{ik_z}{a}\int_s^0\Delta\dot s_T(s'')ds''\right)\exp\left(\frac{ik'_z}{a}\int_0^{s'}\Delta\dot s_T(s'')ds''\right)\theta(-s)\theta(s')$$

$$+\exp\left(\frac{ik_z}{a}\int_0^s\Delta\dot s_T(s'')ds''\right)\exp\left(-\frac{ik'_z}{a}\int_{s'}^0\Delta\dot s_T(s'')ds''\right)\theta(s)\theta(-s')$$

$$\left.+\exp\left(-\frac{ik_z}{a}\int_s^0\Delta\dot s_T(s'')ds''\right)\exp\left(-\frac{ik'_z}{a}\int_{s'}^0\Delta\dot s_T(s'')ds''\right)\theta(-s)\theta(-s')\right]\exp\left(-\frac{l_{st}}{4}\int_{-L/2}^{L/2}\Delta\dot s_T(s'')^2 ds''\right),$$

(B.27)

where

$$\bar Z_\phi = \int D\Delta\dot\phi_T(s)\exp\left(-\frac{l_{tw}}{4}\int_{-L/2}^{L/2}\Delta\dot\phi_T(s'')^2 ds''\right),\tag{B.28}$$

and

$$\bar{Z}_s = \int D\Delta\dot{s}_T(s)\exp\left(-\frac{l_{st}}{4}\int_{-L/2}^{L/2}\Delta\dot{s}_T(s'')^2\,ds''\right). \tag{B.29}$$

In writing Eqs.(B.26) and (B.27), we have also used the theta function, defined as

$$\theta(s) = 1 \quad \text{when} \quad s \geq 0, \tag{B.30}$$

$$\theta(s) = 0 \quad \text{when} \quad s < 0. \tag{B.31}$$

One can do the functional integration in Eq. (B.26) and so obtain

$$\left\langle \exp\left(in\left(\delta\phi_{1,T}(s) - \delta\phi_{2,T}(s)\right)\right)\exp\left(im\left(\delta\phi_{1,T}(s') - \delta\phi_{2,T}(s')\right)\right)\right\rangle_\phi$$

$$= \exp\left(-\frac{1}{l_{tw}}\int_0^\infty \left[n^2\theta(s-s'') + 2mn\theta(s-s'')\theta(s'-s'') + m^2\theta(s'-s'')\right]ds''\right)\theta(s)\theta(s')$$

$$+ \exp\left(-\frac{1}{l_{tw}}\int_{-\infty}^0 \left[n^2\theta(s''-s) + 2mn\theta(s''-s)\theta(s''-s') + m^2\theta(s''-s')\right]ds''\right)\theta(-s)\theta(-s')$$

$$+ \exp\left(-\frac{n^2}{l_{tw}}\int_0^s ds'' - \frac{m^2}{l_{tw}}\int_{s'}^0 ds''\right)\theta(s)\theta(-s') + \exp\left(-\frac{n^2}{l_{tw}}\int_s^0 ds'' - \frac{m^2}{l_{tw}}\int_0^{s'} ds''\right)\theta(-s)\theta(s')$$

$$= \exp\left(-\frac{1}{l_{tw}}\left(n^2 s + 2nm\min(s,s') + m^2 s'\right)\right)\theta(s)\theta(s') + \exp\left(\frac{1}{l_{tw}}\left(n^2 s + 2nm\max(s,s') + m^2 s'\right)\right)\theta(-s)\theta(-s')$$

$$+ \exp\left(-\frac{1}{l_{tw}}\left(n^2 s - m^2 s'\right)\right)\theta(s)\theta(-s') + \exp\left(-\frac{1}{l_{tw}}\left(m^2 s' - n^2 s\right)\right)\theta(-s)\theta(s'). \tag{B.32}$$

The functions $\min(s,s')$ and $\max(s,s')$ are defined as

$$\min(s,s') = s \quad \text{when} \quad s \leq s',$$

$$\min(s,s') = s' \quad \text{when} \quad s' < s, \tag{B.33}$$

$$\max(s,s') = s \quad \text{when} \quad s \geq s',$$

$$\max(s,s') = s' \quad \text{when} \quad s' > s. \tag{B.34}$$

Similarly, we find from Eq. (B.27), following similar steps, that we may write a similar expression for the average over stretching fluctuations. This reads as:

$$\left\langle \exp\left(i\left(\delta s_{1,T}(s)-\delta s_{2,T}(s)\right)\frac{k_z}{a}\right)\exp\left(i\left(\delta s_{1,T}(s')-\delta s_{2,T}(s')\right)\frac{k'_z}{a}\right)\right\rangle_s$$

$$=\exp\left(-\frac{1}{a^2g^2l_{st}}\left(k_z^2 s+2k_zk'_z\min(s,s')+k_z'^2 s'\right)\right)\theta(s)\theta(s')+\exp\left(-\frac{1}{a^2g^2l_{st}}\left(k_z^2 s-k_z'^2 s'\right)\right)\theta(s)\theta(-s')$$

$$+\exp\left(\frac{1}{a^2g^2l_{st}}\left(k_z^2 s+2k_zk'_z\max(s,s')+k_z'^2 s'\right)\right)\theta(-s)\theta(-s')+\exp\left(-\frac{1}{a^2g^2l_{st}}\left(k_z'^2 s'-k_z^2 s\right)\right)\theta(-s)\theta(s').$$

(B.35)

Combining Eqs. (B.23), (B.32) and (B.35) allows us to write

$$\left\langle V_{local}[\delta\phi_{1,T}(s),\delta\phi_{2,T}(s),\delta h_{1,T}(s),\delta h_{2,T}(s)]^2\right\rangle_0 = \Lambda^2 \sum_{n=-\infty}^{\infty}\sum_{m=-\infty}^{\infty}\exp(im(\phi_1-\phi_2))\exp(in(\phi_1-\phi_2))$$

$$\int_{-\infty}^{\infty} dk_z \int_{-\infty}^{\infty} dk'_z (-1)^{n+m}\bar{G}_{n,n}(R/a,k_z)\bar{G}_{m,m}(R/a,k'_z)$$

$$\Xi\left(k_z^2/(a^2g^2l_{st})+n^2/l_{tw},k_zk'_z/(a^2g^2l_{st})+nm/l_{tw},k_z'^2/(a^2g^2l_{st})+m^2/l_{tw}\right),$$

(B.36)

where

$$\Xi\left(k_z^2/(a^2g^2l_{st})+n^2/l_{tw},k_zk'_z/(a^2g^2l_{st})+nm/l_{tw},k_z'^2/(a^2g^2l_{st})+m^2/l_{tw}\right)$$

$$=\int_0^{L/2} ds \int_0^{L/2} ds' \exp\left(-s\left(k_z^2/(a^2g^2l_{st})+n^2/l_{tw}\right)\right)\exp\left(-s'\left(k_z'^2/(a^2g^2l_{st})+m^2/l_{tw}\right)\right)$$

$$\exp\left(-2\min(s,s')\left(k_zk'_z/(a^2g^2l_{st})+nm/l_{tw}\right)\right)$$

$$+\int_{-L/2}^{0} ds \int_{-L/2}^{0} ds' \exp\left(s\left(k_z^2/(a^2g^2l_{st})+n^2/l_{tw}\right)\right)\exp\left(s'\left(k_z'^2/(a^2g^2l_{st})+m^2/l_{tw}\right)\right)$$

$$\exp\left(2\max(s,s')\left(k_zk'_z/(a^2g^2l_{st})+nm/l_{tw}\right)\right)$$

$$+\int_0^{L/2} ds \int_{-L/2}^{0} ds' \exp\left(-\left(\left(n^2/l_{tw}+k_z^2/a^2g^2l_{st}\right)s-\left(m^2/l_{tw}+k_z'^2/a^2g^2l_{st}\right)s'\right)\right)$$

$$+\int_0^{L/2} ds' \int_{-L/2}^{0} ds \exp\left(-\left(\left(m^2/l_{tw}+k_z'^2/a^2g^2l_{st}\right)s'-\left(n^2/l_{tw}+k_z^2/a^2g^2l_{st}\right)s\right)\right).$$

(B.37)

We can evaluate the integrals over $s$ and $s'$, yielding

$$\Xi\left(k_z^2/(a^2g^2l_{st})+n^2/l_{tw},k_zk'_z/(a^2g^2l_{st})+nm/l_{tw},k_z'^2/(a^2g^2l_{st})+m^2/l_{tw}\right)=$$

$$\Xi_1\left(k_z^2/(a^2g^2l_{st})+n^2/l_{tw},k_zk'_z/(a^2g^2l_{st})+nm/l_{tw},k_z'^2/(a^2g^2l_{st})+m^2/l_{tw}\right)$$

$$+\Xi_2\left(k_z^2/(a^2g^2l_{st})+n^2/l_{tw},k_zk'_z/(a^2g^2l_{st})+nm/l_{tw},k_z'^2/(a^2g^2l_{st})+m^2/l_{tw}\right),$$

(B.38)

where

$$\Xi_1\left(k_z^2/(a^2g^2l_{st})+n^2/l_{tw}, k_zk_z'/(a^2g^2l_{st})+nm/l_{tw}, k_z'^2/(a^2g^2l_{st})+m^2/l_{tw}\right)$$

$$=\left[\left(\frac{k_z'^2}{a^2g^2l_{st}}+\frac{m^2}{l_{tw}}+\frac{2k_z'k_z}{a^2g^2l_{st}}+\frac{2nm}{l_{tw}}\right)^{-1}+\left(\frac{k_z^2}{a^2g^2l_{st}}+\frac{n^2}{l_{tw}}+\frac{2k_z'k_z}{a^2g^2l_{st}}+\frac{2nm}{l_{tw}}\right)^{-1}\right]$$

$$\times\frac{2}{\left(\frac{(k_z+k_z')^2}{a^2g^2l_{st}}+\frac{(n+m)^2}{l_{tw}}\right)}\left[\exp\left(-\frac{L}{2}\left(\frac{(k_z+k_z')^2}{a^2g^2l_{st}}+\frac{(n+m)^2}{l_{tw}}\right)\right)-1\right]+$$

$$\frac{2}{\left(\frac{2k_z'k_z}{a^2g^2l_{st}}+\frac{2nm}{l_{tw}}+\frac{k_z'^2}{a^2g^2l_{st}}+\frac{m^2}{l_{tw}}\right)\left(\frac{k_z^2}{a^2g^2l_{st}}+\frac{n^2}{l_{tw}}\right)}\left[1-\exp\left(-\frac{L}{2}\left(\frac{k_z^2}{a^2g^2l_{st}}+\frac{n^2}{l_{tw}}\right)\right)\right]$$

$$+\frac{2}{\left(\frac{2k_z'k_z}{a^2g^2l_{st}}+\frac{2nm}{l_{tw}}+\frac{k_z^2}{a^2g^2l_{st}}+\frac{n^2}{l_{tw}}\right)\left(\frac{k_z'^2}{a^2g^2l_{st}}+\frac{m^2}{l_{tw}}\right)}\left[1-\exp\left(-\frac{L}{2}\left(\frac{k_z'^2}{a^2g^2l_{st}}+\frac{m^2}{l_{tw}}\right)\right)\right].$$

(B.39)

and

$$\Xi_2\left(k_z^2/(a^2g^2l_{st})+n^2/l_{tw}, k_zk_z'/(a^2g^2l_{st})+nm/l_{tw}, k_z'^2/(a^2g^2l_{st})+m^2/l_{tw}\right)$$

$$=\frac{2}{\left(\frac{n^2}{l_{tw}}+\frac{k_z^2}{a^2g^2l_{st}}\right)}\frac{1}{\left(\frac{m^2}{l_{tw}}+\frac{k_z'^2}{a^2g^2l_{st}}\right)}\left[1-\exp\left(-\frac{L}{2}\left(\frac{n^2}{l_{tw}}+\frac{k_z^2}{a^2g^2l_{st}}\right)\right)\right]\left[1-\exp\left(-\frac{L}{2}\left(\frac{m^2}{l_{tw}}+\frac{k_z'^2}{a^2g^2l_{st}}\right)\right)\right].$$

(B.40)

Before looking at plotting $F_{local}$, let's look at the large $L$ limit of Eq. (B.36). For modes where we have that $n\neq -m\neq 0$, in the large $L$ limit, we have that

$$\Xi_1\left(k_z^2/(a^2g^2l_{st})+n^2/l_{tw}, k_zk_z'/(a^2g^2l_{st})+nm/l_{tw}, k_z'^2/(a^2g^2l_{st})+m^2/l_{tw}\right)$$

$$\simeq -\left[\left(\frac{k_z'^2}{a^2g^2l_{st}}+\frac{m^2}{l_{tw}}+\frac{2k_z'k_z}{a^2g^2l_{st}}+\frac{2nm}{l_{tw}}\right)^{-1}+\left(\frac{k_z^2}{a^2g^2l_{st}}+\frac{n^2}{l_{tw}}+\frac{2k_z'k_z}{a^2g^2l_{st}}+\frac{2nm}{l_{tw}}\right)^{-1}\right]\frac{2}{\left(\frac{(k_z+k_z')^2}{a^2g^2l_{st}}+\frac{(n+m)^2}{l_{tw}}\right)}$$

$$+\frac{2}{\left(\frac{2k_z'k_z}{a^2g^2l_{st}}+\frac{2nm}{l_{tw}}+\frac{k_z'^2}{a^2g^2l_{st}}+\frac{m^2}{l_{tw}}\right)\left(\frac{k_z^2}{a^2g^2l_{st}}+\frac{n^2}{l_{tw}}\right)}+\frac{2}{\left(\frac{2k_z'k_z}{a^2g^2l_{st}}+\frac{2nm}{l_{tw}}+\frac{k_z^2}{a^2g^2l_{st}}+\frac{n^2}{l_{tw}}\right)\left(\frac{k_z'^2}{a^2g^2l_{st}}+\frac{m^2}{l_{tw}}\right)},$$

(B.41)

and

$$\Xi_3\left(k_z^2/(a^2g^2l_{st})+n^2/l_{tw}, k_zk_z'/(a^2g^2l_{st})+nm/l_{tw}, k_z'^2/(a^2g^2l_{st})+m^2/l_{tw}\right)$$

$$\simeq \frac{2}{\left(\dfrac{n^2}{l_{tw}}+\dfrac{k_z^2}{a^2g^2l_{st}}\right)\left(\dfrac{m^2}{l_{tw}}+\dfrac{k_z'^2}{a^2g^2l_{st}}\right)}. \tag{B.42}$$

Substitution of Eqs. (B.41) and (B.42) into Eq. (B.36) leads to finite $k$ integrals, and so the contributions to Eq. (B.36) arising from such terms tend off to a constant value when $L \to \infty$. This means that these modes are not the dominant contributions, as we shall see. Now, let's consider the modes for which $m=0$, $n\neq 0$. In this case, we need to consider for large $L$

$$\Xi_1\left(k_z^2/(a^2g^2l_{st})+n^2/l_{tw}, k_zk_z'/(a^2g^2l_{st}), k_z'^2/(a^2g^2l_{st})\right)$$

$$\approx -\left[\left(\frac{k_z'^2}{a^2g^2l_{st}}+\frac{2k_z'k_z}{a^2g^2l_{st}}\right)^{-1}+\left(\frac{k_z^2}{a^2g^2l_{st}}+\frac{n^2}{l_{tw}}+\frac{2k_z'k_z}{a^2g^2l_{st}}\right)^{-1}\right]\frac{2}{\left(\dfrac{(k_z+k_z')^2}{a^2g^2l_{st}}+\dfrac{n^2}{l_{tw}}\right)}$$

$$+\frac{2}{\left(\dfrac{2k_z'k_z}{a^2g^2l_{st}}+\dfrac{k_z'^2}{a^2g^2l_{st}}\right)\left(\dfrac{k_z^2}{a^2g^2l_{st}}+\dfrac{n^2}{l_{tw}}\right)}+\frac{2a^2g^2l_{st}}{k_z'^2\left(\dfrac{2k_z'k_z}{a^2g^2l_{st}}+\dfrac{k_z^2}{a^2g^2l_{st}}+\dfrac{n^2}{l_{tw}}\right)}\left[1-\exp\left(-\frac{Lk_z'^2}{2a^2g^2l_{st}}\right)\right],$$

(B.43)

$$\Xi_2\left(k_z^2/(a^2g^2l_{st})+n^2/l_{tw}, k_zk_z'/(a^2g^2l_{st}), k_z'^2/(a^2g^2l_{st})\right)$$

$$\simeq \frac{2}{\left(\dfrac{n^2}{l_{tw}}+\dfrac{k_z^2}{a^2g^2l_{st}}\right)}\frac{a^2g^2l_{st}}{k_z'^2}\left[1-\exp\left(-\frac{Lk_z'^2}{2a^2g^2l_{st}}\right)\right]. \tag{B.44}$$

In writing Eqs. (B.43) and (B.44), we have taken special care with terms that are singular (leading to divergent integrals) in $k_z'$, if we naively set $L=\infty$ in all the exponentials. To handle the dominant contribution from Eq. (B.43), we look carefully at the small $k_z'$ values for only the terms that are singular at $k_z'$ when we sat $L=\infty$ in Eqs. (B.43) and (B.44). From looking at such $k_z'$ values, we find that the dominant contribution from $\Xi_1$ is

$$\Xi_1\left(k_z^2/(a^2g^2l_{st})+n^2/l_{tw}, k_zk_z'/(a^2g^2l_{st}), k_z'^2/(a^2g^2l_{st})\right) \approx \frac{2a^2g^2l_{st}}{k_z'^2\left(\dfrac{k_z^2}{a^2g^2l_{st}}+\dfrac{n^2}{l_{tw}}\right)}\left[1-\exp\left(-\frac{Lk_z'^2}{2a^2g^2l_{st}}\right)\right].$$

(B.45)

Thus we can write the dominant contribution to $\left\langle V_{local}[\delta\phi_{1,T}(s),\delta\phi_{2,T}(s),\delta h_{1,T}(s),\delta h_{2,T}(s)]^2\right\rangle_0$ from the $m=0$, $n\neq 0$ modes as

$$\langle V_{local}[\delta\phi_{1,T}(s),\delta\phi_{2,T}(s),\delta h_{1,T}(s),\delta h_{2,T}(s)]^2 \rangle_{0,n\neq 0,m=0} \approx 8\Lambda^2 a^2 g^2 l_{st} \bar{G}_{0,0}(R/a,0)\sum_{n=1}^{\infty}\cos(n(\phi_1-\phi_2))(-1)^n$$

$$\int_{-\infty}^{\infty} dk'_z \frac{1}{k'^2_z}\left[1-\exp\left(-\frac{L}{2}(k'^2_z/(a^2g^2l_{st}))\right)\right]\int_{-\infty}^{\infty} dk_z \bar{G}_{n,n}(R/a,k_z)\left(\frac{k_z^2}{a^2g^2l_{st}}+\frac{n^2}{l_{tw}}\right)^{-1}$$

$$=16ag\Lambda^2\sqrt{\frac{\pi L l_{st}}{2}}\bar{G}_{0,0}(R/a,0)\sum_{n=1}^{\infty}\cos(n(\phi_1-\phi_2))(-1)^n\int_{-\infty}^{\infty}dk_z \bar{G}_{n,n}(R/a,k_z)\left(\frac{k_z^2}{a^2g^2l_{st}}+\frac{n^2}{l_{tw}}\right)^{-1}.$$

(B.46)

Here, we have expanded $\bar{G}_{0,0}(R/a,k'_z)$ about $k'_z=0$, as the dominant contribution comes from the $1/k'^2_z$ pole in Eq. (B.45). Similar analysis show that the contribution from the $n=0$, $m\neq 0$ modes is

$$\langle V_{local}[\delta\phi_{1,T}(s),\delta\phi_{2,T}(s),\delta h_{1,T}(s),\delta h_{2,T}(s)]^2 \rangle_{0,n=0,m\neq 0} \approx 16ag\Lambda^2\sqrt{\frac{\pi L l_{st}}{2}}\bar{G}_{0,0}(R/a,0)$$

$$\sum_{m=1}^{\infty}\cos(m(\phi_1-\phi_2))(-1)^m\int_{-\infty}^{\infty} dk_z\left(\frac{k_z^2}{a^2g^2l_{st}}+\frac{m^2}{l_{tw}}\right)^{-1}\bar{G}_{m,m}(R/a,k_z).$$

(B.47)

Next, let us consider the dominant contribution from the modes where $n=-m\neq 0$, by first examining both $\Xi_1$ and $\Xi_2$. These now should be written in large $L$ limit as

$$\Xi_1\left(k_z^2/(a^2g^2l_{st})+n^2/l_{tw},k_zk'_z/(a^2g^2l_{st})-n^2/l_{tw},k'^2_z/(a^2g^2l_{st})+n^2/l_{tw}\right)$$

$$\approx\left[\left(\frac{k'^2_z}{a^2g^2l_{st}}-\frac{n^2}{l_{tw}}+\frac{2k'_zk_z}{a^2g^2l_{st}}\right)^{-1}+\left(\frac{k_z^2}{a^2g^2l_{st}}-\frac{n^2}{l_{tw}}+\frac{2k'_zk_z}{a^2g^2l_{st}}\right)^{-1}\right]\frac{2a^2g^2l_{st}}{(k_z+k'_z)^2}\left[\exp\left(-\frac{L}{2}\left(\frac{(k_z+k'_z)^2}{a^2g^2l_{st}}\right)\right)-1\right]$$

$$+\frac{2}{\left(\frac{2k'_zk_z}{a^2g^2l_{st}}+\frac{k'^2_z}{a^2g^2l_{st}}-\frac{n^2}{l_{tw}}\right)\left(\frac{k_z^2}{a^2g^2l_{st}}+\frac{n^2}{l_{tw}}\right)}+\frac{2}{\left(\frac{2k'_zk_z}{a^2g^2l_{st}}+\frac{k_z^2}{a^2g^2l_{st}}-\frac{n^2}{l_{tw}}\right)\left(\frac{k'^2_z}{a^2g^2l_{st}}+\frac{n^2}{l_{tw}}\right)},$$

(B.48)

$$\Xi_2\left(k_z^2/(a^2g^2l_{st})+n^2/l_{tw},k_zk'_z/(a^2g^2l_{st})-n^2/l_{tw},k'^2_z/(a^2g^2l_{st})+n^2/l_{tw}\right)$$

$$\approx\frac{2}{\left(\frac{n^2}{l_{tw}}+\frac{k_z^2}{a^2g^2l_{tw}}\right)}\frac{1}{\left(\frac{n^2}{l_{st}}+\frac{k'^2_z}{a^2g^2l_{st}}\right)}$$

(B.49)

Here, $\Xi_2$ does not contribute to the leading order term in the limit $L\to\infty$ and can safely be neglected. To deal with $\Xi_1$ we introduce $k'_z=k''_z-k_z$ and look where $k''_z$ is small. So, the dominant contribution to $\Xi_1$ reads as

$$\Xi_1\left(k_z^2/(a^2g^2l_{st})+n^2/l_{tw}, k_zk_z'/(a^2g^2l_{st})-n^2/l_{tw}, k_z'^2/(a^2g^2l_{st})+n^2/l_{tw}\right)$$

$$\approx \frac{4}{\left(\frac{k_z^2}{a^2g^2l_{st}}+\frac{n^2}{l_{tw}}\right)^{-1}} \frac{a^2g^2l_{st}}{k_z''^2}\left[1-\exp\left(-\frac{L}{2}\left(\frac{k_z''^2}{a^2g^2l_{st}}\right)\right)\right]. \tag{B.50}$$

Using Eq. (B.50), we may write the dominant contribution to $\left\langle V_{local}[\delta\phi_{1,T}(s),\delta\phi_{2,T}(s),\delta h_{1,T}(s),\delta h_{2,T}(s)]^2\right\rangle_0$ from the $n=-m\neq 0$ modes as

$$\left\langle V_{local}[\delta\phi_{1,T}(s),\delta\phi_{2,T}(s),\delta h_{1,T}(s),\delta h_{2,T}(s)]^2\right\rangle_{0,n=-m\neq 0}=$$

$$8\Lambda^2 a^2g^2l_{st}\sum_{n=1}^{\infty}\int_{-\infty}^{\infty}dk_z\left(\frac{k_z^2}{a^2g^2l_{st}}+\frac{n^2}{l_{tw}}\right)^{-1}\bar{G}_{n,n}(R/a,k_z)^2\int_{-\infty}^{\infty}\frac{dk_z''}{(k_z'')^2}\left[1-\exp\left(-\frac{L}{2}\left(\frac{k_z''^2}{a^2g^2l_{st}}\right)\right)\right] \tag{B.51}$$

$$=16\Lambda^2 ga\sqrt{\frac{\pi L l_{st}}{2}}\sum_{n=1}^{\infty}\int_{-\infty}^{\infty}dk_z\left(\frac{k_z^2}{a^2g^2l_{st}}+\frac{n^2}{l_{tw}}\right)^{-1}\bar{G}_{n,n}(R/a,k_z)^2.$$

where have expanded out $\bar{G}_{n,n}(R/a,k_z''-k_z)$ to leading order in $k''$ in writing Eq.(B.51), as well as using the fact that $\bar{G}_{n,n}(R/a,k_z')=\bar{G}_{-n,-n}(R/a,k_z')=\bar{G}_{n,n}(R/a,-k_z')$. Last of all, in considering what the large $L$ limit is, we now need to consider the $n=m=0$ mode. Here, we write for both $\Xi_1$ and $\Xi_2$

$$\Xi_1\left(k_z^2/(a^2g^2l_{st}), k_zk_z'/(a^2g^2l_{st}), k_z'^2/(a^2g^2l_{st})\right)$$

$$\approx\left[\left(k_z'^2+2k_z'k_z\right)^{-1}+\left(k_z^2+2k_z'k_z\right)^{-1}\right]\frac{2a^4g^4l_{st}^2}{(k_z+k_z')^2}\left[\exp\left(-\frac{L}{2}\left(\frac{(k_z+k_z')^2}{a^2g^2l_{st}}\right)\right)-1\right] \tag{B.52}$$

$$+\frac{2a^4g^4l_{st}^2}{(2k_z'k_z+k_z'^2)k_z^2}\left[1-\exp\left(-\frac{Lk_z^2}{2a^2g^2l_{st}}\right)\right]+\frac{2a^4g^4l_{st}^2}{(2k_z'k_z+k_z^2)k_z'^2}\left[1-\exp\left(-\frac{Lk_z'^2}{2a^2g^2l_{st}}\right)\right],$$

$$\Xi_2\left(k_z^2/(a^2g^2l_{st}), k_zk_z'/(a^2g^2l_{st}), k_z'^2/(a^2g^2l_{st})\right)$$

$$=\frac{2a^2g^4l_{st}^2}{k_z'^2k_z^2}\left[1-\exp\left(-\frac{Lk_z^2}{2a^2g^2l_{st}}\right)\right]\left[1-\exp\left(-\frac{Lk_z'^2}{2a^2g^2l_{tw}}\right)\right], \tag{B.53}$$

for large $L$. Here, there are poles in the integrand at both $k_z=0$ and $k_z'=0$ in the limit $L\to\infty$, thus to get the leading order behaviour we need to consider small values of both $k_z$ and $k_z'$. Therefore, the contribution to $\left\langle V_{local}[\delta\phi_{1,T}(s),\delta\phi_{2,T}(s),\delta h_{1,T}(s),\delta h_{2,T}(s)]^2\right\rangle_0$ from $n=0$, $m=0$, at large $L$, can be written first as

$$\left\langle V_{local}[\delta\phi_{1,T}(s), \delta\phi_{2,T}(s), \delta h_{1,T}(s), \delta h_{2,T}(s)]^2 \right\rangle_{0,n=m=0} \approx \Lambda^2 \overline{G}_{0,0}(R/a,0)^2$$
$$\int_{-\infty}^{\infty} dk_z \int_{-\infty}^{\infty} dk_z' \Xi\left(k_z^2/(a^2 g^2 l_{st}), k_z k_z'/(a^2 g^2 l_{st}), k_z'^2/(a^2 g^2 l_{st})\right),$$
(B.54)

Thus substituting in Eqs. (B.52) and (B.53) we need to consider the integrals

$$\int_{-\infty}^{\infty} dk \int_{\infty}^{\infty} dk' \Xi_2\left(k_z^2/(a^2 g^2 l_{st}), k_z k_z'/(a^2 g^2 l_{st}), k_z'^2/(a^2 g^2 l_{st})\right)$$
$$\approx 4a^4 g^4 l_{st}^2 \int_{-\infty}^{\infty} dk \int_{\infty}^{\infty} dk' \frac{1}{k_z^2 k_z'^2}\left[1 - \exp\left(-\frac{Lk_z^2}{2a^2 g^2 l_{st}}\right)\right]\left[1 - \exp\left(-\frac{Lk_z'^2}{2a^2 g^2 l_{st}}\right)\right],$$
(B.55)

$$\int_{-\infty}^{\infty} dk \int_{\infty}^{\infty} dk' \Xi_1\left(k_z^2/(a^2 g^2 l_{st}), k_z k_z'/(a^2 g^2 l_{st}), k_z'^2/(a^2 g^2 l_{st})\right)$$
$$\approx 4a^4 g^4 l_{st}^2 \int_{-\infty}^{\infty} dk \int_{\infty}^{\infty} dk' \left[\frac{1}{(2k_z' k_z + k_z'^2)(k_z + k_z')^2}\left[\exp\left(-\frac{L}{2}\left(\frac{(k_z + k_z')^2}{a^2 g^2 l_{st}}\right)\right) - 1\right]\right.$$
$$\left. + \frac{1}{(2k_z' k_z + k_z'^2) k_z^2}\left[1 - \exp\left(-\frac{Lk_z^2}{2a^2 g^2 l_{st}}\right)\right]\right],$$
(B.56)

where we have used the interchange symmetry in $k_z \leftrightarrow k_z'$ under integration in writing Eq.(B.56). In Eq. (B.56) we may make the variable change $k_z' = k_z'' - k_z$, and thus we may write

$$\int_{-\infty}^{\infty} dk \int_{\infty}^{\infty} dk' \Xi_1\left(k_z^2/(a^2 g^2 l_{st}), k_z k_z'/(a^2 g^2 l_{st}), k_z'^2/(a^2 g^2 l_{st})\right)$$
$$\approx 4a^4 g^4 l_{st}^2 \int_{-\infty}^{\infty} dk \int_{\infty}^{\infty} dk' \frac{1}{(k_z''^2 - k_z^2)}\left[\frac{1}{k_z''^2}\left[\exp\left(-\frac{Lk_z''^2}{2a^2 g^2 l_{st}}\right) - 1\right] + \frac{1}{k_z^2}\left[1 - \exp\left(-\frac{Lk_z^2}{2a^2 g^2 l_{st}}\right)\right]\right]$$
$$= 2a^2 g^2 l_{st} L c_1,$$
(B.57)

where

$$c_1 = \int_0^{\infty} ds \int_0^{\infty} ds' \frac{1}{(s'-s)} \frac{1}{s'^{3/2} s^{3/2}}\left[s \exp(-s') - s - s' \exp(-s) + s'\right] = \pi^2.$$
(B.58)

On rescaling Eq. (B.55), one obtains

$$\int_{-\infty}^{\infty} dk \int_{\infty}^{\infty} dk' \Xi_2\left(k_z^2/(a^2 g^2 l_{st}), k_z k_z'/(a^2 g^2 l_{st}), k_z'^2/(a^2 g^2 l_{st})\right) \approx 8a^2 g^2 l_{st} L c_2^2,$$
(B.59)

where

$$c_2 = \int_0^\infty dx \frac{1}{x^2}\left(1-\exp\left(-\frac{x^2}{2}\right)\right) = \frac{1}{2\sqrt{2}}\int_0^\infty dy \frac{1}{y^{3/2}}\left(1-\exp(-y)\right) = \sqrt{\frac{\pi}{2}}. \tag{B.60}$$

Further analysis, subtracting out these apparent poles at $k_z = k_z'' = 0$, leads to the sub-dominant contributions from both $\Xi_1$ and $\Xi_3$, arising when $k_z = 0$, $k_z'' = 0$ or $k_z = -k_z'$. These sub-dominant contributions will go as $\sqrt{L}$. However, as we now see, the dominant contribution to the $n=m=0$ mode scales as $L$, at very large $L$. Thus, the dominant contribution (from the zeroth modes) to Eq. (B.36) is found to be

$$\left\langle V_{local}[\delta\phi_{1,T}(s), \delta\phi_{2,T}(s), \delta h_{1,T}(s), \delta h_{2,T}(s)]^2 \right\rangle_0 \approx \left(2\pi^2 + 4\pi\right)\Lambda^2 a^2 g^2 l_{st} L \bar{G}_{0,0}(R/a, 0)^2. \tag{B.61}$$

All in all, we see that the dominant part of the full second order term in the expansion of the free energy goes as

$$\frac{1}{2}\left[\left\langle V_{local}[\delta\phi_{1,T}(s), \delta\phi_{2,T}(s), \delta h_{1,T}(s), \delta h_{2,T}(s)]^2 \right\rangle_0 - \left\langle V_{local}[\delta\phi_{1,T}(s), \delta\phi_{2,T}(s), \delta h_{1,T}(s), \delta h_{2,T}(s)] \right\rangle_0^2\right]$$
$$\approx \left(\pi^2 - 2\pi\right)\Lambda^2 a^2 g^2 l_{st} L \bar{G}_{0,0}(R/a, 0)^2.$$
$$\tag{B.62}$$

We see that for the zeroth mode, the weak interaction expansion suggests an effective interaction strength $\tilde{\gamma}\sqrt{L}$ for large $L$. To next order, in the expansion (when we come to consider $Z^{(3)}$) there is likely to be three $k_z$ integrations, all with potentially singular terms that should lead to terms proportional to $\tilde{\gamma}^3 L^{3/2}$. This suggests, in fact, that for very large $L$, truncating this expansion is not valid. Instead, for long molecules we shall use a variational approximation described in next appendix, which is more appropriate thing to do.

We now look at some numerical results from the analysis. In Fig. B.1 we show plots of $\left\langle V_{local}\right\rangle_0 / \Lambda L$, the leading order term in the expansion, using the Morse potential form for $\bar{G}_{0,0}(R/a,0)$ (Eq. (2.15) of the main text), which are qualitatively similar to those using the Debye Huckel form (Eq. (2.14) of the main text), presented in Fig. 3 of the main text.

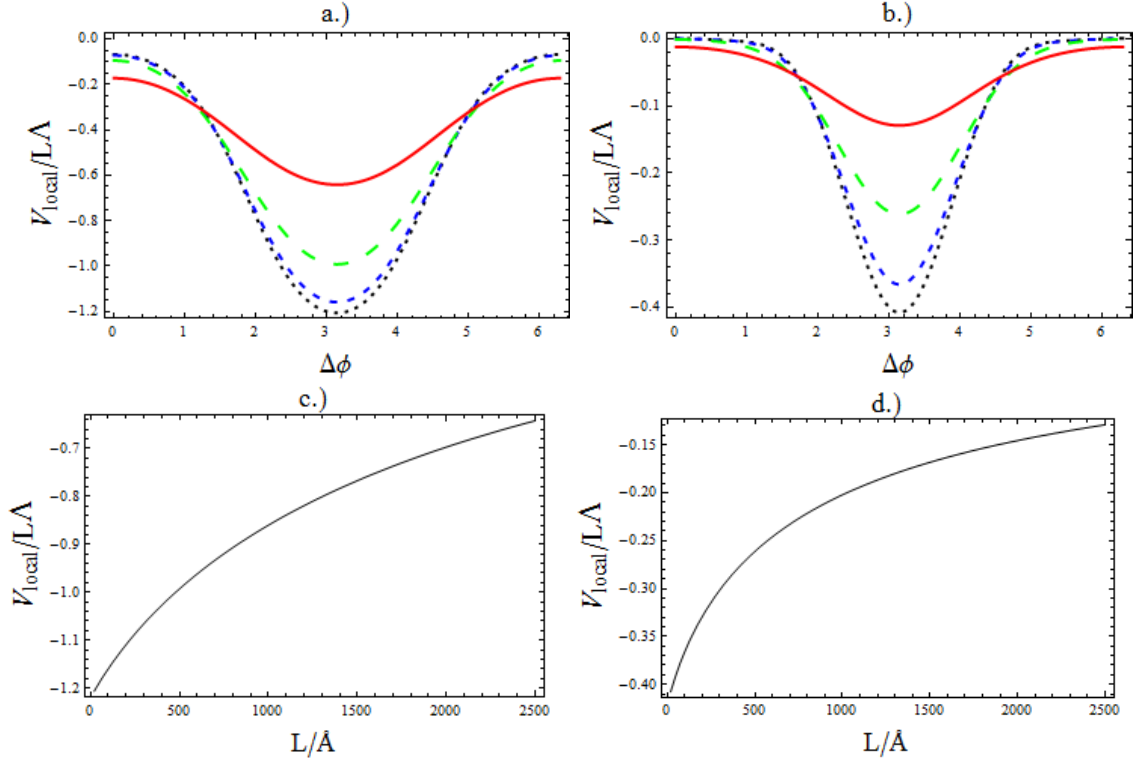

Fig. B.1. The azimuthal dependence of $\langle V_{local}\rangle_0 / \Lambda L$ and the recognition energy per unit length ($\langle V_{local}\rangle_0 / \Lambda L$ at $\phi_1 - \phi_2 = \pi$) with thermal fluctuations unconstrained by the interaction energy, using $\bar{G}_{n,n}(R/a, k_z)$ for the Morse potential interaction (Eq. (2.15) of the main text). In panels a.) and b.), we show the azimuthal dependence of $\langle V_{local}\rangle_0 / \Lambda L$ (in units of Å) in terms of $\Delta\phi = \phi_2 - \phi_1$, for various values of length $L$. In panel a.) we use the value $\kappa_{eff} = 0.25 \text{Å}^{-1}$ and in b.) $\kappa_{eff} = 0.5 \text{Å}^{-1}$. In both plots, the black dotted, blue short dashed, green long dashed and red solid curves correspond to the values $L = 20\text{Å}, 100\text{Å}, 500\text{Å}, 2500\text{Å}$, respectively. In panels c.) and d.), we plot the recognition energy $\langle V_{local}\rangle_0 / \Lambda L$ (where we set $\Delta\phi = \phi_1 - \phi_2 = \pi$) per unit length (units Å) as a function of $L$. For panel c.) we use the value $\kappa_{eff} = 0.25 \text{Å}^{-1}$, for d.) $\kappa_{eff} = 0.5 \text{Å}^{-1}$. In the calculations the values $R = 25\text{Å}$, $a = 11.2\text{Å}$, $l_{tw} = 1000\text{Å}$ and $l_{st} = 700\text{Å}$ were also used.

In Figs. B.2 and B.3 we show plots of the full free energy, given by Eq. (B.3), for parameter values where (B.36) can be considered as a small correction. Both the results utilizing the Debye-Huckel form for $\bar{G}_{n,n}(R/a, k_z)$ (shown in Fig B.2) and Morse form (shown in Fig B.3) look qualitively simular.

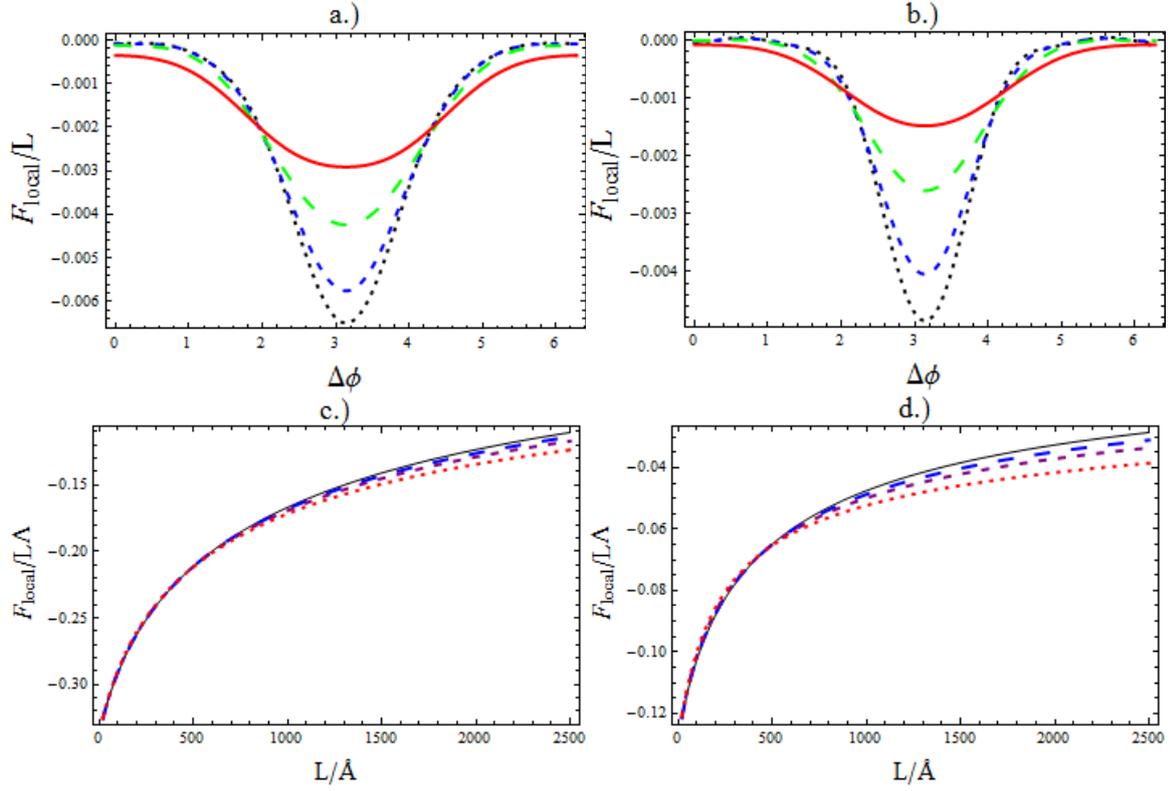

Fig. B.2. Plots showing the length depedence for $F_{local}$ for weak pairing interactions without rigid body rotional fluctuations. In these plots the Deybe-Huckle form for $\bar{G}_{n,n}(R/a,k_z)$ is used, Eq. (2.14) of the main text. Panels a.) and b.) show the dependence of $F_{local}$ on $\Delta\phi = \phi_1 - \phi_2$ (neglecting the unimportant $\ln Z_0$ term ). In a.) the value $\kappa_{eff} = 0.25\text{Å}^{-1}$ is used, and in b.) we use $\kappa_{eff} = 0.5\text{Å}^{-1}$. In both the top plots the black dotted, blue short dashed, green long dashed and red solid curves correspond to values of $L = 20, 100, 500, 2000 \text{Å}$ repectively. In both c.) and d.), we plot $F_{local}/\Lambda L$ (at $\Delta\phi = \pi$) as a function of length for various values of the parameters. In c.), we set $\kappa_{eff} = 0.25\text{Å}^{-1}$ and in d.) we set $\kappa_{eff} = 0.5\text{Å}^{-1}$. In both curves, the black corresponds to simply $\langle V_{local}\rangle_0 / \Lambda L$, the leading order term in $F_{local}$. In c.) and d.), the blue long dashed, purple short dashed, red dotted curves are for the parameter values $\Lambda = 0.0025, 0.005, 0.01 k_B T/\text{Å}$ for c.), and $\Lambda = 0.02, 0.04, 0.08 k_B T/\text{Å}$ for d.), repectively. In the calculations the values $R = 25\text{Å}$, $a = 11.2\text{Å}$, $l_{tw} = 1000\text{Å}$ and $l_{st} = 700\text{Å}$ are used.

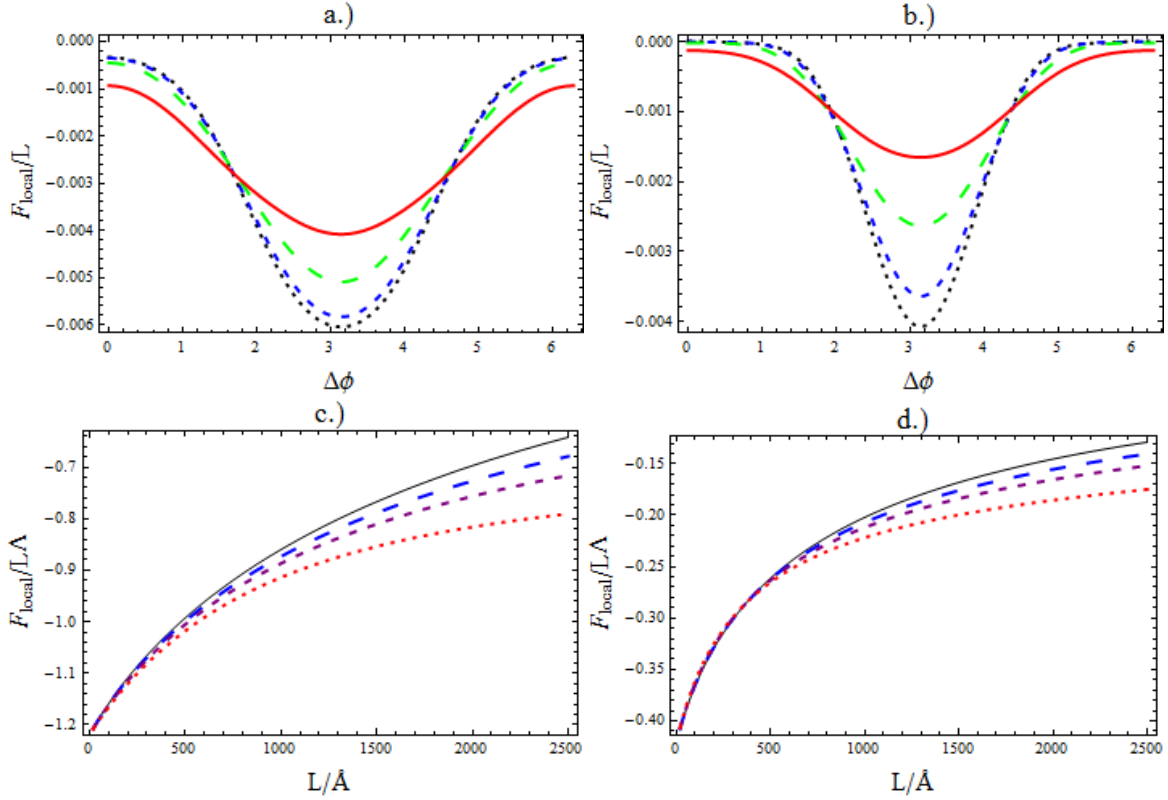

Fig. 3.B. Plots showing the length depedence for $F_{local}$ for weak pairing interactions without rigid body rotional fluctuations. In these plots, the Morse Potential form for $\bar{G}_{n,n}(R/a,k_z)$ is used, Eq. (2.15) of the main text. Panels a.) and b.) show the dependence of $F_{local}$ on $\Delta\phi = \phi_1 - \phi_2$ (neglecting the unimportant $\ln Z_0$ term ). In a.) the value $\kappa_{eff} = 0.25 \text{Å}^{-1}$ is used, and in b.) we use $\kappa_{eff} = 0.5 \text{Å}^{-1}$. In both the top plots the black dotted, blue short dashed, green long dashed and red solid curves correspond to values of $L = 20, 100, 500, 2000 \text{Å}$ repectively. In both c.) and d.), we plot $F_{local}/\Lambda L$ (at $\Delta\phi = \pi$) as a function of length for various values of the parameters. In c.), we set $\kappa_{eff} = 0.25 \text{Å}^{-1}$ and in d.) we set $\kappa_{eff} = 0.5 \text{Å}^{-1}$. In both curves, the black corresponds to simply $\langle V_{local} \rangle_0 / \Lambda L$, the leading order term in $F_{local}$. In c.) and d.), the blue long dashed, purple short dashed, red dotted curves are for the parameter values $\Lambda = 0.00125, 0.0025, 0.005 k_B T/\text{Å}^2$ for c.), and $\Lambda = 0.005, 0.01, 0.02 k_B T/\text{Å}^2$ for d.), repectively. In the calculations the values $R = 25\text{Å}, a = 11.2\text{Å}, l_{tw} = 1000 \text{Å}$ and $l_{st} = 700 \text{Å}$ are used.

In Figs. B.4 and B.5 we show plots, for Debye Huckel and Morse forms of $\bar{G}_{n,n}(R/a,k_z)$ (Eqs. (2.14) and (2.15) of the main text) of the free energy $F_{local}$ averaged over rigid body rotations, fluctuations in $\Delta\phi = \phi_1 - \phi_2$ . This is when the helices are allowed to rotate freely around their principle axes. Here, all terms dependent on $\Delta\phi$ average out to zero, and the magnitude of the free energy and recognition energy are dramatically reduced. Again, the plots for both sets of graphs look qualitatively similar, but the Morse potential ones show a greater sensitivity on $\Lambda$.

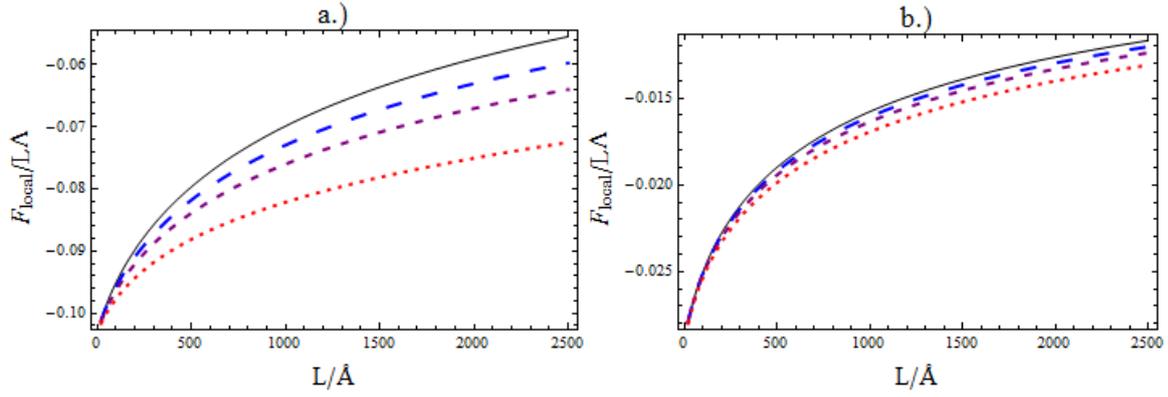

Fig. B.4. Plots showing the free energy, $F_{local}$ averaged over $\Delta\phi = \phi_1 - \phi_2$ for free rigid body rotations. Here we use the Debye-Huckel form for $\bar{G}_{n,n}(R/a, k_z)$ given by Eq. (2.14) of the main text. In panel a.) we set $\kappa_{eff} = 0.25 \text{Å}^{-1}$ and b.) $\kappa_{eff} = 0.5 \text{Å}^{-1}$. In both graphs, the black solid curves corresponds to the result for $\langle V_{local} \rangle_0 / \Lambda L$ averaged also over the rigid body fluctuations. In a.), the blue long dashed, purple short dashed, red dotted curves are for the parameter values $\Lambda = 0.0025, 0.005, 0.01 k_B T / \text{Å}$, repectively. In b.), the blue long dashed, purple short dashed, red dotted curves are for the parameter values $\Lambda = 0.02, 0.04, 0.08 k_B T / \text{Å}$. In the calculations the values $R = 25\text{Å}$, $a = 11.2\text{Å}$, $l_{tw} = 1000\text{Å}$ and $l_{st} = 700\text{Å}$ are used.

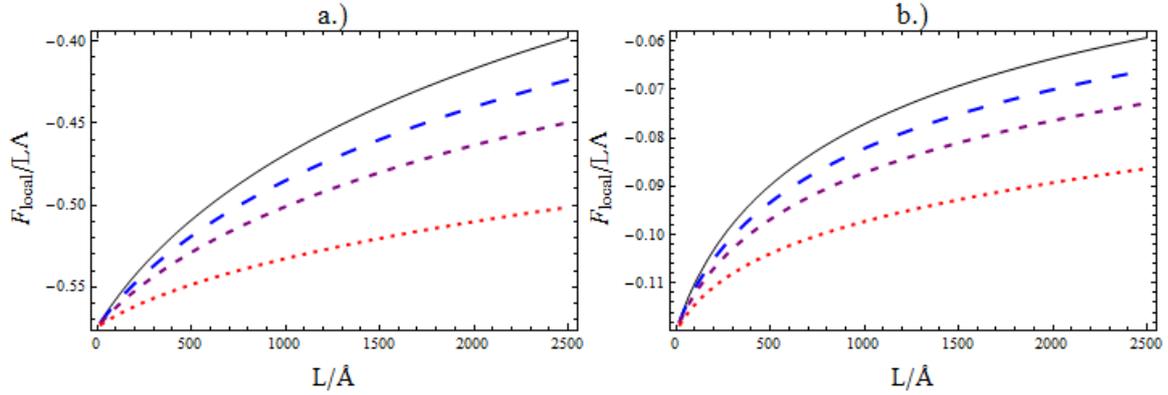

Fig.B.5 Plots showing the free energy averaged over $\Delta\phi = \phi_1 - \phi_2$ for free rigid body rotations using Morse potential (using Eq. (2.15) of the main text). In a.) we set $\kappa_{eff} = 0.25 \text{Å}^{-1}$ and in b.) $\kappa_{eff} = 0.5 \text{Å}^{-1}$. In both graphs, the black corresponds to $\langle V_{local} \rangle_0 / \Lambda L$, averaged over rigid body fluctuations so that only the $n=0$ mode survives. In the bottom left hand panel, the blue, purple, red curves are for the parameter values $\Lambda = 0.000625, 0.00125, 0.0025 k_B T / \text{Å}^2$, repectively. In the bottom right panel, the blue, purple, red curves are for the parameter values $\Lambda = 0.0025, 0.005, 0.01 k_B T / \text{Å}^2$. In the calculations the values $R = 25\text{Å}$, $a = 11.2\text{Å}$, $l_{tw} = 1000\text{Å}$ and $l_{st} = 700\text{Å}$ are used.

## Appendix C Thermal fluctuations for strong local pairing interactions or long helices

Our starting point here is Eqs. (2.31)-(2.36) of the main text. First of all, we may write Eq. (2.35) of the main text as

$$\langle E_{total}[\delta\phi_{1,T}(s),\delta\phi_{2,T}(s),\delta s_{1,T}(s),\delta s_{2,T}(s)] - E_{tr}[\delta\phi_{1,T}(s),\delta\phi_{2,T}(s),\delta s_{1,T}(s),\delta s_{2,T}(s)]\rangle_{tr} =$$
$$= \langle V_{local}[\delta\phi_{1,T}(s),\delta\phi_{2,T}(s),\delta s_{1,T}(s),\delta s_{2,T}(s)]\rangle_{tr} \quad \text{(C.1)}$$
$$-\frac{1}{2}\int_{-L/2}^{L/2} ds\left[\alpha\langle(\delta\phi_{1,T}(s)-\delta\phi_{2,T}(s))^2\rangle_{tr} + \beta\langle(\delta s_{1,T}(s)-\delta s_{2,T}(s))^2\rangle_{tr}\right].$$

Here, the $tr$ subscript on the averaging brackets refers to thermal averaging using Eqs. (2.31)-(2.33) of the main text for energy functional in the Boltzmann weight. Let us first consider the average of the local pairing interaction energy (again only considering diagonal $n = n'$ modes). This reads as, using Eq. (2.19) of the main text,

$$\langle V_{local}[\delta\phi_{1,T}(s),\delta\phi_{2,T}(s),\delta s_{1,T}(s),\delta s_{2,T}(s)]\rangle_{tr} = -\Lambda \int_{-L/2}^{L/2} ds \sum_{n=-\infty}^{\infty} \exp(in(\phi_1-\phi_2))\langle \exp(in(\delta\phi_{1,T}(s)-\delta\phi_{2,T}(s)))\rangle_{tr,\phi}$$
$$\int_{-\infty}^{\infty} dk_z (-1)^n \bar{G}_{n,n'}(R/a,k_z)\left\langle \exp\left(i\frac{(\delta s_{1,T}(s)-\delta s_{2,T}(s))}{a}k_z\right)\right\rangle_{tr,s},$$
$$\text{(C.2)}$$

where

$$\langle \exp(in(\delta\phi_{1,T}(s)-\delta\phi_{2,T}(s)))\rangle_{tr,\phi} =$$
$$\frac{1}{Z_{tr,\phi}}\int D\delta\phi_{1,T}(s)\int D\delta\phi_{2,T}(s)\exp(in(\delta\phi_{1,T}(s)-\delta\phi_{2,T}(s)))\exp(-E_{tr,\phi}[\delta\phi_{1,T}(s),\delta\phi_{2,T}(s)]), \quad \text{(C.3)}$$

$$\left\langle \exp\left(i\frac{(\delta s_{1,T}(s)-\delta s_{2,T}(s))}{a}k_z\right)\right\rangle_{tr,s}$$
$$= \frac{1}{Z_{tr,s}}\int D\delta s_{1,T}(s)\int D\delta s_{2,T}(s)\exp\left(i\frac{(\delta s_{1,T}(s)-\delta s_{2,T}(s))}{a}k_z\right)\exp(-E_{tr,s}[\delta s_{1,T}(s),\delta s_{2,T}(s)]),$$
$$\text{(C.4)}$$

and

$$Z_{tr,\phi} = \int D\delta\phi_{1,T}(s)\int D\delta\phi_{2,T}(s)\exp(-E_{tr,\phi}[\delta\phi_{1,T}(s),\delta\phi_{2,T}(s)]), \quad \text{(C.5)}$$

$$Z_{tr,s} = \int D\delta s_{1,T}(s)\int D\delta s_{2,T}(s)\exp(-E_{tr,s}[\delta s_{1,T}(s),\delta s_{2,T}(s)]). \quad \text{(C.6)}$$

In this approximation, we will deal with the case when the molecules are very long and neglect any finite size effects. Thus, following from previous work (for instance see Ref. [1]) we may evaluate

$$\left\langle \exp\left(in\left(\delta\phi_{1,T}(s)-\delta\phi_{2,T}(s)\right)\right)\right\rangle_{tr,\phi} = \exp\left(-\frac{n^2 d_\phi^2}{2}\right), \tag{C.7}$$

$$\left\langle \exp\left(i\frac{(\delta s_{1,T}(s)-\delta s_{2,T}(s))}{a}k_z\right)\right\rangle_{tr,s} = \exp\left(-\frac{k_z^2 d_s^2}{2a^2}\right), \tag{C.8}$$

where

$$d_\phi^2 = \frac{1}{\pi l_{tw}}\int_{-\infty}^{\infty}\frac{dk}{k^2 + 2\frac{\alpha}{l_{tw}}} = \frac{1}{l_{tw}}\left(\frac{l_{tw}}{2\alpha}\right)^{1/2} \equiv \frac{\lambda_{tw}}{2l_{tw}}, \tag{C.9}$$

$$d_s^2 = \frac{1}{\pi g^2 l_{st}}\int_{-\infty}^{\infty}\frac{dk}{k^2 + 2\frac{\beta}{g^2 l_{st}}} = \frac{1}{g^2 l_{st}}\left(\frac{g^2 l_{st}}{2\beta}\right)^{1/2} \equiv \frac{\lambda_{st}}{2g^2 l_{st}}. \tag{C.10}$$

Note that also $d_\phi$ and $d_s$ are defined as $d_\phi^2 = \left\langle \left(\delta\phi_{1,T}(s)-\delta\phi_{2,T}(s)\right)^2\right\rangle_{tr,\phi}$ and $d_s^2 = \left\langle \left(\delta s_{1,T}(s)-\delta s_{2,T}(s)\right)^2\right\rangle_{tr,s}$, respectively. The lengths $\lambda_{tw}$ and $\lambda_{st}$ are the adaptation lengths for twisting and stretching fluctuations; the physical meaning of these is explained in the main text. Essentially, for length scales smaller than these lengths thermal fluctuations in base pair rises and twists are weakly affected by interactions, and the results considered in Appendix B apply for molecules for which $L \ll \lambda_{tw}, \lambda_{st}$. On the other hand, when length scales become comparable to these adaptation lengths, thermal fluctuations are affected greatly by interactions. The approximations that we consider in this section should hold when $L \gg \lambda_{tw}, \lambda_{st}$, where the interaction energy per unit length saturates to a constant value (as we shall see). Using Eq. (C.7)-(C.10) allow us to write Eq. (C.2) as

$$\left\langle V_{local}[\delta\phi_{1,T}(s),\delta\phi_{2,T}(s),\delta s_{1,T}(s),\delta s_{2,T}(s)]\right\rangle_{tr} = -\Lambda L \sum_{n=-\infty}^{\infty}\exp\left(in(\phi_1-\phi_2)\right)\exp\left(-\frac{n^2\lambda_{tw}}{4l_{tw}}\right)$$
$$\int_{-\infty}^{\infty}dk_z(-1)^n \bar{G}_{n,n}(R/a,k_z)\exp\left(-\frac{k_z^2\lambda_{st}}{4a^2 g^2 l_{st}}\right). \tag{C.11}$$

Next, we evaluate:

$$\frac{1}{2}\int_{-L/2}^{L/2}ds\left[\left\langle \alpha\left(\delta\phi_{1,T}(s)-\delta\phi_{2,T}(s)\right)^2\right\rangle + \left\langle \beta\left(\delta h_{1,T}(s)-\delta h_{2,T}(s)\right)^2\right\rangle\right]$$
$$= \frac{L}{2}\left[\alpha d_\phi^2 + \beta d_s^2\right] = \frac{L}{2\lambda_{tw}} + \frac{L}{2\lambda_{st}}. \tag{C.12}$$

Also, we may write (from Eqs. (C.5) and (C.6), as well Eqs. (2.32) and (2.33) of the main text) with $Z_{tr} = Z_{tr,\phi} Z_{tr,s}$

$$-\frac{\partial \ln Z_{tr}}{\partial \alpha} = \frac{L d_\phi^2}{2} = \frac{L}{2} \frac{1}{l_{tw}} \left( \frac{l_{tw}}{2\alpha} \right)^{1/2}, \quad -\frac{\partial \ln Z_{tr}}{\partial \beta} = \frac{L d_s^2}{2} = \frac{L}{2} \frac{1}{g^2 l_{st}} \left( \frac{g^2 l_{st}}{2\beta} \right)^{1/2}. \quad \text{(C.13)}$$

Integrating Eq. (C.13) up we obtain

$$-\ln Z_{tr} = L \left( \frac{\alpha}{2l_{tw}} \right)^{1/2} + L \left( \frac{\beta}{2g^2 l_{st}} \right)^{1/2} = L \left[ \frac{1}{\lambda_{tw}} + \frac{1}{\lambda_{st}} \right]. \quad \text{(C.14)}$$

Thus, combining Eqs. (C.11), (C.12) and (C.14), we can write the total free energy as

$$\frac{F_T}{k_B T} = L \left[ \frac{1}{2\lambda_{tw}} + \frac{1}{2\lambda_{st}} \right] - \frac{L\Lambda}{k_B T} \sum_{n=-\infty}^{\infty} \exp(in(\phi_1 - \phi_2)) \exp\left( -\frac{n^2 \lambda_{tw}}{4 l_{tw}} \right)$$
$$\int_{-\infty}^{\infty} dk_z (-1)^n \bar{G}_{n,n}(R/a, k_z) \exp\left( -\frac{k_z^2 \lambda_{st}}{4 a^2 g^2 l_{st}} \right). \quad \text{(C.15)}$$

Eq. (C.15) is minimized by $\langle \Delta\phi(s) \rangle = \phi_1 - \phi_2 = \pi$ substituting this value gives

$$\frac{F_T}{k_B T} = L \left[ \frac{1}{2\lambda_{tw}} + \frac{1}{2\lambda_{st}} \right] - \frac{L\Lambda}{k_B T} \sum_{n=-\infty}^{\infty} \exp\left( -\frac{n^2 \lambda_{tw}}{4 l_{tw}} \right) \int_{-\infty}^{\infty} dk_z \bar{G}_{n,n}(R/a, k_z) \exp\left( -\frac{k_z^2 \lambda_{st}}{4 a^2 g^2 l_{st}} \right).$$

(C.16)

Which is precisely Eq. (2.37) of the main text. The values of $\lambda_{tw}$ and $\lambda_{st}$ that minimise Eq. (C.16) are obtained from Eqs. (2.38) and (2.39) of the main text.

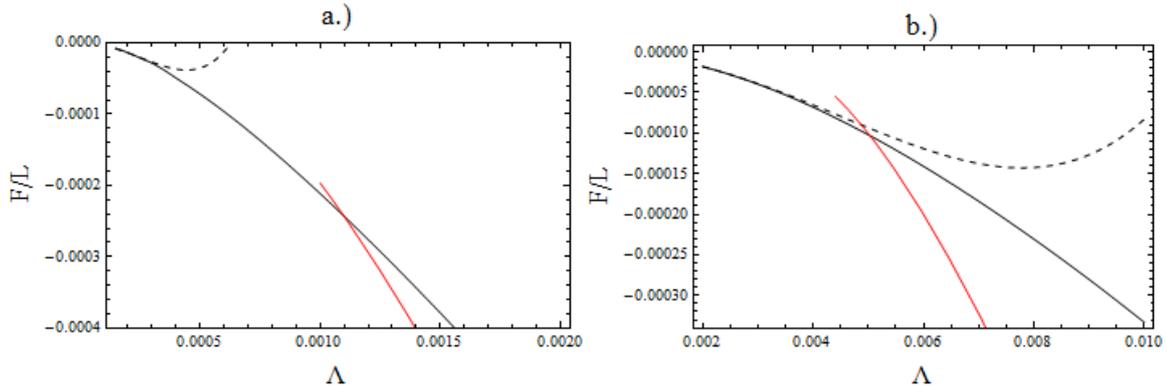

Fig. C.1 Free energies per unit length for the various pairing states at large $L$ using the Morse form for $\bar{G}_{n,n}(R/a, k_z)$ (Eq. (2.15) of the main text), as a function of $\Lambda$ in units of $k_B T$. For panel a.) the value $\lambda_{eff} = 4\text{Å}$ is used and for panel b.) the value $\lambda_{eff} = 2\text{Å}$. The red curves correspond to the state where $\lambda_{tw}$ is finite in Eq. (C.16), the solid black lines correspond to the state where $\lambda_{tw} = \infty$ (the free energy desribed by Eq. (2.43) of the main text). The black dashed lines correspond to the anaylitical form for the free energy for small $\Lambda$ given by Eq. (2.44) of the main text, where Eq. (2.15) of the main text is also used.

In Fig C.1, we show the Free energies of the two states around the point of transition between them; the one where $\lambda_{tw} = \infty$ and the one where $\lambda_{tw}$ is finite. The point of transition is where the energies of the two states are equal. The small $\Lambda$ analytical approximation of the free energy for the $\lambda_{tw} = \infty$ state (Eq. (2.44) of the main text) fails to yield a transition between the two states at $\lambda_{eff} = 4\text{Å}$. On the other hand, for $\lambda_{eff} = 2\text{Å}$, the analytical approximation becomes more accurate. When $\lambda_{eff}$ is reduced, the transition is pushed to larger values of $\Lambda$. In Fig. C.2 we present graphs for both $\lambda_{st}$ and $\lambda_{tw}$. These graphs qualitatively similar to those calculated using the Debye-Huckel form for $\bar{G}_{n,n}(R/a, k_z)$ presented in the main text (c.f. Fig. 4). However, the transition occurs at smaller values of $\Lambda$.

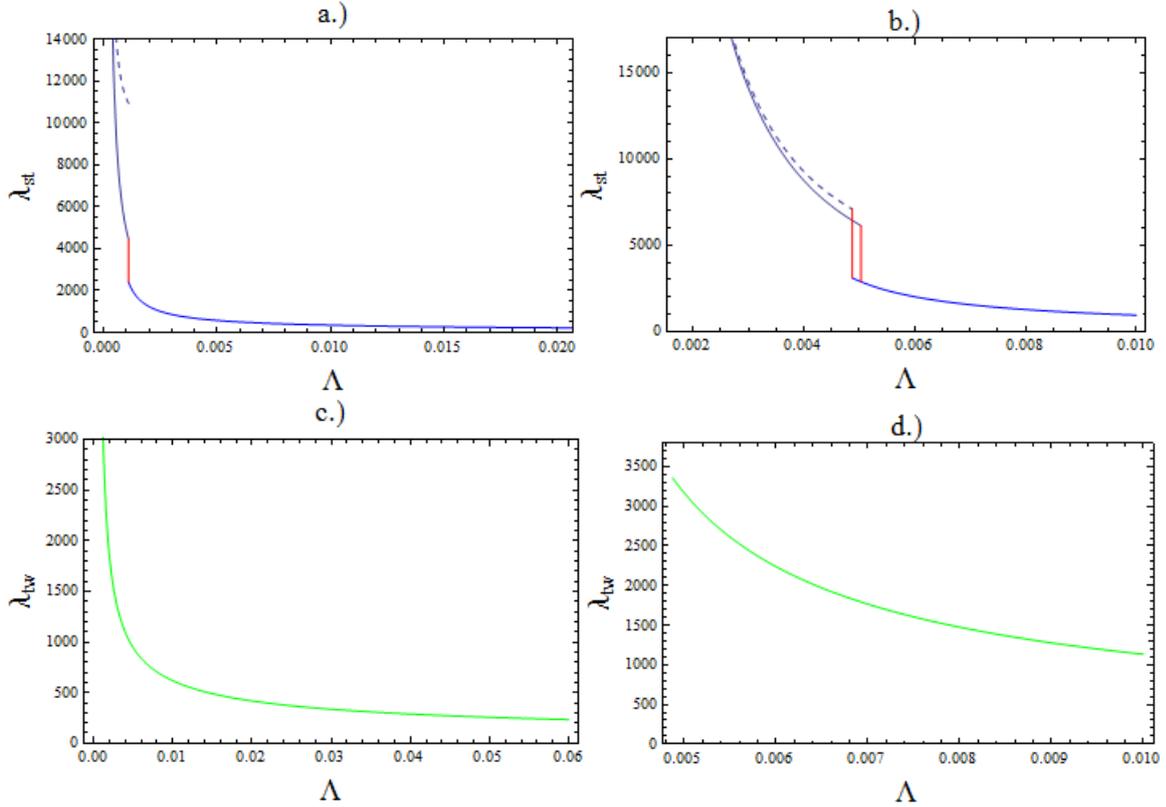

Fig. C.2 Plots of $\lambda_{st}$ and $\lambda_{tw}$ using $\bar{G}_{n,n}(R/a, k_z)$ for the Morse potential (Eq. (2.15) of the main text) as functions of $\Lambda$ in units of $k_B T$. Panels a.) and b.) show $\lambda_{st}$ calculated for the values $\lambda_{eff} = 4\text{Å}$ and $\lambda_{eff} = 2\text{Å}$, respectively. Here, the red lines mark a discontinuity between the state where $\lambda_{tw} = \infty$ (for small values of $\Lambda$) to a state where $\lambda_{tw}$ is finite. The dashed lines are the analytical form for $\lambda_{st}$ for small $\Lambda$ (Eq. (2.44) of the main text). For panel a.), the analytical form for $\lambda_{st}$ does not give a transition between the two states. In panels c.) and d.) graphs for $\lambda_{tw}$ are shown for the values $\lambda_{eff} = 4\text{Å}$ and $\lambda_{eff} = 2\text{Å}$, respectively.

Finally, in Appendix C, we present the graphs for the free energies of the two states, calculated using the Debye-Huckel form for $\bar{G}_{n,n}(R/a, k_z)$. Again, the point of transition is determined where the free energy of the $\lambda_{tw} = \infty$ and that of the $\lambda_{tw}$ state are equal. We see that, here, the small $\Lambda$ analytical approximation works much better than when the Morse form for $\bar{G}_{n,n}(R/a, k_z)$ is used.

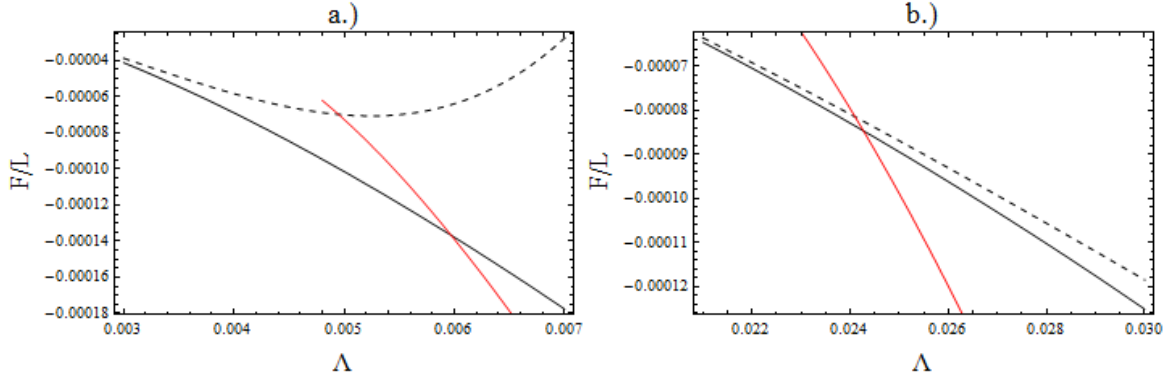

Fig C.3 Free energies per unit length (units of $k_B T / Å$) for the various pairing states at large $L$ using the Debye-Huckel form for $\bar{G}_{n,n}(R/a, k_z)$ (Eq. (2.14) of the main text), as functions of $\Lambda$ in units of $k_B T$. For panel a.) the value $\lambda_{eff} = 4Å$ is used and for panel b.) the value $\lambda_{eff} = 2Å$. The red curves correspond to the state where $\lambda_{tw}$ is finite in Eq. (C.16), the solid black lines correspond to the state where $\lambda_{tw} = \infty$ (the free energy desribed by Eq. (2.43) of the main text). The black dashed lines correspond to the anaylitical form for the free energy for small $\Lambda$ given by Eq. (2.44) of the main text, where Eq. (2.14) of the main text is also used.

## Appendix D Corrections to the variational approximation for small $\Lambda$

Here, we consider corrections to the free energy for the finite $\lambda_{st}$, $\lambda_{tw} = \infty$ state arising from residual correlations between fluctuations in the azimuthal angles $\delta\phi_{1,T}(s)$ and $\delta\phi_{2,T}(s)$. We start by writing the total energy functional as

$$E_{total}[\delta\phi_{1,T}(s), \delta\phi_{2,T}(s), \delta s_{1,T}(s), \delta s_{2,T}(s)] = E_{el}[\delta\phi_{1,T}(s), \delta\phi_{2,T}(s), \delta s_{1,T}(s), \delta s_{2,T}(s)]$$
$$+ V_0[\delta s_{1,T}(s), \delta s_{2,T}(s)] + \Delta V_L[\delta\phi_{1,T}(s), \delta\phi_{2,T}(s), \delta s_{1,T}(s), \delta s_{2,T}(s)],$$

(D.1)

where

$$V_0[\delta s_{1,T}(s), \delta s_{2,T}(s)] = -\Lambda \int_{-L/2}^{L/2} ds \int_{-\infty}^{\infty} dk_z \exp\left(i\frac{(\delta s_{1,T}(s) - \delta s_{2,T}(s))}{a} k_z\right) \bar{G}_{0,0}(R/a, k_z),$$

(D.2)

and

$$\Delta V_L[\delta\phi_{1,T}(s),\delta\phi_{2,T}(s),\delta s_{1,T}(s),\delta s_{2,T}(s)] = -\Lambda \int_{-L/2}^{L/2} ds \sum_{n=-\infty}^{\infty} (1-\delta_{n,0}) \exp(in(\phi_1-\phi_2))$$
$$\exp(in(\delta\phi_{1,T}(s)-\delta\phi_{2,T}(s))) \int_{-\infty}^{\infty} dk_z (-1)^{n'} \bar{G}_{n,n}(R/a,k_z) \exp\left(i\frac{(\delta s_{1,T}(s)-\delta s_{2,T}(s))}{a}k_z\right).$$

(D.3)

We then expand out the partition function in powers of $\Delta V_L$ so that

$$Z = \sum_{n=0}^{\infty} \int D\delta s_{1,T}(s) \int D\delta s_{2,T}(s) \exp\left(-\frac{E_0[\delta s_{1,T}(s),\delta s_{2,T}(s)] + E_{st}[\delta s_{1,T}(s),\delta s_{2,T}(s)]}{k_B T}\right) \tilde{Z}^{(n)}[\delta s_{1,T}(s),\delta s_{2,T}(s)],$$

(D.4)

where

$$\tilde{Z}^{(n)}[\delta s_{1,T}(s),\delta s_{2,T}(s)] = \frac{(-1)^n}{n!} \int D\delta\phi_{1,T}(s) \int D\delta\phi_{2,T}(s)$$
$$\left(\frac{\Delta V_L[\delta\phi_{1,T}(s),\delta\phi_{2,T}(s),\delta s_{1,T}(s),\delta s_{2,T}(s)]}{k_B T}\right)^n \exp\left(-\frac{E_{tw}[\delta\phi_{1,T}(s),\delta\phi_{2,T}(s)]}{k_B T}\right),$$

(D.5)

and $E_{st}$ and $E_{tw}$ are given by Eqs. (2.20) and (2.21) of the main text. We can rewrite Eq. (D.4) in terms of an effective energy functional as

$$Z = \int D\delta s_{1,T}(s) \int D\delta s_{2,T}(s) \exp\left(-\frac{E_{eff}[\delta s_{1,T}(s),\delta s_{2,T}(s)]}{k_B T}\right),$$

(D.6)

where

$$E_{eff}[\delta s_{1,T}(s),\delta s_{2,T}(s)] = V_0[\delta s_{1,T}(s),\delta s_{2,T}(s)] + E_{st}[\delta s_{1,T}(s),\delta s_{2,T}(s)] - k_B T \ln\left(\sum_{n=0}^{\infty} \tilde{Z}^{(n)}[\delta s_{1,T}(s),\delta s_{2,T}(s)]\right).$$

(D.7)

In this analysis, we retain in the sum up to $n=2$, and thus we may approximate

$$E_{eff}[\delta s_{1,T}(s),\delta s_{2,T}(s)] \approx -k_B T \ln \tilde{Z}^{(0)}$$
$$+ E_0[\delta s_{1,T}(s),\delta s_{2,T}(s)] + E_{st}[\delta s_{1,T}(s),\delta s_{2,T}(s)] - k_B T \frac{\tilde{Z}^{(1)}[\delta s_{1,T}(s),\delta s_{2,T}(s)]}{\tilde{Z}^{(0)}} +$$
$$-k_B T \left(\frac{\tilde{Z}^{(2)}[\delta s_{1,T}(s),\delta s_{2,T}(s)]}{\tilde{Z}^{(0)}} - \frac{1}{2}\left(\frac{\tilde{Z}^{(1)}[\delta s_{1,T}(s),\delta s_{2,T}(s)]}{\tilde{Z}^{(0)}}\right)^2\right).$$

(D.8)

We then use the variational trial functional described by Eq. (2.33) of the main text and write for the approximate free energy

$$F_T = -k_B T \ln Z_{tr,s} + \left\langle E_{eff}[\delta s_{1,T}(s), \delta s_{2,T}(s)] - E_{tr,s}[\delta s_{1,T}(s), \delta s_{2,T}(s)] \right\rangle_{tr,s}, \tag{D.9}$$

where

$$\left\langle E_{eff}[\delta s_{1,T}(s), \delta s_{2,T}(s)] - E_{tr,s}[\delta s_{1,T}(s), \delta s_{2,T}(s)] \right\rangle_{tr,s}$$
$$= \frac{1}{Z_{tr,s}} \int D\delta s_{1,T}(s) \int D\delta s_{2,T}(s) \exp\left(-\frac{E_{tr,s}[\delta s_{1,T}(s), \delta s_{2,T}(s)]}{k_B T}\right) \tag{D.10}$$
$$(E_{eff}[\delta s_{1,T}(s), \delta s_{2,T}(s)] - E_{tr,s}[\delta s_{1,T}(s), \delta s_{2,T}(s)]),$$

where $Z_{tr,s}$ is given by Eq. (C.6). The first term in Eq. (D.8) which contributes to Eq. (D.10) depends on $\ln \tilde{Z}^{(0)}$ is an unimportant constant that can be got rid of by subtracting off the free energy at $R = \infty$. The next contribution from Eq. (D.8) to Eqs. (D.9) and (D.10) is

$$F^{(0)} = -k_B T \ln Z_{tr,s} + \left\langle E_{st}[\delta s_{1,T}(s), \delta s_{2,T}(s)] + E_0[\delta s_{1,T}(s), \delta s_{2,T}(s)] \right\rangle_{tr,s}. \tag{D.11}$$

The terms described by Eq. (D.11) are simply the set of terms for the variational approximation for the finite $\lambda_{st}$, $\lambda_{tw} = \infty$ state. Thus, following the analysis of Appendix C, we may write

$$\frac{F^{(0)}}{L k_B T} = \frac{1}{2\lambda_{st}} - \frac{\Lambda}{k_B T} \int_{-\infty}^{\infty} dk_z \bar{G}_{0,0}(R/a, k_z) \exp\left(-\frac{k_z^2 \lambda_{st}}{4a^2 g^2 l_{st}}\right) \tag{D.12}$$

Next we consider the term

$$F^{(1)} = -\frac{k_B T}{\tilde{Z}^{(0)}} \left\langle \tilde{Z}^{(1)}[\delta s_{1,T}(s), \delta s_{2,T}(s)] \right\rangle_{tr,s}$$
$$= \frac{1}{Z_\phi Z_{tr,s}} \int D\delta s_{1,T}(s) \int D\delta s_{2,T}(s) \int D\delta\phi_{1,T}(s) \int D\delta\phi_{2,T}(s) \Delta V_L[\delta\phi_{1,T}(s), \delta\phi_{2,T}(s), \delta s_{1,T}(s), \delta s_{2,T}(s)]$$
$$\exp\left(-\frac{E_{tw}[\delta\phi_{1,T}(s), \delta\phi_{2,T}(s)]}{k_B T}\right) \exp\left(-\frac{E_{tr,s}[\delta s_{1,T}(s), \delta s_{2,T}(s)]}{k_B T}\right).$$

$$\tag{D.13}$$

Using Eq. (D.3) we can write Eq. (D.13) as

$$F^{(1)} = -\Lambda \int_{-L/2}^{L/2} ds \sum_{n=-\infty}^{\infty} (1-\delta_{n,0}) \exp(in(\phi_1 - \phi_2)) \left\langle \exp(in(\delta\phi_{1,T}(s) - \delta\phi_{2,T}(s))) \right\rangle_\phi$$
$$\int_{-\infty}^{\infty} dk_z (-1)^n \bar{G}_{n,n}(R/a, k_z) \left\langle \exp\left(\frac{ik_z(\delta s_{1,T}(s) - \delta s_{2,T}(s))}{a}\right) \right\rangle_{tr,s}.$$

(D.14)

Evaluation of the averages has already been performed before and their expressions are given by Eqs. (B.17) (with $n = n'$) and (C.8), and substitution of these results and integration over $s$ gives

$$F^{(1)} = -2\Lambda l_{tw} \sum_{n=-\infty}^{\infty} (1-\delta_{n,0}) \frac{1}{n^2} \exp(in(\phi_1 - \phi_2)) \left[1 - \exp\left(-\frac{n^2 L}{2l_{tw}}\right)\right]$$

$$\int_{-\infty}^{\infty} dk_z (-1)^n \bar{G}_{n,n}(R/a, k_z) \exp\left(-\frac{k_z^2 \lambda_{st}}{4a^2 g^2 l_{st}}\right).$$

(D.15)

For $L \gg l_{tw}$ (it is expected that $\lambda_{st} > l_{tw}$ for the $\lambda_{tw} = \infty$ state, and for the following analysis we require $L \gg \lambda_{st}$) this simply becomes

$$F^{(1)} = -2\Lambda l_{tw} \sum_{n=-\infty}^{\infty} (1-\delta_{n,0}) \frac{1}{n^2} \exp(in(\phi_1 - \phi_2)) \int_{-\infty}^{\infty} dk_z (-1)^n \bar{G}_{n,n}(R/a, k_z) \exp\left(-\frac{k_z^2 \lambda_{st}}{4a^2 g^2 l_{st}}\right),$$

(D.16)

and so can be neglected from the free energy at large $L$, as it scales as a constant with respect to $L$. The next term in Eq.(D.8) that contributes to Eq. (D.10) is

$$F^{(2)} = -\frac{k_B T}{\tilde{Z}^{(0)}} \left\langle \tilde{Z}^{(2)}[\delta s_{1,T}(s), \delta s_{2,T}(s)] \right\rangle_{tr,s}$$

$$= -\frac{1}{2Z_\phi Z_{tr,s} k_B T} \int D\delta s_{1,T}(s) \int D\delta s_{2,T}(s) \int D\delta\phi_{1,T}(s) \int D\delta\phi_{2,T}(s) \Delta V_L [\delta\phi_{1,T}(s), \delta\phi_{2,T}(s), \delta s_{1,T}(s), \delta s_{2,T}(s)]^2$$

$$\exp\left(-\frac{E_{tw}[\delta\phi_{1,T}(s), \delta\phi_{2,T}(s)]}{k_B T}\right) \exp\left(-\frac{E_{tr,s}[\delta s_{1,T}(s), \delta s_{2,T}(s)]}{k_B T}\right).$$

(D.17)

Using Eq. (D.17) this can be written as

$$F^{(2)} \approx -\frac{\Lambda^2}{2k_B T} \int_{-L/2}^{L/2} ds \int_{-L/2}^{L/2} ds' \sum_{m=-\infty}^{\infty} \sum_{n=-\infty}^{\infty} \left\langle \exp(i(n\Delta\phi_T(s) + m\Delta\phi_T(s'))) \right\rangle_\phi (1-\delta_{n,0})(1-\delta_{m,0})$$

$$\int_{-\infty}^{\infty} dk_z \int_{-\infty}^{\infty} dk_z' \bar{G}_{n,n}(R/a, k_z) \bar{G}_{m,m}(R/a, k_z') \left\langle \exp(i\Delta s_T(s) k_z) \exp(i\Delta s_T(s') k_z') \right\rangle_{tr,s},$$

(D.18)

where $\Delta\phi_T(s) = \delta\phi_{1,T}(s) - \delta\phi_{2,T}(s)$ and $\Delta s_T(s) = \delta s_{1,T}(s) - \delta s_{2,T}(s)$. From Eq. (D.18), we consider only the $m = -n$ modes that will predominate in the $L$ limit, so that

$$F^{(2)} \approx -\frac{\Lambda^2}{2k_B T} \int_{-L/2}^{L/2} ds \int_{-L/2}^{L/2} ds' \sum_{n=-\infty}^{\infty} \left\langle \exp\left(in(\Delta\phi_T(s) - \Delta\phi_T(s'))\right)\right\rangle_\phi (1-\delta_{n,0})$$

$$\int_{-\infty}^{\infty} dk_z \int_{-\infty}^{\infty} dk'_z \, \bar{G}_{n,n}(R/a, k_z) \bar{G}_{n,n}(R/a, k'_z) \left\langle \exp(i\Delta s_T(s)k_z) \exp(i\Delta s_T(s')k'_z)\right\rangle_{tr,s}.$$

(D.19)

Let us consider the averages in Eq. (D.19). First of all, by noting that

$$\left\langle \exp\left(in(\Delta\phi_T(s) - \Delta\phi_T(s'))\right)\right\rangle_\phi = \left\langle \exp\left(in\left(\int_{s'}^{s} ds'' \Delta\dot\phi_T(s'')\right)\right)\right\rangle_\phi$$

$$= \frac{1}{Z_\phi} \int D\Delta\dot\phi_T(s'') \exp\left(in\int_{s'}^{s} ds'' \Delta\dot\phi_T(s'')\right) \exp\left(-\frac{l_{tw}}{4} \int_{-\infty}^{\infty} ds'' (\Delta\dot\phi_T(s''))^2\right),$$

(D.20)

we evaluate that average to be

$$\left\langle \exp\left(in(\Delta\phi_T(s) - \Delta\phi_T(s'))\right)\right\rangle_\phi = \exp\left(-\frac{n^2 |s-s'|}{l_{tw}}\right).$$

(D.21)

The other average we may express as

$$\left\langle \exp(i\Delta s_T(s)k_z) \exp(i\Delta s_T(s')k'_z)\right\rangle_s$$

$$= \frac{\int D\Delta s_T(s) \exp(i\Delta s_T(s)k_z) \exp(i\Delta s_T(s')k'_z) \exp\left(-\dfrac{\tilde{E}_{tr,s}[\Delta s_T(s)]}{k_B T}\right)}{\int D\Delta s_T(s) \exp\left(-\dfrac{\tilde{E}_{tr,s}[\Delta s_T(s)]}{k_B T}\right)}$$

(D.22)

where

$$\tilde{E}_{tr,s}[\Delta s_T(s)] = \frac{1}{2} \int_{-L/2}^{L/2} ds \int_{-L/2}^{L/2} ds' \Delta s_T(s) G^{-1}(s-s') \Delta s_T(s')$$

(D.23)

and

$$G^{-1}(s-s') = \left(-\frac{l_p}{2}\frac{d^2}{ds^2} + \beta\right)\delta(s-s').$$

(D.24)

Evaluation of the average given by Eq. (D.22) yields

$$\left\langle \exp(i\Delta s_T(s)k_z)\exp(i\Delta s_T(s')k'_z)\right\rangle = \exp\left(-\frac{k_z^2 d_s^2}{2a^2}\right)\exp\left(-\frac{k_z'^2 d_s^2}{2a^2}\right)\exp\left(-\frac{k_z k'_z}{a^2}G(s-s')\right)$$

$$= \exp\left(-\frac{k_z^2 \lambda_{st}}{4a^2 g^2 l_{st}}\right)\exp\left(-\frac{k_z'^2 \lambda_{st}}{4a^2 g^2 l_{st}}\right)\exp\left(-\frac{k_z k'_z}{a^2}G(s-s')\right),$$

(D.25)

and

$$G(s-s') = \frac{1}{\pi g^2 l_{st}} \int_{-\infty}^{\infty} \frac{dk}{k^2 + \frac{4}{\lambda_{st}^2}} \exp(-k(s-s')) = \frac{\lambda_{st}}{2g^2 l_{st}} \exp\left(-\frac{2|s-s'|}{\lambda_{st}}\right). \tag{D.26}$$

Thus, we can write Eq. (D.19) as

$$F^{(2)} \approx \frac{2\Lambda^2 L}{k_B T} \int_0^{\infty} dx \sum_{n=1}^{\infty} \exp\left(-\frac{n^2 x}{l_{tw}}\right) \int_{-\infty}^{\infty} dk_z \int_{-\infty}^{\infty} dk_z' \bar{G}_{n,n}(R/a, k_z) \bar{G}_{n,n}(R/a, k_z')$$

$$\exp\left(-\frac{k_z^2 \lambda_{st}}{4a^2 g^2 l_{st}}\right) \exp\left(-\frac{k_z'^2 \lambda_{st}}{4a^2 g^2 l_{st}}\right) \exp\left(-\frac{k_z k_z' \lambda_{st}}{2a^2 g^2 l_{st}} \exp\left(-\frac{2x}{\lambda_{st}}\right)\right). \tag{D.27}$$

Next, one can expand out Eq. (D.27), so obtaining

$$F^{(2)} \approx -\frac{2\Lambda^2 L}{k_B T} \sum_{n=1}^{\infty} \sum_{j=0}^{\infty} \frac{(-1)^j}{j!} \frac{\lambda_{st}^j}{\left(2a^2 g^2 l_{st}\right)^j} \int_0^{\infty} dx \exp\left(-\frac{n^2 x}{l_{tw}}\right) \exp\left(-\frac{2jx}{\lambda_{st}}\right)$$

$$\int_{-\infty}^{\infty} dk_z \int_{-\infty}^{\infty} dk_z' \bar{G}_{n,n}(R/a, k_z) \bar{G}_{n,n}(R/a, k_z') k_z^j k_z'^j \exp\left(-\frac{k_z^2 \lambda_{st}}{4a^2 g^2 l_{st}}\right) \exp\left(-\frac{k_z'^2 \lambda_{st}}{4a^2 g^2 l_{st}}\right). \tag{D.28}$$

Then, on evaluation of each $x$ integral in Eq. (D.28), we obtain

$$F^{(2)} \approx -\frac{2\Lambda^2 L}{k_B T} \sum_{n=1}^{\infty} \sum_{j=0}^{\infty} \frac{1}{(2j)!} \frac{\lambda_{st}^{2j}}{\left(2a^2 g^2 l_{st}\right)^{2j}} \frac{1}{\frac{n^2}{l_{tw}} + \frac{4j}{\lambda_{st}}}$$

$$\int_{-\infty}^{\infty} dk_z \int_{-\infty}^{\infty} dk_z' \bar{G}_{n,n}(R/a, k_z) \bar{G}_{n,n}(R/a, k_z') k_z^{2j} k_z'^{2j} \exp\left(-\frac{k_z^2 \lambda_{st}}{4a^2 g^2 l_{st}}\right) \exp\left(-\frac{k_z'^2 \lambda_{st}}{4a^2 g^2 l_{st}}\right). \tag{D.29}$$

In writing Eq. (D.29), we have taken account that the odd powers in $j$ vanish as the integrands are odd in both $k_z$ and $k_z'$.

When $\lambda_{st}$ is very large, we can approximate both the $k_z$ and $k_z'$ integrals to integrals that can be performed analytically, and Eq. (D.29) simplifies to

$$F^{(2)} \approx -\frac{2\Lambda^2 L}{k_B T} \frac{\left(4a^2 g^2 l_{st}\right)}{\lambda_{st}} \sum_{n=1}^{\infty} \sum_{j=0}^{\infty} \frac{4^j}{(2j)!} \frac{1}{\frac{n^2}{l_{tw}} + \frac{4j}{\lambda_{st}}} \bar{G}_{n,n}(R/a, 0)^2 D_j^2, \tag{D.30}$$

where

$$D_j = \int_{-\infty}^{\infty} x^{2j} \exp(-x^2) = \Gamma(j+1/2), \tag{D.31}$$

and $\Gamma(j+1/2)$ is the standard gamma function with argument $j+1/2$. Let's estimate the series over $j$ contained in Eq. (D.30), namely

$$S = \sum_{j=0}^{\infty} \frac{4^j}{\frac{n^2}{l_{tw}} + \frac{4j}{\lambda_{st}}} \frac{\Gamma(j+1/2)^2}{(2j)!}. \tag{D.32}$$

Firstly, for large $j$, we observe that

$$\frac{4^j \Gamma(j+1/2)^2}{(2j)!} = \frac{4^j \Gamma(j+1/2)^2}{\Gamma(2j+1)} \approx \frac{\sqrt{\pi}}{\left(j+\frac{1}{2}\right)^{1/2}}. \tag{D.33}$$

We have used Stirling's formula for large arguments of the Gamma function that reads as $\Gamma(z) \approx \sqrt{2\pi} z^{z-1/2} e^{-z}$ in approximating the terms in Eq.(D.33). The summing $j$ over only Eq. (D.33) is not a convergent sum; the additional $1/(n^2/l_{tw} + 4j/\lambda_{st})$ term in Eq. (D.32) is needed to insure that the sum is convergent. Thus, when $\lambda_{st}$ is large, the sum is dominated by large values of $j$. The upshot is that we may estimate the sum (Eq. (D.32) for large $\lambda_{st}$) by

$$S \approx \lambda_{st}^{-1/2} \sum_{j=0}^{\infty} \frac{1}{\left(\frac{n^2}{l_{tw}} + \frac{4j}{\lambda_{st}}\right)} \frac{\sqrt{\pi}}{\left(\frac{j}{\lambda_{st}}\right)^{1/2}} \approx \frac{\lambda_{st}^{1/2}}{2} \int_0^{\infty} dy \frac{1}{n^2/(4l_{tw}) + y^2} = \frac{\lambda_{st}^{1/2} l_{tw}^{1/2}}{|n|} \frac{\pi}{2}. \tag{D.34}$$

where we replace $j/\lambda_{st}$ in the sum by the integration variable $y$. Thus, in the limit of large $\lambda_{st}$, we have that for the correction

$$F^{(2)} \approx -\frac{4\pi\Lambda^2 a^2 g^2 l_{st} L}{k_B T} \left(\frac{l_{tw}}{\lambda_{st}}\right)^{1/2} \sum_{n=1}^{\infty} \frac{1}{n} \bar{G}_{n,n}(R/a,0)^2. \tag{D.35}$$

In the limit of large $\lambda_{st}$ we may substitute the leading order term of the analytic form for $\lambda_{st}$ for small $\Lambda$ (Eq. (2.41) of the main text, for large $\lambda_{st}$) this yields

$$F^{(2)} \approx -8\pi^{3/2} a^3 g^3 l_{st}^{3/2} l_{tw}^{1/2} \frac{L\Lambda^3 \bar{G}_{0,0}(R/a,0)}{(k_B T)^2} \sum_{n=1}^{\infty} \frac{1}{n} \bar{G}_{n,n}(R/a,0)^2. \tag{D.36}$$

Last of all, let us consider

$$F^{(3)} = \frac{k_B T}{2(\tilde{Z}^{(0)})^2} \left\langle \tilde{Z}^{(1)}[\delta s_{1,T}(s), \delta s_{2,T}(s)]^2 \right\rangle_{tr,s}$$

$$\approx \frac{\Lambda^2}{2k_B T} \int_{-L/2}^{L/2} ds \int_{-L/2}^{L/2} ds' \sum_{m=-\infty}^{\infty} \sum_{n=-\infty}^{\infty} \left\langle \exp(in\Delta\phi_T(s)) \right\rangle_\phi \left\langle \exp(im\Delta\phi_T(s')) \right\rangle_\phi (1-\delta_{n,0})(1-\delta_{m,0}) \quad \text{(D.37)}$$

$$\int_{-\infty}^{\infty} dk_z \int_{-\infty}^{\infty} dk_z' \bar{G}_{n,n}(R/a, k_z) \bar{G}_{m,m}(R/a, k_z') \left\langle \exp(i\Delta s_T(s) k_z) \exp(i\Delta s_T(s') k_z') \right\rangle_{tr,s},$$

Inserting the various averages (Eqs. (B.17) and (D.25)), and retaining only the dominant $n=-m$ modes, this becomes for very large $L$

$$F^{(3)} \approx \frac{\Lambda^2}{2k_B T} \int_{-\infty}^{\infty} ds' \int_{-\infty}^{\infty} ds \sum_{n=1}^{\infty} \exp\left(-\frac{n^2|s|}{2l_{tw}}\right) \exp\left(-\frac{n^2|s'|}{2l_{tw}}\right) \int_{-\infty}^{\infty} dk_z \int_{-\infty}^{\infty} dk_z' \bar{G}_{n,n}(R/a, k_z) \bar{G}_{n,n}(R/a, k_z')$$

$$\exp\left(-\frac{k_z^2 \lambda_{st}}{4a^2 g^2 l_{st}}\right) \exp\left(-\frac{k_z'^2 \lambda_{st}}{4a^2 g^2 l_{st}}\right) \exp\left(-\frac{k_z k_z' \lambda_{st}}{2a^2 g^2 l_{st}} \exp\left(-\frac{2|s-s'|}{\lambda_{st}}\right)\right).$$

(D.38)

As opposed to evaluating the $s$ integrals Eq. (D.38), in similar fashion to Eq. (D.27), will simply estimate its maximum size. We notice that it is bounded such that $0 \leq F^{(3)} \leq F_B$ where

$$F_B \approx \frac{\Lambda^2}{2k_B T} \int_{-\infty}^{\infty} ds' \int_{-\infty}^{\infty} ds \sum_{n=1}^{\infty} \exp\left(-\frac{n^2|s|}{2l_{tw}}\right) \exp\left(-\frac{n^2|s'|}{2l_{tw}}\right) \int_{-\infty}^{\infty} dk_z \int_{-\infty}^{\infty} dk_z' \bar{G}_{n,n}(R/a, k_z) \bar{G}_{n,n}(R/a, k_z')$$

$$\approx \frac{8\Lambda^2 l_{tw}^2}{k_B T} \sum_{n=1}^{\infty} \frac{1}{n^4} \int_{-\infty}^{\infty} dk_z \int_{-\infty}^{\infty} dk_z' \bar{G}_{n,n}(R/a, k_z) \bar{G}_{n,n}(R/a, k_z')$$

(D.39)

and this is the maximum size $F^{(3)}$ can take (this is when $\lambda_{st} = 0$). This suggests that $L \gg l_{tw}$, which we are considering, we can neglect $F^{(3)}$.

Let us now combine the results for $F^{(0)}$ and $F^{(2)}$. This yields the following result for the free energy

$$\frac{F}{k_B T} \approx \frac{L}{2\lambda_{st}} - \frac{\Lambda L}{k_B T} \int_{-\infty}^{\infty} dk_z G_{0,0}(R/a, k_z) \exp\left(-\frac{k_z^2 \lambda_{st}}{4a^2 g^2 l_{st}}\right)$$

$$-2L\Lambda^2 \sum_{n=1}^{\infty} \sum_{j=0}^{\infty} \frac{1}{(2j)!} \frac{\lambda_{st}^{2j}}{\left(2a^2 g^2 l_{st}\right)^{2j}} \frac{1}{\frac{n^2}{l_{tw}} + \frac{4j}{\lambda_{st}}} \quad \text{(D.40)}$$

$$\int_{-\infty}^{\infty} dk_z \int_{-\infty}^{\infty} dk_z' G_{n,n}(R/a, k_z) G_{n,n}(R/a, k_z') k_z^{2j} k_z'^{2j} \exp\left(-\frac{k_z^2 \lambda_{st}}{4a^2 g^2 l_{st}}\right) \exp\left(-\frac{k_z'^2 \lambda_{st}}{4a^2 g^2 l_{st}}\right).$$

When $\tilde{\gamma}$ is very small (using Eq. (2.44) of the main text and Eq. (D.36)) we can write up to $\Lambda^4$

$$\frac{F}{k_B TL} \approx -\frac{2\pi a^2 g^2 l_{st} \Lambda^2}{(k_B T)^2} G_{0,0}(R/a, a\kappa_{eff}, 0)^2 - \frac{32\pi^2 a^6 g^6 l_{st}^3 \Lambda^4}{(k_B T)^4} G_{0,0}(R/a, a\kappa_{eff}, 0)^3 G'_{0,0}(R/a, a\kappa_{eff}, 0)$$

$$-8\pi^{3/2} a^3 g^3 l_{st}^{3/2} l_{tw}^{1/2} \frac{L\Lambda^3 \overline{G}_{0,0}(R/a, 0)}{(k_B T)^3} \sum_{n=1}^{\infty} \frac{1}{n} \overline{G}_{n,n}(R/a, 0)^2$$

(D.41)

We may now examine the three possible ways to handle the correction from the residual correlations between $\delta\phi_{1,T}(s)$ and $\delta\phi_{2,T}(s)$. The first way is to minimize the whole free energy, Eq. (D.40) with respect to $\lambda_{st}$. The second is to minimise only the leading term (Eq.(D.12)) with respect to $\lambda_{st}$. Last of all is to use the small $\Lambda$ expansion given by Eq.(D.41). The latter provides the most systematic treatment, as it is a power series expansion truncated to a particular order in $\Lambda$. On the other hand, in principle, retaining enough terms in the expansion for a weak interaction (Eq. (D.40)), and completely minimizing the free energy with respect to $\lambda_{st}$, should provide the most accurate approach.

Firstly, in Fig. D.1, we present the values of the free energy (calculated using the Debye-Huckel form of $\overline{G}_{n,n}(R/a, k_z)$) using these three ways of dealing with the correction $F^{(2)}$, compared with the result of simply only retaining $F^{(0)}$ (given by Eq. (D.12)). One thing to note is that the correction due to the residual correlations between $\delta\phi_{1,T}(s)$ and $\delta\phi_{2,T}(s)$ gives a considerable change in the free energy of the finite $\lambda_{st}$, $\lambda_{tw} = \infty$ state, lowering it. Also, fully minimizing Eq. (D.40) with respect to $\lambda_{st}$ and using Eq. (D.41) both have problems associated with not including a sufficient number of terms in the expansion (Eq.(D.4)). We see that for $\lambda_{eff} = 4\text{Å}$ the full minimization of Eq. (D.40) fails to give a transition to the finite $\lambda_{tw}$ state; if we included next correction arsing from the expansion (Eq. (D.4)) we would expect this to change, as this term would more than likely be positive. On the other hand using the small $\Lambda$ result, Eq.(D.41), doesn't seem that accurate, as the energy calculated does not agree that well with the other two approaches, when we consider $\lambda_{eff} = 4\text{Å}$. If we were to include higher order terms in the expansion given by Eq.(D.40) (laborious to calculate), and in Eq.(D.41) for small $\Lambda$, this problem undoubtedly would be corrected. For $\lambda_{eff} = 2\text{Å}$, however, all three approaches seem to work reasonably well, but fully minimizing Eq. (D.40) does not agree quite so well with the other two approaches. Thus, it seems, to this order of the calculation, the most reliable thing to do in finding the transition between the two states is to use the value of $\lambda_{st}$ that minimizes only Eq. (D.12), which we adopt later.

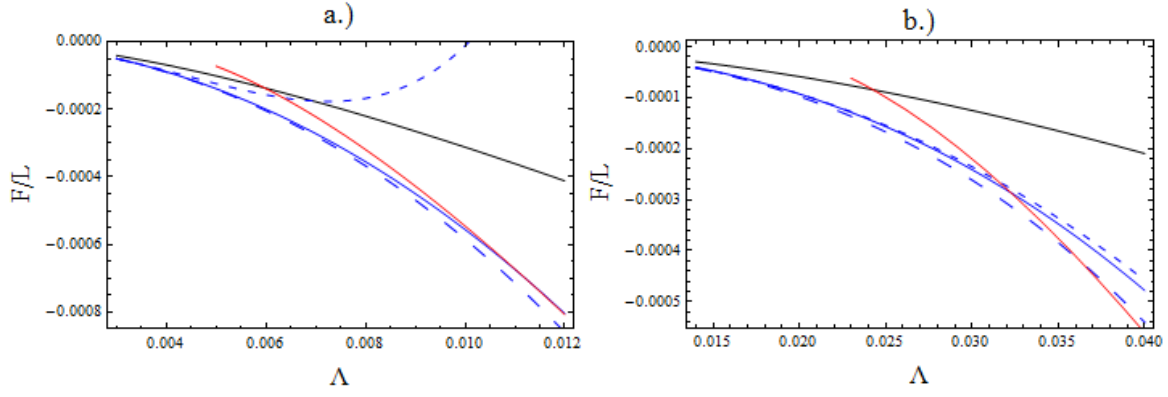

Fig. D.1. We show plots of the free energies with respect to the interaction strength, where corrections due to residual correlations in $\Delta\phi = \phi_1 - \phi_2$ have been considered. Here, we have used the Debye-Huckel form of $\bar{G}_{n,n}(R/a, k_z)$ (Eq. (2.14) of the main text). Both the interaction strength, $\Lambda$ and $F/L$ have been given in units of $k_B T$ and $k_B T / \text{Å}$, respectively. In panel a.) we use $\lambda_{eff} = 4\text{Å}$ and for b.) we use $\lambda_{eff} = 2\text{Å}$. The black curve corresponds to the result for the finite $\lambda_{st}$, $\lambda_{tw} = \infty$ state without the corrections, calculated retaining only $F^{(0)}$ (using only Eq. (D.12)). The red curve corresponds again to the finite $\lambda_{tw}$ free energy state calculated with Eq. (C.16), which depends on the azimuthal orientation of the molecules $\Delta\phi = \phi_1 - \phi_2$. The blue curves correspond to including to the free energy due to residual correlations in $\Delta\phi$. For the solid blue curve we calculate $\lambda_{st}$ through only Eq. (D.12) and substitute that value into Eq. (D.40). For the long dashed blue curve we have minimized Eq. (D.40) completely with respect to $\lambda_{st}$. For the medium dashed blue curve we use the small $\Lambda$ result for the free energy Eq. (D.41). In the calculations the values $R = 25\text{Å}$, $a = 11.2\text{Å}$, $l_{tw} = 1000\text{Å}$ and $l_{st} = 700\text{Å}$ are used.

Next, we replot $\lambda_{st}$ using the transition point calculated by using the value of $\lambda_{st}$ that minimizes only Eq. (D.12) in the full free energy (Eq. (D.40)) in Fig. D.2. Comparing with Fig. 4 of the main text we see that the effect of the corrections is to shift up transition between the two states to larger interaction strengths and to reduce the size of the jump in $\lambda_{st}$, at the transition.

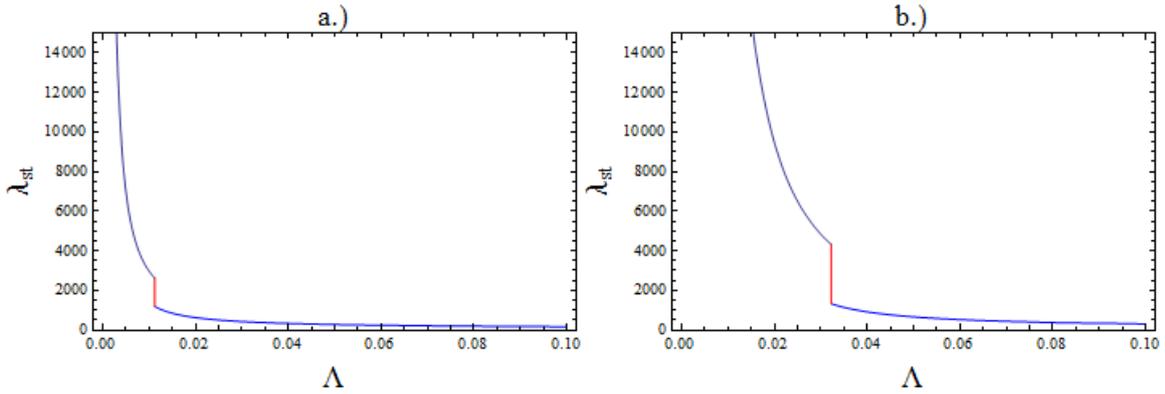

Fig. D.2. Plots showing $\lambda_{st}$ calculated through minimizing either Eqs. (C.16) or (D.12) and using the Debye-Huckel form for $\bar{G}_{n,n}(R/a, k_z)$, depending on the state of the molecular pair, as in $\lambda_{st}$ plots presented in the main text. The interaction strength $\Lambda$ is given in units of $k_B T$. However, now, we use Eq.(D.40) instead of only Eq. (D.12) for the free energy to determine the value of $\Lambda$ at which the transition occurs. In the calculations the values $R = 25\text{Å}$, $a = 11.2\text{Å}$, $l_{tw} = 1000\text{Å}$ and $l_{st} = 700\text{Å}$ are used.

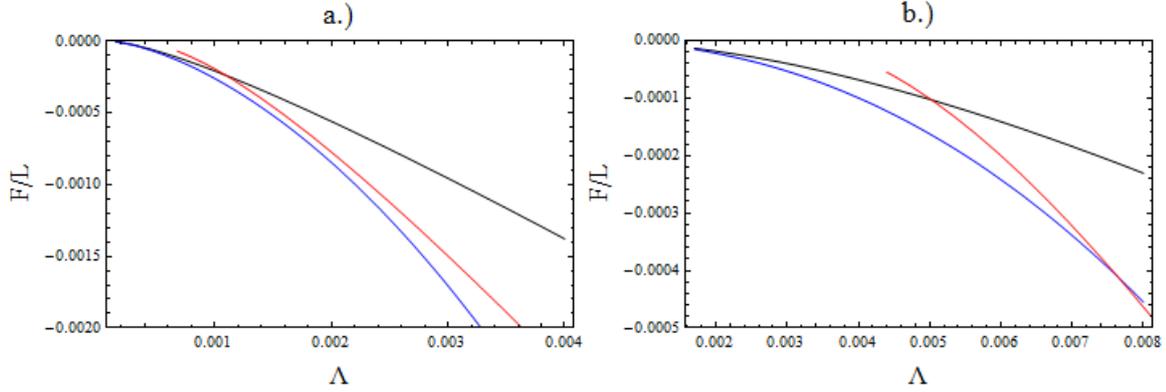

Fig. D.3. We show plots of the free energies with respect to the interaction strength, including the term $F^{(2)}$. Here, we have used the Morse potential form of $\bar{G}_{n,n}(R/a, k_z)$ (Eq. (2.15) of the main text). Again, the interaction strength $\Lambda$ and $F$ are given in units of $k_B T$. In panel a.) we use $\lambda_{eff} = 4\text{Å}$ and for b.) we use $\lambda_{eff} = 2\text{Å}$. The black curve corresponds to the result for the finite $\lambda_{st}$, $\lambda_{tw} = \infty$ state without the corrections, calculated retaining only $F^{(0)}$ (using only Eq. (D.12)). The red curve corresponds again to the finite $\lambda_{tw}$ free energy state calculated with Eq. (C.16), which depends on the azimuthal orientation of the molecules $\Delta\phi = \phi_1 - \phi_2$. The blue curve corresponds to including to the free energy due to residual correlations in $\Delta\phi$. For the solid blue curve we calculate $\lambda_{st}$ through only Eq. (D.12) and substitute that value into Eq. (D.40). In the calculations the values $R = 25\text{Å}$, $a = 11.2\text{Å}$, $l_{tw} = 1000\text{Å}$ and $l_{st} = 700\text{Å}$ are used.

In Fig. D.3 we plot free energy curves using $\bar{G}_{n,n}(R/a, k_z)$ for the Morse potential. Here, when considering the correction $F^{(2)}$, we only minimize $F^{(0)}$ with respect to $\lambda_{st}$. Unfortunately the correction is too large to get a transition between the $\lambda_{tw} = \infty$ state and the finite $\lambda_{tw}$ state for $\lambda_{eff} = 4\text{Å}$ and changes considerably the value at which the transition occurs with $\lambda_{eff} = 2\text{Å}$. This problem would be corrected if we included the next term in the expansion (Eq. (D.4)), but this would be rather laborious. Thus, we have refrained, in this case from plotting $\lambda_{st}$.

## Part 2 Models of global homology pairing through helix distortions

### Appendix E Rigid identical helices

Let us again start by considering rigid identical helices. For models based on non-pair specific interaction potentials between base-pairs, we start from Eq. (2.47) of the main text. We first make the continuum approximation, such that for identical sequences

$$\langle V_{global}(R)\rangle_{\Omega} = \frac{1}{h^2} \int_{-L/2}^{L/2} ds \int_{-L/2}^{L/2} ds' \int \frac{d^3k}{(2\pi)^3} G_{eff}(\mathbf{k}) \exp(-Rk_r \cos\phi_k) \exp(ik_z(s-s'))$$

$$\langle \exp(-iak_r \cos(gs' + \phi_2 + \delta\phi(s') - \phi_k)) \exp(iak_r \cos(gs + \phi_1 + \delta\phi(s) - \phi_k))\rangle_{\Omega}.$$

(E.1)

Going through exactly the same steps as presented in Appendix A. We obtain

$$\langle V_{global}(R)\rangle_{\Omega} = -\frac{\Lambda_g}{2\pi} \int_{-L/2}^{L/2} ds \int_{-L/2}^{L/2} ds' \sum_{n=-\infty}^{\infty} \sum_{n'=-\infty}^{\infty} \exp(ig(sn-s'n')) \exp(i(\phi_1 n - \phi_2 n')) \langle \exp(i(n\delta\phi(s) - n'\delta\phi(s')))\rangle_{\Omega}$$

$$\int_{-\infty}^{\infty} dk_z (-1)^{n'} \bar{G}_{n,n'}(R/a, ak_z) \exp(ik_z(s-s')),$$

(E.2)

where the function $\bar{G}_{n,n'}(R, ak_z)$ is exactly the same, as it was in the local models for same interaction potential (i.e. in general, it is given by Eqs. (A.13) or (A.14)), and where $\Lambda_g = \gamma_{eff} / h^2 (2\pi)$. We can evaluate the average (in similar fashion to the average considered in Eq. (B.24))

$$\langle \exp(i(n\delta\phi(s) - n'\delta\phi(s')))\rangle_{\Omega} = \exp\left(-\frac{(n^2 s - 2nn' \min(s,s') + n'^2 s')}{2\lambda_{tw}^{(0)}}\right) \theta(s)\theta(s')$$

$$+ \exp\left(\frac{(n^2 s - 2nn' \max(s,s') + n'^2 s')}{2\lambda_{tw}^{(0)}}\right) \theta(-s)\theta(-s') + \exp\left(-\frac{(n^2 s - n'^2 s')}{2\lambda_{tw}^{(0)}}\right) \theta(s)\theta(-s')$$

$$+ \exp\left(-\frac{(n'^2 s' - n^2 s)}{2\lambda_{tw}^{(0)}}\right) \theta(-s)\theta(s').$$

(E.3)

Where the functions $\theta(s)$ is the Heaviside theta function and the functions $\min(s,s')$ and $\max(s,s')$ were defined previously through Eqs. (B.33) and (B.34). Thus, using Eq. (E.3), we may express Eq. (E.2) as

$$\langle V_{global}(R)\rangle_{\Omega} = -\Lambda_g \sum_{n=-\infty}^{\infty} \sum_{n'=-\infty}^{\infty} \exp(i(\phi_1 n - \phi_2 n')) \int_{-\infty}^{\infty} dk_z (-1)^{n'} \bar{G}_{n,n'}(R/a, k_z) \Theta_{n,n'}(L, k_z),$$

(E.4)

where

$$\Theta_{n,n'}(L,k_z) = \int_0^{L/2} ds \int_0^s ds' \exp(ig(sn - s'n')) \exp(ik_z(s-s')) \exp\left(-\frac{1}{2\lambda_{tw}^{(0)}}\left(n^2 s - 2nn's' + n'^2 s'\right)\right)$$

$$+ \int_0^{L/2} ds' \int_0^{s'} ds \exp(ig(sn - s'n')) \exp(ik_z(s-s')) \exp\left(-\frac{1}{2\lambda_{tw}^{(0)}}\left(n^2 s - 2nn's + n'^2 s'\right)\right)$$

$$+ \int_{-L/2}^{0} ds \int_{s'}^{0} ds' \exp(ig(sn - s'n')) \exp(ik_z(s-s')) \exp\left(\frac{1}{2\lambda_{tw}^{(0)}}\left(n^2 s - 2nn's' + n'^2 s'\right)\right)$$

$$+ \int_{-L/2}^{0} ds' \int_{s}^{0} ds \exp(ig(sn - s'n')) \exp(ik_z(s-s')) \exp\left(\frac{1}{2\lambda_{tw}^{(0)}}\left(n^2 s - 2nn's + n'^2 s'\right)\right)$$

$$+ \int_{-L/2}^{0} ds' \int_{0}^{L/2} ds \exp(ig(sn - s'n')) \exp(ik_z(s-s')) \exp\left(-\frac{1}{2\lambda_{tw}^{(0)}}\left(n^2 s - n'^2 s'\right)\right)$$

$$+ \int_{-L/2}^{0} ds \int_{0}^{L/2} ds' \exp(ig(sn - s'n')) \exp(ik_z(s-s')) \exp\left(\frac{1}{2\lambda_{tw}^{(0)}}\left(n^2 s - n'^2 s'\right)\right).$$

(E.5)

Performing the integrations, Eq. (E.5) evaluates to the expression

$$\Theta_{n,n'}(L,k_z) = 2\Theta_{n,n'}^{(1)}(L,k_z) + 2\Theta_{n',n}^{(1)}(L,k_z) + 2\Theta_{n,n'}^{(2)}(L,k_z),$$

(E.6)

where

$$\Theta_{n,n'}^{(1)}(L,k_z) = \left[\left(\frac{n^2(n'^2 - 2nn')}{4(\lambda_{tw}^{(0)})^2} + (k_z + ng)(k_z + n'g)\right)^2 + \left(\frac{n^2(k_z + n'g)}{2\lambda_{tw}^{(0)}} - \frac{(n'^2 - 2nn')(k_z + ng)}{2\lambda_{tw}^{(0)}}\right)\right]^{-1}$$

$$\left\{\left[\frac{n^2(n'^2 - 2nn')}{4(\lambda_{tw}^{(0)})^2} + (k_z + ng)(k_z + n'g)\right]\left(1 - \cos\left(\frac{(k_z + ng)L}{2}\right)\exp\left(-\frac{n^2 L}{4\lambda_{tw}^{(0)}}\right)\right)\right.$$

$$\left. + \left[\frac{n^2(k_z + n'g)}{2\lambda_{tw}^{(0)}} - \frac{(n'^2 - 2nn')(k_z + ng)}{2\lambda_{tw}^{(0)}}\right]\sin\left(\frac{(k_z + ng)L}{2}\right)\exp\left(-\frac{n^2 L}{4\lambda_{tw}^{(0)}}\right)\right\}$$

$$+ \left[\left(\frac{(n'^2 - 2nn')(n - n')^2}{4(\lambda_{tw}^{(0)})^2} - (k_z + n'g)g(n - n')\right)^2 + \left(\frac{g(n-n')(n'^2 - 2nn')}{2\lambda_{tw}^{(0)}} + \frac{(k_z + n'g)(n - n')^2}{2\lambda_{tw}^{(0)}}\right)^2\right]^{-1}$$

$$\left\{\left[\frac{(n'^2 - 2nn')(n-n')^2}{4(\lambda_{tw}^{(0)})^2} - (k_z + n'g)g(n - n')\right]\left(1 - \cos\left(\frac{(n-n')gL}{2}\right)\exp\left(-\frac{L(n-n')^2}{4\lambda_{tw}^{(0)}}\right)\right)\right.$$

$$\left. + \left[\frac{g(n-n')(n'^2 - 2nn')}{2\lambda_{tw}^{(0)}} + \frac{(k_z + n'g)(n-n')^2}{2\lambda_{tw}^{(0)}}\right]\sin\left(\frac{(n-n')gL}{2}\right)\exp\left(-\frac{L(n-n')^2}{4\lambda_{tw}^{(0)}}\right)\right\},$$

(E.7)

and

$$\Theta^{(2)}_{n,n'}(L,k_z) = \left[\left(\frac{n^2 n'^2}{4(\lambda^{(0)}_{tw})^2} - (gn+k_z)(gn'+k_z)\right)^2 + \left(\frac{n^2(gn'+k_z)}{2\lambda^{(0)}_{tw}} + \frac{n'^2(gn+k_z)}{2\lambda^{(0)}_{tw}}\right)^2\right]^{-1}$$

$$\left\{\left[\frac{n^2 n'^2}{4(\lambda^{(0)}_{tw})^2} - (gn+k_z)(gn'+k_z)\right]\left[1 - \cos\left(\frac{L(gn'+k_z)}{2}\right)\exp\left(-\frac{Ln'^2}{4\lambda^{(0)}_{tw}}\right) - \cos\left(\frac{L(gn+k_z)}{2}\right)\exp\left(-\frac{Ln^2}{4\lambda^{(0)}_{tw}}\right)\right.\right.$$

$$\left.+ \cos\left(\frac{L(gn'+gn+2k_z)}{2}\right)\exp\left(-\frac{L(n^2+n'^2)}{4\lambda^{(0)}_{tw}}\right)\right] + \left[\frac{n^2(gn'+k_z)}{2\lambda^{(0)}_{tw}} + \frac{n'^2(gn+k_z)}{2\lambda^{(0)}_{tw}}\right]\left(\sin\left(\frac{L(gn'+k_z)}{2}\right)\right.$$

$$\left.\left.\exp\left(-\frac{Ln'^2}{4\lambda^{(0)}_{tw}}\right) + \sin\left(\frac{L(gn+k_z)}{2}\right)\exp\left(-\frac{Ln^2}{4\lambda^{(0)}_{tw}}\right) + \sin\left(\frac{L(gn'+gn+2k_z)}{2}\right)\exp\left(-\frac{L(n^2+n'^2)}{4\lambda^{(0)}_{tw}}\right)\right)\right\}.$$

(E.8)

The modes that dominate, for $Lg \gg 1$, are the $n=n'$ modes for which we extract from Eqs. (E.7) and (E.8). For $n=n'$ we have that

$$\Theta^{(1)}_{n,n}(L,k_z) = \frac{1}{\left[(k_z+ng)^2 + \frac{n^4}{4(\lambda^{(0)}_{tw})^2}\right]^2}\left[\left[(k_z+ng)^2 - \frac{n^2}{4(\lambda^{(0)}_{tw})^2}\right]\left[1 - \cos\left(\frac{(k_z+ng)L}{2}\right)\exp\left(-\frac{Ln^2}{4\lambda^{(0)}_{tw}}\right)\right]\right.$$

$$\left.+ \frac{(k_z+ng)n^2}{\lambda^{(0)}_{tw}}\sin\left(\frac{(k_z+ng)L}{2}\right)\exp\left(-\frac{Ln^2}{4\lambda^{(0)}_{tw}}\right)\right] + \frac{\frac{n^2 L}{4\lambda^{(0)}_{tw}}}{(k_z+ng)^2 + \frac{n^4}{4(\lambda^{(0)}_{tw})^2}},$$

(E.9)

and

$$\Theta^{(2)}_{n,n}(L,k_z) = \frac{1}{\left[(k_z+ng)^2 + \frac{n^4}{4(\lambda^{(0)}_{tw})^2}\right]^2}\left\{\left[\frac{n^4}{4(\lambda^{(0)}_{tw})^2} - (gn+k_z)^2\right]\left[1 - 2\cos\left(\frac{L(gn+k_z)}{2}\right)\exp\left(-\frac{n^2 L}{4\lambda^{(0)}_{tw}}\right)\right.\right.$$

$$\left.+ \cos\left(L(gn+k_z)\right)\exp\left(-\frac{n^2 L}{2\lambda^{(0)}_{tw}}\right)\right] - \frac{n^2(gn+k_z)}{\lambda^{(0)}_{tw}}\left[2\sin\left(\frac{L(gn+k_z)}{2}\right)\exp\left(-\frac{n^2 L}{4\lambda^{(0)}_{tw}}\right)\right.$$

$$\left.\left.- \sin\left(L(gn+k_z)\right)\exp\left(-\frac{n^2 L}{2\lambda^{(0)}_{tw}}\right)\right]\right\}.$$

(E.10)

We consider when $\lambda^{(0)}_{tw} g \gg 1$, and firstly we look when $L/\lambda^{(0)}_{tw} \ll 1$. Then, Eqs. (E.6),(E.9) and (E.10) reduce to

$$\Theta_{n,n}(L,k_z) \approx L^2 \delta_{k_z,-ng} = 2\pi L \delta(k_z + ng) \tag{E.11}$$

We also look when $L/\lambda_{tw}^{(0)} \gg 1$. In this case, we obtain the Lorentzian

$$\Theta_{n,n}(L,k_z) \approx 2L\left[\frac{\frac{n^2}{2\lambda_{tw}^{(0)}}}{\frac{n^4}{4(\lambda_{tw}^{(0)})^2}+(k_z+ng)^2}\right] \approx 2\pi L\delta(k_z+ng). \tag{E.12}$$

This analysis suggests that for $\lambda_{tw}^{(0)}g \gg 1$ we may approximate the interaction energy by choosing $\langle\exp(i(n\delta\phi(s)-n'\delta\phi(s')))\rangle_\Omega \approx 1$. This limit is satisfied, as the value of $\lambda_{tw}^{(0)}$ is estimated to be $\lambda_{tw}^{(0)} \approx 500\text{Å}$. Thus, we may rewrite Eq. (E.2) as

$$\langle V_{global}(R)\rangle_\Omega \approx V_{global}(R) \approx -\frac{\Lambda_g}{2\pi}\int_{-L/2}^{L/2}ds\int_{-L/2}^{L/2}ds'\sum_{n=-\infty}^{\infty}\sum_{n'=-\infty}^{\infty}\exp(ig(sn-s'n'))\exp(i(\phi_1 n-\phi_2 n'))$$
$$\int_{-\infty}^{\infty}dk_z(-1)^{n'}\overline{G}_{n,n'}(R/a,k_z)\exp(ik_z(s-s')), \tag{E.13}$$

which, on approximating the integrations in the limit when $Lg \gg 1$, reduces to

$$\langle V_{global}(R)\rangle_\Omega \approx -\Lambda_g L\sum_{n=-\infty}^{\infty}\exp(in(\phi_1-\phi_2))(-1)^n\overline{G}_{n,n}(R/a,-ang), \tag{E.14}$$

which is indeed Eq. (2.48) of the main text.

## Appendix F Thermal fluctuations for weak non-base pair specific pairing interactions

Now, let us consider thermal fluctuations so that we use Eq. (2.18) of the main text to describe the position of interaction sites on the two helices. Then, we must write

$$V_{global}(R) \approx -\frac{\Lambda_g}{2\pi}\int_{-L/2}^{L/2}ds\int_{-L/2}^{L/2}ds'\sum_{n=-\infty}^{\infty}\sum_{n'=-\infty}^{\infty}\exp(ig(sn-s'n'))\exp(i((\phi_1+\delta\phi_{1,T}(s))n-(\phi_2+\delta\phi_{2,T}(s'))n'))$$
$$\int_{-\infty}^{\infty}dk_z(-1)^{n'}\overline{G}_{n,n'}(R/a,k_z)\exp(ik_z(s+\delta s_{1,T}(s)-s'-\delta s_{2,T}(s'))).$$

$$\tag{F.1}$$

As we are after a simplified energy functional, valid for the case that $\lambda_{tw}^{(0)}g \gg 1$, we can make derivative expansions of both $\delta\phi_{2,T}(s')$ and $\delta s_{2,T}(s')$. To do this, we write Taylor expansions

$$\delta\phi_{2,T}(s') \approx \delta\phi_{2,T}(s)+\frac{d\delta\phi_{2,T}(s)}{ds}(s'-s)+..., \tag{F.2}$$

$$\delta s_{2,T}(s') \approx \delta s_{2,T}(s) + \frac{d\delta s_{2,T}(s)}{ds}(s'-s) + \dots \tag{F.3}$$

If we retain only the leading order terms in the derivative expansion, we obtain

$$V_{global}(R) \approx -\frac{\Lambda_g}{2\pi} \sum_{n=-\infty}^{\infty} \sum_{n'=-\infty}^{\infty} \int_{-L/2}^{L/2} ds \exp\left(i\left((\phi_1 + \delta\phi_{1,T}(s))n - (\phi_2 + \delta\phi_{2,T}(s))n'\right)\right) \exp(igs(n-n'))$$

$$\int_{-L/2}^{L/2} ds' \exp(ign'(s-s')) \int_{-\infty}^{\infty} dk_z (-1)^{n'} \bar{G}_{n,n'}(R/a, k_z) \exp(ik_z(s + \delta s_{1,T}(s) - s' - \delta s_{2,T}(s))). \tag{F.4}$$

We will suppose that $Lg \gg 1$, so that we may approximate

$$V_{global}(R) \approx -\frac{\Lambda_g}{2\pi} \sum_{n=-\infty}^{\infty} \sum_{n'=-\infty}^{\infty} \int_{-L/2}^{L/2} ds \exp\left(i\left((\phi_1 + \delta\phi_{1,T}(s))n - (\phi_2 + \delta\phi_{2,T}(s))n'\right)\right) \exp(igs(n-n'))$$

$$\int_{-\infty}^{\infty} dx \exp(ign'x) \int_{-\infty}^{\infty} dk_z (-1)^{n'} \bar{G}_{n,n'}(R/a, k_z) \exp(ik_z(x + \delta s_{1,T}(s) - \delta s_{2,T}(s)))$$

$$\approx -\Lambda_g \sum_{n=-\infty}^{\infty} \sum_{n'=-\infty}^{\infty} \int_{-L/2}^{L/2} ds \exp\left(i\left((\phi_1 + \delta\phi_{1,T}(s))n - (\phi_2 + \delta\phi_{2,T}(s))n'\right)\right) \exp(igs(n-n'))$$

$$(-1)^{n'} \bar{G}_{n,n'}(R/a, -n'g) \exp(-in'g(\delta s_{1,T}(s) - \delta s_{2,T}(s))). \tag{F.5}$$

If $l_{st}g, l_{tw}g \gg 1$, only the $n = n'$ diagonal modes dominate, thus we approximate Eq. (F.5) with

$$V_{global}(R) \approx -\Lambda_g \sum_{n=-\infty}^{\infty} \int_{-L/2}^{L/2} ds \exp\left(in\left((\phi_1 + \delta\Phi_{1,T}(s)) - (\phi_2 + \delta\Phi_{2,T}(s))\right)\right)(-1)^{n'} \bar{G}_{n,n'}(R/a, -ng),$$

$$\tag{F.6}$$

where the helical phases $\delta\Phi_{1,T}(s)$ and $\delta\Phi_{2,T}(s)$ are defined through the relationship $\delta\Phi_{\mu,T}(s) = \delta\phi_{\mu,T}(s) - g\delta s_{\mu,T}$. Then, let us consider the partition function for this model. For it, we can first write

$$Z_g = \int D\Delta\Phi_T(s) \exp\left(-\frac{V_{global}[\Delta\Phi_T(s)]}{k_B T}\right) \exp\left(-\frac{l_h}{4} \int_{-L/2}^{L/2} \left(\frac{d\Delta\Phi_T(s)}{ds}\right)^2\right), \tag{F.7}$$

where $\Delta\Phi_T(s) = \delta\Phi_{1,T}(s) - \delta\Phi_{2,T}(s)$ and we have the combined persistence length

$$l_h = \frac{l_{tw} l_{st}}{l_{st} + l_{tw}}. \tag{F.8}$$

The second term is the contribution from the elastic energies $E_{tw}$ and $E_{st}$ given by Eqs. (2.20) and (2.21) of the main text. It is arrived at by integrating out the degree of freedom, $\bar{\Phi}_T(s) = \delta\Phi_{1,T}(s) + \delta\Phi_{2,T}(s)$ independent of $\Delta\Phi(s)$ (to see how this is done see the supplemental material of Ref. [2]). We can perform a similar type of expansion for weak interactions as that which was considered in Appendix B. Thus, we write the partition function as the following sum

$$Z_g = \sum_{n=0}^{\infty} Z_g^{(n)} \exp\left(-\frac{V_g^{(0)}}{k_B T}\right), \tag{F.9}$$

with

$$Z_g^{(n)} = \frac{(-1)^n}{n!} \int D\Delta\Phi_T(s) \left(\frac{\Delta V_{global}[\Delta\Phi_T(s)]}{k_B T}\right)^n \exp\left(-\frac{l_h}{4}\int_{-L/2}^{L/2}\left(\frac{d\Delta\Phi_T(s)}{ds}\right)^2\right), \tag{F.10}$$

where we have written

$$V_{global}[\Delta\Phi_T(s)] = V_g^{(0)} + \Delta V_{global}[\Delta\Phi_T(s)], \tag{F.11}$$

with

$$V_g^{(0)} \approx -\Lambda_g L \bar{G}_{0,0}(R/a,0), \tag{F.12}$$

$$\Delta V_{global}[\Delta\Phi_T(s)] \approx -\Lambda_g \sum_{n=-\infty}^{\infty} \int_{-L/2}^{L/2} ds (1-\delta_{n,0}) \exp\left(in(\phi_1 - \phi_2 + \Delta\Phi_T(s))\right)(-1)^n \bar{G}_{n,n}(R/a,-ng).$$

(F.13)

Retaining up to order $n = 2$, in the expansion (Eq. (F.9)), we can approximate the free energy for global interaction as

$$F_g \approx -k_B T \ln\left(Z_g^{(0)} + Z_g^{(1)} + Z_g^{(2)}\right) \approx -k_B T \ln Z_g^{(0)} + \left\langle V_{global}[\Delta\Phi_T(s)]\right\rangle_{g,0}$$
$$-\frac{1}{2k_B T}\left[\left\langle \Delta V_{global}[\Delta\Phi_T(s)]^2\right\rangle_{g,0} - \left\langle \Delta V_{global}[\Delta\Phi_T(s)]\right\rangle_{g,0}^2\right], \tag{F.14}$$

where

$$\left\langle V_{global}[\Delta\Phi_T(s)]\right\rangle_{g,0} = -k_B T Z_g^{(1)}/Z_g^{(0)} + V_g, \qquad \left\langle \Delta V_{global}[\Delta\Phi_T(s)]^2\right\rangle_{g,0} = (k_B T)^2 Z_g^{(2)}/Z_g^{(0)}.$$

(F.15)

Again, $-k_B T \ln Z_g^{(0)}$ is a unimportant constant term that can be discarded by subtracting off the free energy at $R = \infty$. Let us consider the first of the two averages given by Eq. (F.15). Writing it out explicitly, it reads as

$$\left\langle V_{global}[\Delta\Phi_T(s)]\right\rangle_{g,0} \approx -\Lambda_g \sum_{n=-\infty}^{\infty} \int_{-L/2}^{L/2} ds \exp\left(in(\phi_1 - \phi_2)\right)\left\langle \exp(in\Delta\Phi_T(s))\right\rangle_{g,0} (-1)^n \bar{G}_{n,n}(R/a,-ng),$$

(F.16)

where the average $\langle \exp(in\Delta\Phi_T(s))\rangle_{g,0}$ is given by

$$\langle \exp(in\Delta\Phi_T(s))\rangle_{g,0} = \frac{1}{Z_g^{(0)}} \int D\Delta\Phi_T(s) \exp(in\Delta\Phi_T(s)) \exp\left(-\frac{l_h}{4} \int_{-L/2}^{L/2} \left(\frac{d\Delta\Phi_T(s)}{ds}\right)^2\right)$$
$$= \frac{1}{Z_\Phi} \int D\Delta\dot{\Phi}_T(s) \exp\left(in\int_0^s ds' \Delta\dot{\Phi}_T(s')\right) \exp\left(-\frac{l_h}{4} \int_{-L/2}^{L/2} \Delta\dot{\Phi}_T(s')^2\right),$$

(F.17)

and

$$Z_\Phi = \int D\Delta\dot{\Phi}_T(s) \exp\left(-\frac{l_h}{4} \int_{-L/2}^{L/2} \Delta\dot{\Phi}_T(s')^2\right).$$

(F.18)

The average defined by Eq. (F.17) evaluates to

$$\langle \exp(in\Delta\Phi_T(s))\rangle_{g,0} = \exp\left(-\frac{n^2|s|}{l_h}\right).$$

(F.19)

On substitution of Eq. (F.19) into Eq. (F.16) and performing the $s$ integral, we obtain

$$\langle V_{global}[\Delta\Phi_T(s)]\rangle_{g,0} \approx -2\Lambda_g l_h \sum_{n=-\infty}^{\infty} (-1)^n \exp(in(\phi_1-\phi_2)) \frac{1}{n^2}\left(1-\exp\left(-\frac{Ln^2}{2l_h}\right)\right) \overline{G}_{n,n}(R/a,-ng).$$

(F.20)

This is namely Eq. (2.49) of the main text. For large $L$, the dominant mode is $n=0$, one for which we have simply

$$\langle V_{global}[\Delta\Phi_T(s)]\rangle_{g,0} \approx -\Lambda_g L \overline{G}_{0,0}(R/a,0).$$

(F.21)

Let us now consider the second average in Eq. (F.15). This can be written as

$$\langle V_{global}[\Delta\Phi_T(s)]^2\rangle_{g,0} \approx \frac{\Lambda_g^2}{k_B T} \sum_{m=-\infty}^{\infty} \sum_{n=-\infty}^{\infty} \int_{-L/2}^{L/2} ds \exp(in(\phi_1-\phi_2))\exp(im(\phi_1-\phi_2))$$
$$\langle \exp(in\Delta\Phi_T(s)+im\Delta\Phi_T(s'))\rangle_{g,0} (1-\delta_{n,0})(1-\delta_{m,0})(-1)^{n+m} \overline{G}_{n,n}(R/a,-ng)\overline{G}_{m,m}(R/a,-mg).$$

(F.22)

For the thermal average in Eq. (F.22), we have the following expression

$$\langle \exp(in\Delta\Phi_T(s))\exp(im\Delta\Phi_T(s'))\rangle_{g,0}$$

$$= \frac{1}{Z_\Phi}\int D\Delta\dot{\Phi}_T(s)\left[\exp\left(in\int_0^s \Delta\dot{\Phi}_T(s'')ds''\right)\exp\left(im\int_0^{s'}\Delta\dot{\Phi}_T(s'')ds''\right)\theta(s)\theta(s')\right.$$

$$+\exp\left(-in\int_s^0 \Delta\dot{\Phi}_T(s'')ds''\right)\exp\left(im\int_0^{s'}\Delta\dot{\Phi}_T(s'')ds''\right)\theta(-s)\theta(s')$$

$$+\exp\left(in\int_0^s \Delta\dot{\Phi}_T(s'')ds''\right)\exp\left(-im\int_{s'}^0\Delta\dot{\Phi}_T(s'')ds''\right)\theta(s)\theta(-s')$$

$$\left.+\exp\left(-in\int_s^0 \Delta\dot{\Phi}_T(s'')ds''\right)\exp\left(-im\int_{s'}^0\Delta\dot{\Phi}_T(s'')ds''\right)\theta(-s)\theta(-s')\right]\exp\left(-\frac{l_{tw}}{4}\int_{-L/2}^{L/2}\Delta\dot{\Phi}_T(s'')^2 ds''\right).$$

(F.23)

This is evaluated in the same way as Eq. (B.26) of Appendix B. This leads to the expression

$$\langle \exp(in\Delta\Phi_T(s))\exp(im\Delta\Phi_T(s'))\rangle_{g,0} = \exp\left(-\frac{1}{l_h}(n^2 s + 2nm\min(s,s') + m^2 s')\right)\theta(s)\theta(s')$$

$$+\exp\left(\frac{1}{l_h}(n^2 s + 2nm\max(s,s') + m^2 s')\right)\theta(-s)\theta(-s') + \exp\left(-\frac{1}{l_h}(n^2 s - m^2 s')\right)\theta(s)\theta(-s')$$

$$+\exp\left(-\frac{1}{l_h}(m^2 s' - n^2 s)\right)\theta(-s)\theta(s').$$

(F.24)

Thus, in similar fashion to the analysis in Appendix B, we obtain from Eq. (F.23) the expression

$$\langle V_{global}[\Delta\Phi_T(s)]^2\rangle_{g,0} = \frac{\Lambda_g^2}{k_B T}\sum_{n=-\infty}^{\infty}\sum_{m=-\infty}^{\infty}\exp(im(\phi_1-\phi_2))\exp(in(\phi_1-\phi_2))$$

$$(1-\delta_{n,0})(1-\delta_{m,0})(-1)^{n+m}\overline{G}_{n,n}(R/a,-ng)\overline{G}_{m,m}(R/a,-mg)\Xi(n^2/l_h, nm/l_h, m^2/l_h),$$

(F.25)

Here $\Xi = \Xi_1 + \Xi_2$ is the same function defined by Eqs.(B.38)-(B.40). An explicit form can be obtained for $\Xi(n^2/l_h, nm/l_h, m^2/l_h)$ by replacing, in Eqs. (B.38)- (B.40), $l_{tw}$ with $l_h$ and setting $k_z = k'_z = 0$

For $L \gg l_h$ the dominant behaviour comes from $\Xi_1$ for the modes where $n = -m$. In this limit

$$\Xi_1\left(n^2/l_h, nm/l_h, m^2/l_h\right) \approx \frac{2l_h L}{n^2}\delta_{n,-m},$$

(F.26)

so that for $L \gg l_h$

$$F - F_0 \approx -\Lambda_g L \bar{G}_{0,0}(R/a,0) - \frac{1}{2k_b T}\left\langle V_{local}[\Delta\Phi_T(s)]^2 \right\rangle_0 \approx$$
$$-\Lambda_g L \bar{G}_{0,0}(R/a,0) - \frac{2l_h L \Lambda_g^2}{k_B T}\sum_{n=1}^{\infty}\frac{1}{n^2}\bar{G}_{n,n}(R/a,-ng)\bar{G}_{n,n}(R/a,ng).$$
(F.27)

where we have used the fact that $\bar{G}_{n,n}(R/a,ng) = \bar{G}_{-n,-n}(R/a,-ng)$.

As the interactions between base pairs do not discriminate between base pairs, we can also need to consider a functional for the interaction energy between two helices with non-alike sequences. This is given by the expression

$$V_{global}^{NL}(R) \approx -\frac{\Lambda_g}{2\pi}\int_{-L/2}^{L/2}ds\int_{-L/2}^{L/2}ds'\sum_{n=-\infty}^{\infty}\sum_{n'=-\infty}^{\infty}\exp(ig(sn-s'n'))$$
$$\exp\left(i\left(n(\phi_1 + \delta\phi_1(s) + \delta\phi_{1,T}(s)) - n'(\phi_2 + \delta\phi_2(s') + \delta\phi_{2,T}(s'))\right)\right)$$
$$\int_{-\infty}^{\infty}dk_z(-1)^{n'}\bar{G}_{n,n'}(R/a,k_z)\exp\left(ik_z(s + \delta s_1(s) + \delta s_{1,T}(s) - s' - \delta s_2(s') - \delta s_{2,T}(s'))\right).$$
(F.28)

As $\delta\phi(s)$ for identical helices, both $\delta\phi_1(s)$ and $\delta\phi_2(s)$ are accumulation in the base pair dependent patterns of distortions in the twist, now different from each other. Similarly, we have $\delta s_1(s)$ and $\delta s_2(s)$ that are the accumulations in the different patterns of base pair dependent axial distortions, just as we had $\delta s(s)$ for identical sequences. We have that

$$\delta\phi_\mu(s) = \frac{1}{h}\int_0^s ds'\delta\Omega_\mu(s') \qquad \text{and} \qquad \delta s_\mu(s) = \frac{1}{h}\int_0^s ds'\delta h_\mu(s')$$
(F.29)

where $\delta\Omega_\mu(s')$ and $\delta h_\mu(s')$ are the variations in twist and rise from their average values. Going through same steps as Eq. (F.1)-(F.6), it is possible to approximate the energy functional given by Eq. (F.28) as

$$V_{global}^{NL}(R) \approx -\Lambda_g\sum_{n=-\infty}^{\infty}\int_{-L/2}^{L/2}ds\exp\left(in\left((\phi_1 + \delta\Phi_1(s) + \delta\Phi_{1,T}(s)) - (\phi_2 + \delta\Phi_2(s) + \delta\Phi_{2,T}(s))\right)\right)$$
$$(-1)^n\bar{G}_{n,n'}(R/a,-ng),$$
(F.30)

where $\delta\Phi_\mu(s) = \delta\phi_\mu(s) - g\delta s_\mu$. We can also write

$$\delta\Phi_\mu(s) = \int_0^s ds'\delta g_\mu(s).$$
(F.31)

Over large length scales it is possible to show [3]

$$\left\langle \delta g_\mu(s)\delta g_\nu(s') \right\rangle_\Omega = \frac{\delta_{\mu,\nu}}{\lambda_c^{(0)}}\delta(s-s'),$$
(F.32)

and the pattern of helix distortions, $\delta g_\mu(s)$ may be supposed to be Gaussian distributed. Here $\lambda_c^{(0)}$ is the helical coherence length, the combined coherence length of both rise and twist distortions. Essentially, $\lambda_c^{(0)}$ is the distance that two of the DNA helices in their ground state stay roughly in register with each other.

In the case of non-alike molecules, we may write for the partition function

$$Z_g^{NL} = \int D\Delta\Phi_T(s) \exp\left(-\frac{V_{global}^{NL}[\Delta\Phi(s)+\Delta\Phi_T(s)]}{k_B T}\right) \exp\left(-\frac{l_h}{4}\int_{-L/2}^{L/2}\left(\frac{d\Delta\Phi_T(s)}{ds}\right)^2\right). \qquad (F.33)$$

where $\Delta\Phi(s) = \delta\Phi_1(s) - \delta\Phi_2(s)$. Again, to consider weak pairing interactions, we can write

$$\Delta V_{global}^{NL}[\Delta\Phi_T(s)] = V_{global}^{NL}[\Delta\Phi_T(s)] - V_g^{(0)}, \qquad (F.34)$$

with

$$\Delta V_{global}^{NL}[\Delta\Phi(s)+\Delta\Phi_T(s)] \approx -\Lambda_g \sum_{n=-\infty}^{\infty} \int_{-L/2}^{L/2} ds(1-\delta_{n,0})(-1)^n$$
$$\bar{G}_{n,n}(R/a,-ng)\exp\left(in\left(\phi_1-\phi_2+\Delta\Phi(s)+\Delta\Phi_T(s)\right)\right). \qquad (F.35)$$

We can expand out the partition and free energy in a similar way to that for the interaction between like helices with the same pattern of distortions (c.f. Eqs. (F.9), (F.10) and (F.14)). Therefore, we can approximate the ensemble averaged free energy as

$$\left\langle F_g^{NL} \right\rangle_\Omega \approx -k_B T \ln Z_g^{(0)} + \left\langle\left\langle V_{global}^{NL}[\Delta\Phi(s)+\Delta\Phi_T(s)]\right\rangle_{g,0}\right\rangle_\Omega$$
$$-\frac{1}{2k_B T}\left\langle\left[\left\langle[\Delta V_{global}^{NL}[\Delta\Phi(s)+\Delta\Phi_T(s)]]^2\right\rangle_{g,0} - \left\langle\Delta V_{global}^{NL}[\Delta\Phi(s)+\Delta\Phi_T(s)]\right\rangle_{g,0}^2\right]\right\rangle_\Omega. \qquad (F.36)$$

Let us consider the first average in Eq. (F.36). This reads as

$$\left\langle\left\langle V_{global}^{NL}[\Delta\Phi_T(s)]\right\rangle_{g,0}\right\rangle_\Omega \approx -\Lambda_g \sum_{n=-\infty}^{\infty}\int_{-L/2}^{L/2} ds \exp\left(in(\phi_1-\phi_2)\right)\left\langle\exp(in\Delta\Phi(s))\right\rangle_\Omega$$
$$\left\langle\exp(in\Delta\Phi_T(s))\right\rangle_{g,0}(-1)^n \bar{G}_{n,n}(R/a,-ng). \qquad (F.37)$$

The ensemble average contained in Eq. (F.37) is written as

$$\left\langle\exp(in\Delta\Phi(s))\right\rangle_\Omega = \frac{\int D\Delta g(s)\exp\left(in\int_0^s ds'\Delta g(s')\right)\exp\left(-\frac{1}{4\lambda_c^{(0)}}\int_{-L/2}^{L/2} ds\Delta g(s)^2\right)}{\int D\Delta g(s)\exp\left(-\frac{1}{4\lambda_c^{(0)}}\int_{-L/2}^{L/2} ds\Delta g(s)^2\right)}, \qquad (F.38)$$

where $\Delta g(s) = \delta g_1(s) - \delta g_2(s)$. Eq. (F.38) evaluates to

$$\langle \exp(in\Delta\Phi(s)) \rangle_\Omega = \exp\left(-\frac{n^2|s|}{\lambda_c^{(0)}}\right). \tag{F.39}$$

Using both Eqs. (F.19), (F.39) and integrating over $s$ we may write

$$\langle V_{global}^{NH}[\Delta\Phi_T(s)] \rangle_{g,0} \approx -2\Lambda_g \lambda_c \sum_{n=-\infty}^{\infty} (-1)^n \exp(in(\phi_1-\phi_2))\frac{1}{n^2}\left(1-\exp\left(-\frac{Ln^2}{2\lambda_c}\right)\right)\bar{G}_{n,n}(R/a,-ng), \tag{F.40}$$

where $\lambda_c = \lambda_c^{(0)} l_h / (l_h + \lambda_c^{(0)})$. This is Eq. (2.52) of the main text. For large $L$, this again becomes $-\Lambda_g L \bar{G}_{0,0}(R/a,0)$, as Eq. (F.21).

Let's now consider the next term in Eq. (F.36). We have that

$$\langle V_{global}^{NL}[\Delta\Phi_T(s)]^2 \rangle_{g,0} \approx \frac{\Lambda_g^2}{k_B T}\sum_{m=-\infty}^{\infty}\sum_{n=-\infty}^{\infty}\int_{-L/2}^{L/2}ds\int_{-L/2}^{L/2}ds'\exp(in(\phi_1-\phi_2))\exp(im(\phi_1-\phi_2))\langle\exp(in\Delta\Phi_T(s)+im\Delta\Phi_T(s'))\rangle_{g,0}$$
$$\langle\exp(in\Delta\Phi(s)+im\Delta\Phi(s'))\rangle_\Omega (1-\delta_{n,0})(1-\delta_{m,0})(-1)^{n+m}\bar{G}_{n,n}(R/a,-ng)\bar{G}_{m,m}(R/a,-mg). \tag{F.41}$$

We find that the average $\langle \exp(in\Delta\Phi(s))\exp(im\Delta\Phi(s'))\rangle_\Omega$ evaluates to Eq. (F.24) but now with $l_h$ replaced by $\lambda_c^{(0)}$. Thus, we can write the average as

$$\left\langle \langle V_{global}^{NL}[\Delta\Phi_T(s)]^2 \rangle_{g,0}\right\rangle_\Omega = \frac{\Lambda_g^2}{k_B T}\sum_{n=-\infty}^{\infty}\sum_{m=-\infty}^{\infty}\exp(im(\phi_1-\phi_2))\exp(in(\phi_1-\phi_2))$$
$$(1-\delta_{n,0})(1-\delta_{m,0})(-1)^{n+m}\bar{G}_{n,n}(R/a,-ng)\bar{G}_{m,m}(R/a,-mg)\Xi\left(n^2/\lambda_c, nm/\lambda_c, m^2/\lambda_c\right), \tag{F.42}$$

Last of all, in Eq. (F.36), we need to consider the average

$$\left\langle \langle \Delta V_{global}^{NL}[\Delta\Phi(s)+\Delta\Phi_T(s)]\rangle^2_{g,0}\right\rangle_\Omega \approx \frac{\Lambda_g^2}{k_B T}\sum_{m=-\infty}^{\infty}\sum_{n=-\infty}^{\infty}\int_{-L/2}^{L/2}ds\int_{-L/2}^{L/2}ds'\exp(in(\phi_1-\phi_2))\exp(im(\phi_1-\phi_2))$$
$$\exp\left(-\frac{n^2|s|+m^2|s'|}{l_h}\right)\langle\exp(in\Delta\Phi(s)+im\Delta\Phi(s'))\rangle_\Omega (1-\delta_{n,0})(1-\delta_{m,0})(-1)^{n+m}$$
$$\bar{G}_{n,n}(R/a,-ng)\bar{G}_{m,m}(R/a,-mg) \tag{F.43}$$

This evaluates to

$$\left\langle \langle V_{global}[\Delta\Phi_T(s)]^2 \rangle_{g,0}\right\rangle_\Omega = \frac{\Lambda_g^2}{k_B T}\sum_{n=-\infty}^{\infty}\sum_{m=-\infty}^{\infty}\exp(im(\phi_1-\phi_2))\exp(in(\phi_1-\phi_2))$$
$$(1-\delta_{n,0})(1-\delta_{m,0})(-1)^{n+m}\bar{G}_{n,n}(R/a,-ng)\bar{G}_{m,m}(R/a,-mg)\Xi\left(n^2/\lambda_c, nm/\lambda_c^{(0)}, m^2/\lambda_c\right), \tag{F.44}$$

where

$$\Xi\left(n^2/\lambda_c, nm/\lambda_c^{(0)}, m^2/\lambda_c\right) = \Xi_1\left(n^2/\lambda_c, nm/\lambda_c^{(0)}, m^2/\lambda_c\right) + \Xi_2\left(n^2/\lambda_c, nm/\lambda_c^{(0)}, m^2/\lambda_c\right)$$

(F.45)

Here for the purposes of orientation we give explicit form of the functions contained in Eq.(F.45). These read as

$$\Xi_1\left(n^2/\lambda_c, nm/\lambda_c^{(0)}, m^2/l_h\right) = \left[\left(m^2 + \frac{2\lambda_c nm}{\lambda_c^{(0)}}\right)^{-1} + \left(n^2 + \frac{2\lambda_c nm}{\lambda_c^{(0)}}\right)^{-1}\right] \frac{2\lambda_c^2}{\left(n^2 + 2nm\lambda_c/\lambda_c^{(0)} + m^2\right)}$$

$$\left[\exp\left(-\frac{L}{2}\left(\frac{n^2+m^2}{\lambda_c} + \frac{2nm}{\lambda_c^{(0)}}\right)\right) - 1\right] + \frac{2\lambda_c^2}{\left(2nm\lambda_c/\lambda_c^{(0)} + m^2\right)n^2}\left[1 - \exp\left(-\frac{Ln^2}{2\lambda_c}\right)\right]$$

$$+ \frac{2\lambda_c^2}{\left(2nm\lambda_c/\lambda_c^{(0)} + n^2\right)m^2}\left[1 - \exp\left(-\frac{Lm^2}{2\lambda_c}\right)\right],$$

(F.46)

$$\Xi_2\left(n^2/\lambda_c, nm/\lambda_c^{(0)}, m^2/\lambda_c\right) = \frac{\lambda_c^2}{n^2 m^2}\left[1 - \exp\left(-\frac{Ln^2}{2\lambda_c}\right)\right]\left[1 - \exp\left(-\frac{Lm^2}{2\lambda_c}\right)\right].$$

(F.47)

For $L \gg \lambda_c, \lambda_c^{(0)}$, Eq. (F.44) tends to a constant value with respect to $L$ as $\lambda_c < \lambda_c^{(0)}$. Therefore, for large $L$, we can neglect Eqs. (F.40) and (F.44). The upshot is that at $L \gg \lambda_c, \lambda_c^{(0)}$ the free energy is approximated by

$$F_g^{NL} - F_0 \approx -L\Lambda_g G_{0,0}(R/a, 0) - \frac{1}{2k_B T}\left\langle\left\langle V_{global}^{NL}[\Delta\Phi(s) + \Delta\Phi_T(s)]^2\right\rangle_{g,0}\right\rangle_\Omega$$

$$\approx -L\Lambda_g G_{0,0}(R/a, 0) - \frac{2\lambda_c L\Lambda_g^2}{k_B T}\sum_{n=1}^{\infty}\frac{1}{n^2}\bar{G}_{n,n}(R/a, -ng)\bar{G}_{n,n}(R/a, ng).$$

(F.48)

using the fact that for the large $L$ behaviour of Eq. (F.42), we can simply replace $l_h$ with $\lambda_c$ in Eq. (F.26).

In Fig. F.1, we plot the azimuthal ($\Delta\phi = \phi_1 - \phi_2$) dependence of the free energy for interactions between alike helices (described by Eqs.(F.14),(F.20) and (F.25)) and non-alike helices (described by Eqs. (F.36), (F.40),(F.42) and (F.44)). As expected, the azimuthal dependence diminishes faster with length for the non-alike helices than the ones that are alike, as the length that controls this is $\lambda_c$, and $\lambda_c < \lambda_c^{(0)}$. In Fig. F.2, we plot the length dependence of the free energies $F_g$ and $F_g^{NL}$ for various values of $\Lambda_c = \Lambda_g V_{global}(R; \Delta\phi = \pi)/V_{local}(R; \Delta\phi = \pi)$ as well as the recognition energy per unit length $E_R/L = (F_g - F_g^{NL})/L$. Importantly, in the absence of corrections, i.e. using only Eqs. (F.20) and (F.40) to describe the free energy, the recognition energy per unit length diminishes to zero as $L \to \infty$. However, when residual azimuthal correlations between helices are considered (i.e. all terms in Eqs. (F.14) and (F.36)), the recognition energy per unit length tends to a constant.

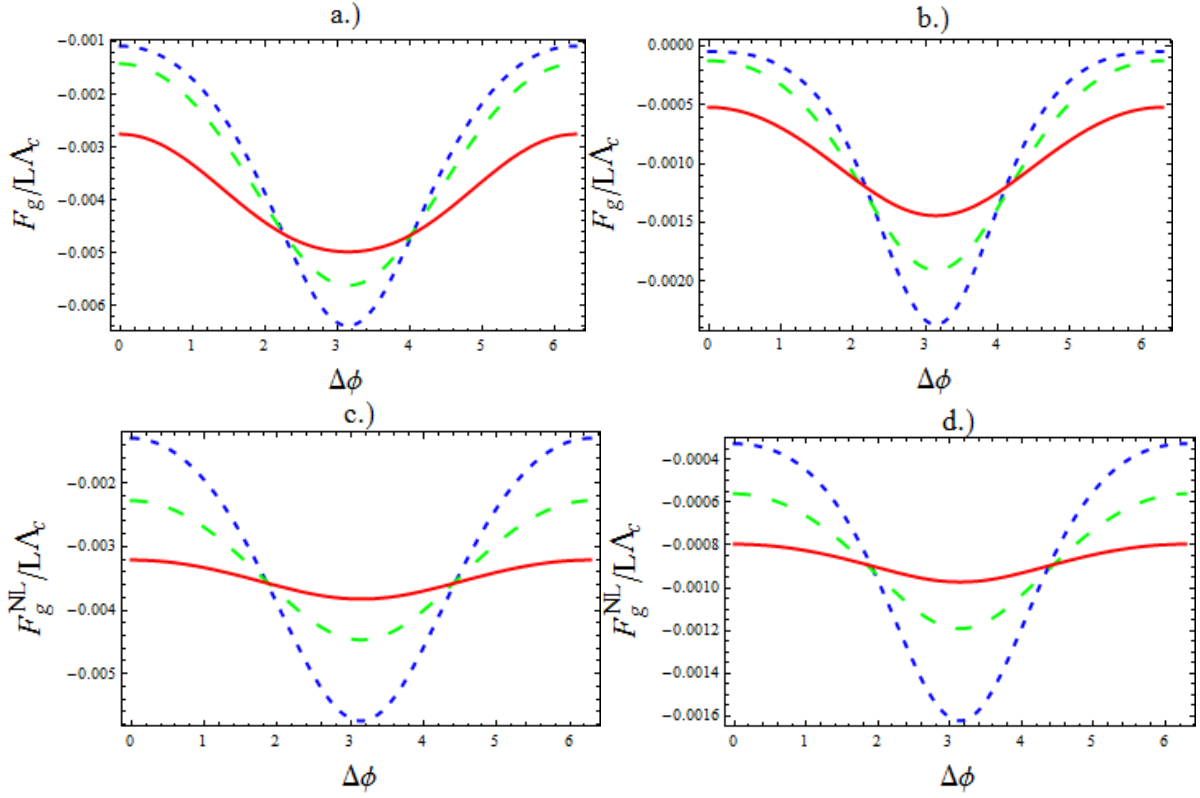

Fig. F.1 Plots showing the $\Delta\phi = \phi_1 - \phi_2$ dependence of the free energies $F_g$ and $F_g^{NL}$ for various helix lengths. Panels a.) and b.) are plots of $F_g$ for the values $\lambda_{eff} = 4\text{Å}$ and $\lambda_{eff} = 2\text{Å}$, respectively. Panels c.) and d.) are plots of $F_g^{NL}$, again for the values $\lambda_{eff} = 4\text{Å}$ and $\lambda_{eff} = 2\text{Å}$, respectively. The short dashed blue, long dashed green and solid red curves are for the values $L = 100\text{Å}, 500\text{Å}, 2000\text{Å}$, respectively. An interaction strength of $\Lambda_c = 0.02 k_B T / \text{Å}$ is used. Also, in the calculations, the values $R = 25\text{Å}$, $a = 11.2\text{Å}$, $l_{tw} = 1000\text{Å}$, $l_{st} = 700\text{Å}$ and $\lambda_c^{(0)} = 150\text{Å}$ are used.

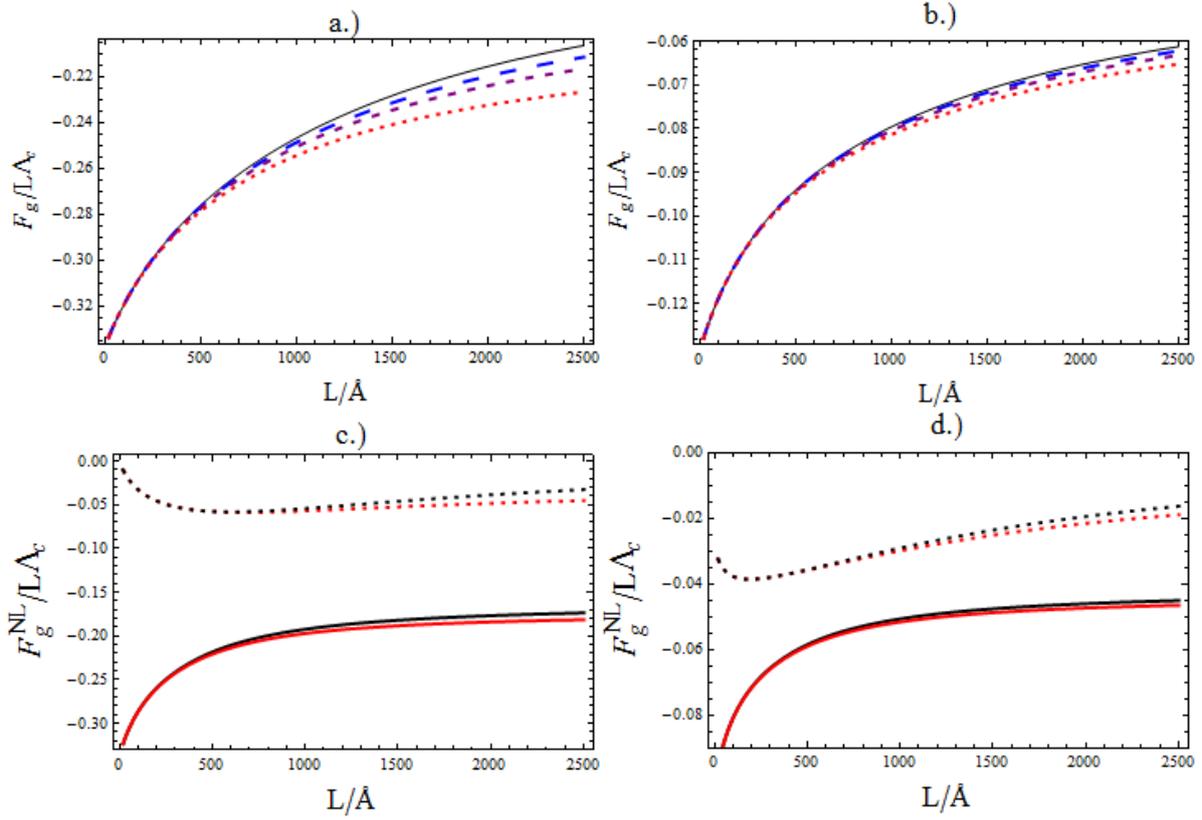

Fig. F.2 Plots showing the length dependence of the free energies $F_g$ and $F_g^{NL}$ for various helix lengths, as well as the recognition energy per unit length $E_R/L$, where $E_R = F_g - F_g^{NL}$ for global pairing interactions. In panels a.) and b.), $F_g/L\Lambda_c$ is plotted as a function of length $L$ for the values $\lambda_{eff} = 4\text{Å}$ and $\lambda_{eff} = 2\text{Å}$, respectively. Here, the solid black curves correspond to only including Eq. (F.20) in the free energy. The other curves correspond to the full form of the free energy given by Eq. (F.14). The long dashed blue curves, short dashed purple curves, and dotted red curves correspond to the values $\Lambda_c = 0.0025, 0.005, 0.01 k_B T/\text{Å}$, respectively. In panels c.) and d.), $F_g^{NL}/L\Lambda_c$ is plotted as a function of length $L$ for the values $\lambda_{eff} = 4\text{Å}$ and $\lambda_{eff} = 2\text{Å}$, respectively. Here, the solid black curves correspond to only including Eq. (F.40) in the free energy. The solid red curves are for the full form of the ensemble averaged free energy, Eq. (F.36) evaluated at $\Lambda_c = 0.01 k_B T/\text{Å}$. Also in panels c.) and d.) we plot $E_R/L\Lambda_c$, for the values $\lambda_{eff} = 4\text{Å}$ and $\lambda_{eff} = 2\text{Å}$, respectively, depicted by dotted curves. The black dotted curve corresponds to the recognition energy just calculated using the free energies containing only Eqs. (F.20) and (F.40). The red dotted curve corresponds to the recognition energy calculated using the full expressions for the free energy, Eqs. (F.14) and (F.36), using a value of $\Lambda_c = 0.01 k_B T/\text{Å}$.

## Appendix G Thermal fluctuations for strong global pairing interactions

Let's now consider when the global pairing interactions are strong (further details of such analysis for the Kornyshev- Leikin theory can be found in Refs. [2,4]). In this case, we use a variational approximation. For the thermal fluctuations, we use the trial functional

$$E_{tr,\Phi}[\Delta\Phi_T(s)] = \frac{k_B T}{2} \int_{-L/2}^{L/2} ds \left[ \frac{l_h}{2}\left(\frac{d\Delta\Phi_T(s)}{ds}\right)^2 + \alpha_\Phi (\Delta\Phi_T(s))^2 \right], \tag{G.1}$$

where $\alpha_\Phi$ is the parameter that minimizes the variational free energy. Here, for the moment, we will consider the variational free energy for non-alike helices (the derivation for identical helices is simply got by setting $\Delta\Phi(s) = 0$ or $\lambda_c^{(0)} = \infty$). This reads as

$$F_{g,T}^{NL} = -k_B T \ln Z_{tr,\Phi} + k_B T \frac{l_h}{4} \int_{-L/2}^{L/2} ds \left[ \left(\frac{d\Delta\Phi(s)}{ds} - \Delta g(s)\right)^2 + \left\langle \left(\frac{d\Delta\Phi_T(s)}{ds}\right)^2 \right\rangle_{tr,\Phi} \right]$$
$$+ \left\langle V_{global}^{NL}[\Delta\Phi(s) + \Delta\Phi_T(s)] - E_{tr,\Phi}[\Delta\Phi_T(s)] \right\rangle_{tr,\Phi}. \tag{G.2}$$

Here, we have included the term $(d\Delta\Phi/ds - \Delta g)^2$ in the elastic energy (the second term in Eq. (G.2)) that allows for interactions to deform $d\Delta\Phi(s)/ds$ away from $\Delta g(s) = \delta g_1(s) - \delta g_2(s)$.

The thermal averages in Eq. (G.2) evaluate to

$$\left\langle \left(\frac{d\Delta\Phi_T(s)}{ds}\right)^2 \right\rangle_{tr,\Phi} - \left\langle E_{tr,\Phi}[\Delta\Phi_T(s)] \right\rangle_{tr,\Phi} = -\frac{L}{4\lambda_h}, \tag{G.3}$$

and

$$\left\langle V_{global}^{NL}[\Delta\Phi(s) + \Delta\Phi_T(s)] \right\rangle_{tr,\Phi} =$$
$$-\Lambda_g \sum_{n=-\infty}^{\infty} \int_{-L/2}^{L/2} ds (-1)^n \exp(in(\phi_1 - \phi_2)) \exp(in\Delta\Phi(s)) \exp\left(-\frac{n^2 \lambda_h}{2l_h}\right) \bar{G}_{n,n}(R/a, -ng), \tag{G.4}$$

where $\lambda_h = (\lambda_c / 2\alpha_\Phi)^{1/2}$. We also find that

$$-k_B T \ln Z_{tr,\Phi} = \frac{L}{2\lambda_h} \tag{G.5}$$

We next want to perform the ensemble average of the variational free energy to do this we use a variational trial function based on the linear response theory [5], this is

$$\Delta\Phi(s) = \frac{1}{2} \int_{-\infty}^{\infty} \frac{(s-s')}{|s-s'|} \Delta g(s') \exp\left(-\frac{|s-s'|}{\lambda_h}\right) ds'. \tag{G.6}$$

Where the optimal orientation $\langle \Delta\phi \rangle = \phi_1 - \phi_2$ is also considered as a variational parameter and have already assumed that the same adaptation length for thermal fluctuations, $\lambda_h$ can be used in Eq.(G.6), for a proof see Refs. [1,3]. Thus, using Eq. (G.6), we may approximate the ensemble average

$$\langle F_{g,T}^{NL}\rangle_\Omega = \frac{k_B TL}{4\lambda_h} + k_B T \frac{l_h}{4} \int_{-L/2}^{L/2} ds \left[ \left\langle \left( \frac{d\Delta\Phi(s)}{ds} - \Delta g(s) \right)^2 \right\rangle_\Omega \right]$$
$$+ \left\langle \langle V_{global}^{NL}[\Delta\Phi(s) + \Delta\Phi_T(s)]\rangle_{tr,\Phi} \right\rangle_\Omega.$$
(G.7)

Using Eq. (G.6), we find that the ensemble averages evaluate such that

$$\left\langle \left( \frac{d\Delta\Phi(s)}{ds} - \Delta g(s) \right)^2 \right\rangle_\Omega = \frac{1}{2\lambda_c^{(0)} \lambda_h},$$
(G.8)

$$\left\langle \langle V_{global}^{NL}[\Delta\Phi(s) + \Delta\Phi_T(s)]\rangle_{tr,\Phi} \right\rangle_\Omega =$$
$$-\Lambda_g \sum_{n=-\infty}^{\infty} \int_{-L/2}^{L/2} ds \exp\left(in(\phi_1 - \phi_2)\right) \exp\left(-\frac{n^2 \lambda_h}{2}\left(\frac{1}{l_h} + \frac{1}{2\lambda_c^{(0)}}\right)\right)(-1)^n \bar{G}_{n,n}(R/a, -ng).$$
(G.9)

We can then rewrite the ensemble averaged variational trial function as

$$\frac{\langle F_{g,T}^{NL}\rangle_\Omega}{k_B TL} = \frac{(l_c + \lambda_c)^2}{16\lambda_h^* \lambda_c l_h} - \Lambda_g' \sum_{n=-\infty}^{\infty} (-1)^n \exp\left(in(\phi_1 - \phi_2)\right) \exp\left(-\frac{n^2 \lambda_h^*}{2\lambda_c}\right) \bar{G}_{n,n}(R/a, -ng),$$
(G.10)

where $\Lambda_g' = \Lambda_g / k_B T$ and $\lambda_h^* = \lambda_h \lambda_c (1/l_c + 1/(2\lambda_c^{(0)}))$. For the Debye-Huckel and Morse potential forms for $\bar{G}_{n,n}(R/a, -ng)$ given by Eq. (2.14) and (2.15) of the text, the mean optimal azimuthal orientation is $(\phi_1 - \phi_2) = \pi$. Thus, we have

$$\frac{\langle F_{g,T}^{NL}\rangle_\Omega}{k_B TL} = \frac{(l_c + \lambda_c)^2}{16\lambda_h^* \lambda_c l_h} - \Lambda_g' \sum_{n=-\infty}^{\infty} \exp\left(-\frac{n^2 \lambda_h^*}{2\lambda_c}\right) \bar{G}_{n,n}(R/a, -ng).$$
(G.11)

For two identical helices, the result is simply obtained from Eq. (G.11) by setting $\lambda_c = l_h$, namely

$$\frac{F_{g,T}}{k_B TL} = \frac{1}{4\lambda_h^*} - \Lambda_g' \sum_{n=-\infty}^{\infty} \exp\left(-\frac{n^2 \lambda_h^*}{2l_h}\right) \bar{G}_{n,n}(R/a, -ng).$$
(G.12)

Minimization of Eqs. (G.11) and (G.12), with respect to $\lambda_h^*$, yields the following equations; for non-alike helices

$$\frac{(l_c + \lambda_c)^2}{8(\lambda_h^*)^2 l_h} = \Lambda_g' \sum_{n=-\infty}^{\infty} n^2 \exp\left(-\frac{n^2 \lambda_h^*}{2\lambda_c}\right) \bar{G}_{n,n}(R/a, -ng),$$
(G.13)

and for alike helices

$$\frac{l_h}{2(\lambda_h^*)^2} = \Lambda_g' \sum_{n=-\infty}^{\infty} n^2 \exp\left(-\frac{n^2 \lambda_h^*}{2l_h}\right) \bar{G}_{n,n}(R/a, -ng).$$
(G.14)

In Fig. G.1 we plot the free energies described by Eqs. (G.11) and (G.12) along with the free energy that describes the state where there is no-adaptation over large length scales for which the free energy is calculated in Appendix F. For very long molecules, the free energy for this state is simply given by (in both cases) $F_g = F_g^{NL} = -L\Lambda_g \bar{G}_{0,0}(R/a, k_z)$, when the correction due to azimuthal correlations is neglected. Here, we find a transition between the two states when the free energies of the two states are equal, this is used for generating the plots of $\lambda_h$ presented in the main text.

When the leading order correction due to correlations in $\Delta\Phi_T(s)$ is included, the free energy is given by either Eq. (F.27) (alike helices) or (F.48) (non-alike helices). Here, unfortunately we see no transition between the two states, as the correction is too large. However, it is expected that when the next to leading order correction is included that we will again see a transition between the two states.

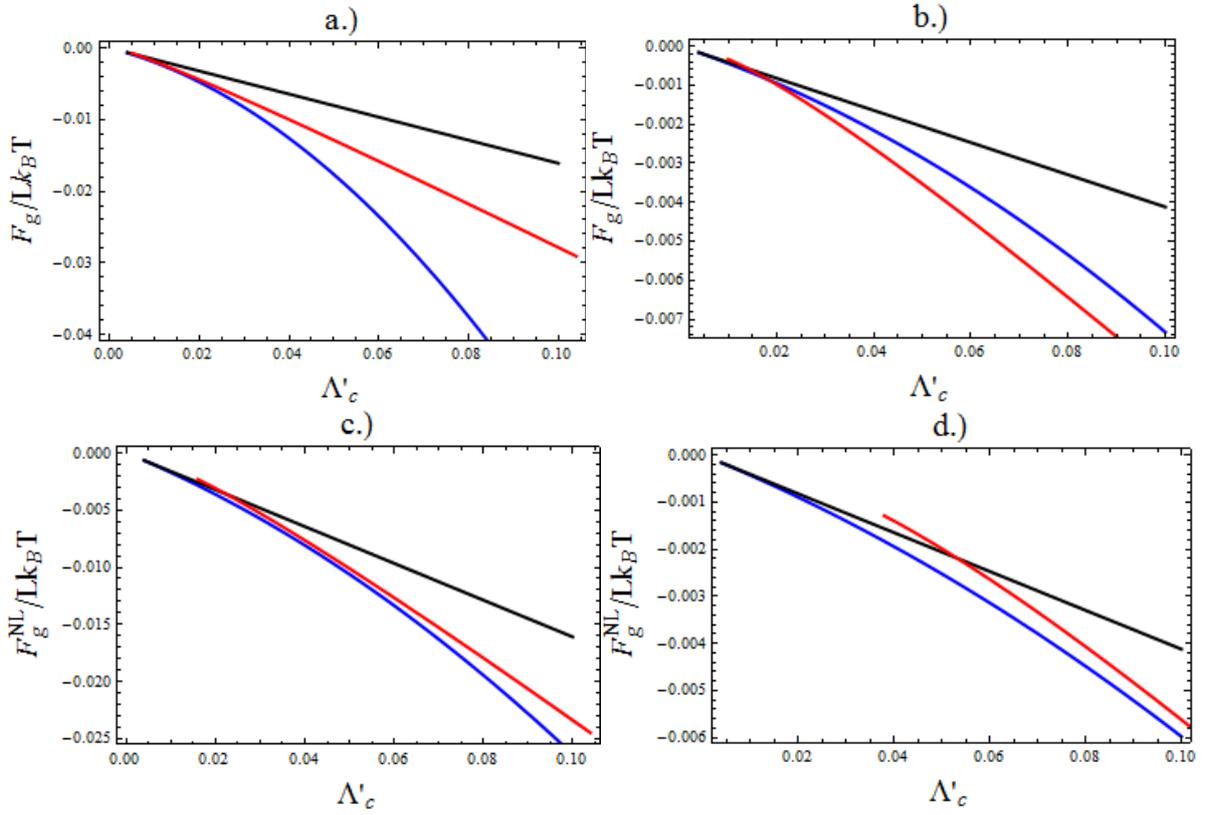

Fig. G.1. Plots of the Free energies per unit length for very long helices for the state where $\lambda_h$ is finite and the state where there is no adaptation ($\lambda_h = \infty$). Panels a.) and b.) are plots of $F_g/L$, the free energy for alike helices, for the decay lengths $\lambda_{eff} = 4\text{Å}$ and $\lambda_{eff} = 2\text{Å}$, respectively. Panels c.) and d.) are plots of $F_g^{NL}/L$, the free energy for non-alike helices, for the decay lengths $\lambda_{eff} = 4\text{Å}$ and $\lambda_{eff} = 2\text{Å}$, respectively. The red curve is the values of the free energy for the finite $\lambda_h$ state, calculated through Eqs. (G.12) and (G.14) for alike helices and Eqs. (G.11) and (G.13) for non-alike ones. The black curves are the free energy in the state where there is no adaptation, without the correction due to residual azimuthal correlations, where $F_g = F_g^{NL} = -L\Lambda_g G_{0,0}(R/a, 0)$. In the blue curves, we consider the correction to the free energies describing this state from the residual correlations in $\Delta\Phi_T(s)$. For alike helices this is Eq. (F.27), and for non-alike ones we use Eq. (F.48).